\documentclass[%
 reprint,
 amsmath,amssymb, 
 aps,
 prd
tightenlines
]{revtex4-2}

\usepackage{graphicx}
\usepackage{dcolumn}
\usepackage{bm}
\usepackage[hidelinks,colorlinks=true,allcolors=blue]{hyperref}
\usepackage{orcidlink}     

\usepackage{color}
\usepackage{slashed}
\usepackage[utf8]{inputenc}
\usepackage{amsmath,amssymb,multirow,graphicx, subfigure}
\usepackage{amsthm}
\usepackage{braket}
\usepackage[T1]{fontenc}
\usepackage{enumitem}

\theoremstyle{definition}

\theoremstyle{remark}

\newcommand{\calC}{\mathcal{C}}
\newcommand{\calD}{\mathcal{D}}

\newcommand{\calJ}{\mathcal{J}}

\newcommand{\calO}{\mathcal{O}}

\newcommand{\calR}{\mathcal{R}}
\newcommand{\calS}{\mathcal{S}}

\newcommand{\e}{\mathrm{e}}
\renewcommand{\i}{\mathrm{i}}
\renewcommand{\d}{\mathrm{d}}

\newcommand{\tr}{\mathrm{tr}}
\newcommand{\Z}{\mathbb{Z}}

\newcommand{\R}{\mathbb{R}}

\usepackage{stackengine}

\usepackage{titlesec}  

\newcounter{subsubsubsection}[subsubsection]

\titleformat{\paragraph}[block]{\normalfont\normalsize\bfseries}{\theparagraph}{1em}{}

\titlespacing*{\paragraph}{0pt}{1.5ex plus 0.5ex minus .2ex}{0.8ex plus .2ex}

\begin{document}

\title{
Does hot QCD have a conformal manifold in the chiral limit?
}

\author{Shi Chen \orcidlink{0000-0002-8554-5098}}
\email{s.chern.phys@gmail.com}
\affiliation{School of Physics and Astronomy, University of Minnesota, Minneapolis, MN 55455, USA}
\author{Aleksey Cherman \orcidlink{0000-0002-1039-8476}}
\email{acherman@umn.edu}
\affiliation{School of Physics and Astronomy, University of Minnesota, Minneapolis, MN 55455, USA}
\author{Robert D. Pisarski \orcidlink{0000-0002-7862-4759}}
\email{pisarski@bnl.gov}
\affiliation{Physics Department, Brookhaven National Laboratory,
Upton, NY 11973-5000}

\date{\today}

\begin{abstract}
Recent lattice evidence may suggest the chiral phase transition in QCD is second-order for $N_f \ge 2$ massless flavors. 
We constrain CFT descriptions of a critical line in
temperature $T$ and imaginary baryon chemical potential $\theta_B =
i\mu_B/T$.
An 't~Hooft anomaly at
general $\theta_B$ constrains the transition even at $\theta_B = 0$,
leaving only three minimal scenarios. The best-motivated scenario for
$N_f\ge3$, and perhaps also $N_f = 2$, is beyond Ginzburg-Landau,
featuring a conformal manifold of $\theta_B$-dependent universality
classes with an exactly marginal operator related to baryon density.
\end{abstract}

\maketitle

{\bf Introduction.}
Understanding the phase diagram of quantum chromodynamics (QCD) as a
function of temperature $T$ and baryon chemical potential $\mu_B$ is a
central challenge in high-energy physics, with implications for
heavy-ion collisions, astrophysics, and the early universe.  
At $\mu_B=0$, increasing $T$ restores the spontaneously broken
approximate chiral symmetry through a crossover at $T \sim 155\,
\textrm{MeV}$~\cite{HotQCD:2018pds,Borsanyi:2020fev}. If $N_f$ quarks
are massless, chiral symmetry becomes exact, and the crossover becomes a
genuine phase transition. The standard approach to this nonperturbative
phase transition appeals to Landau universality. A Ginzburg-Landau
analysis based on the chiral condensate $\Phi = \bar{\psi}_L
\psi_R$~\cite{Pisarski:1983ms} suggests a generic first-order transition for
$N_f\geq3$, and a second-order transition for $N_f\!=\!2$, in the $O(4)$
Wilson-Fisher universality class.


Early lattice QCD simulations seemed to agree with this prediction~\cite{Brown:1990ev,Karsch:1994hm,Iwasaki:1995ij,JLQCD:1998qth, Karsch:2001nf,Liao:2001en,deForcrand:2003vyj,Jin:2014hea}, but recent improved simulations~\cite{Kuramashi:2020meg,Cuteri:2021ikv,Dini:2021hug,Nakamura:2022abk,Zhang:2022kzb,Zhang:2024ldl,Zhang:2025vns,Klinger:2025xxb,DAmbrosio:2025ldv,Klinger:2026pbe} do not support first order chiral transitions in the chiral limit  for any $N_f$ from $N_f = 2$ up to the onset of the conformal window, at $\mu_B = 0$ and along a line to imaginary $\mu_B$.  These simulations can thus be interpreted as suggesting that the chiral transition is either second-order or perhaps very weakly first-order for all $N_f$.  Both of these scenarios would need an explanation beyond the original Ginzburg-Landau analysis.

\begin{figure*}[ht]
  \centering
  \includegraphics[width=0.3\textwidth]{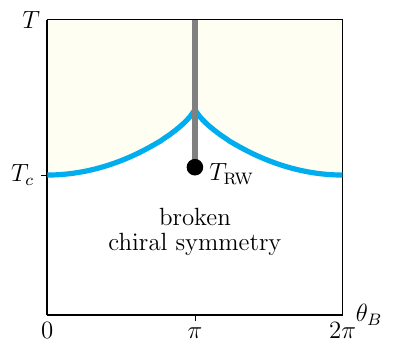}%
  \hfill
  \includegraphics[width=0.3\textwidth]{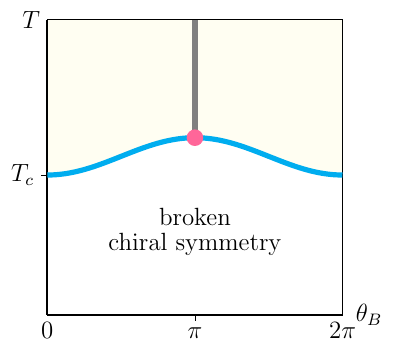}%
  \hfill
  \includegraphics[width=0.3\textwidth]{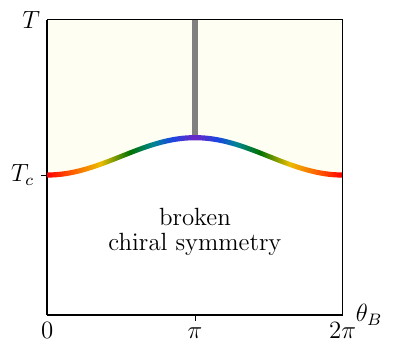}%
  \caption{Minimal phase diagrams of massless QCD consistent with a
   natural second-order line of chiral phase transitions. The gray
   vertical line represents Roberge-Weiss phase transitions of first
   order, while
   $T_c$ is the chiral phase transition temperature at $\theta_B = 0$.
   \textbf{Left}: Landau scenario, where $T_{\rm RW}$ is the
   Roberge-Weiss Ising critical point. \textbf{Middle}: Landau-DQCP
   scenario, where the Roberge-Weiss line terminates at a beyond-Landau
   critical point (pink dot) analogous to a DQCP. \textbf{Right}:
   Conformal-manifold scenario, where the rainbow line represents a
   continuous family of CFTs.
   The conformal-manifold scenario is plausible for all $N_f \ge 2$, while the Landau and the
   Landau-DQCP scenarios are only possible for $N_f=2$ due to the absence of candidate Ginzburg-Landau CFTs for $N_f >2$. 
   }
  \label{fig:allowed_scenarios}  
\end{figure*}

Various theoretical interpretations of the recent lattice results have been proposed~\cite{Fejos:2022mso,Bernhardt:2023hpr,Fejos:2024bgl,Giacosa:2024orp,Pisarski:2024esv}, but compelling solutions remain elusive.
Here we systematically constrain the 3D conformal field theories (CFTs) that can appear on a natural critical line of chiral transitions in the $\theta_B$-$T$ phase diagram, with real $\theta_B= i\mu_B / T$. 
We find only three minimal scenarios compatible with the ’t Hooft anomaly~\cite{Yonekura:2019vyz,Nishimura:2019umw,Kobayashi:2023ajk}, high-$T$ perturbative calculations~\cite{Roberge:1986mm}, and low-$T$ QCD phenomenology, shown in Fig.~\ref{fig:allowed_scenarios}.

In both the Landau (left) and Landau-DQCP (middle) scenarios, the
critical line flows to a single Ginzburg-Landau CFT for generic
$\theta_B$, but the Landau-DQCP scenario also requires an analog of condensed-matter
deconfined quantum critical point
(DQCP)~\cite{Senthil:2003eed,Senthil:2004fuw,Senthil:2023vqd} to appear
at $\theta_B = \pi$. In the conformal-manifold scenario (right), the
critical line is a continuum of distinct beyond-Landau 3D CFTs. The
Landau and Landau-DQCP scenarios require a natural fixed point in the
Ginzburg-Landau description of Ref.~\cite{Pisarski:1983ms} that has not
been found despite extensive searches for $N_f \ge 3$, and thus are
disfavored for generic $N_f$. In contrast, we find that the conformal
manifold scenario is not excluded by existing constraints, and the
existence of a suitable 3d CFT is supported by a $(2+\epsilon)$D
expansion and existing conformal bootstrap bounds for $N_f \ge 2$. So
the answer to the title's question could be \emph{yes}.

{\bf Symmetry and anomaly.}
We work in Euclidean signature throughout. Massless QCD, a 4D $SU(N_c)$
gauge theory for $N_c\ge3$ with $N_f$ massless fundamental Dirac
fermions, has the continuous global symmetry
\begin{align}
\begin{split}
    G = &\:\frac{U(1)_B \times SU(N_f)_L \times SU(N_f)_R }{\Z_{N_f}},
\end{split}
\label{eq:QCD_symmetry}
\end{align}
where $\Z_{N_f}$ sets $(\e^{2\pi i N_c/N_f}, \e^{-2\pi i/N_f}
\mathbf{1}_{N_f}, \e^{-2\pi i/N_f} \mathbf{1}_{N_f})$ to  the identity.
Here $U(1)_B=U(1)_V/\Z_{N_c}$ is the baryon symmetry, while $SU(N_f)_{L}
\times SU(N_f)_R$ is the chiral symmetry. The $U(1)_A$ symmetry of the
classical theory is destroyed by the axial anomaly, and so it is not
part of $G$. The discrete global symmetry acts non-trivially on $G$ and
is generated by charge conjugation $\calC$ and geometric reflections,
including time reversal.

The symmetry $G$ has an 't Hooft anomaly~\cite{tHooft:1979rat}. Turning
on background gauge fields $A_B$, $A_L$, $A_R$ for $U(1)_B$,
$SU(N_f)_L$, $SU(N_f)_R$, respectively, leads to an anomalous background
gauge variation.  Mathematically, this can be described by a 5D
invertible field theory,\footnote{An invertible field theory evaluates
to a pure phase on closed manifolds and has an anomalous variation on
boundaries, encoding the 't~Hooft anomaly of the theory living there.}
see
Refs.~\cite{Freed:2014iua,Kapustin:2014lwa,Witten:2015aba,Yonekura:2016wuc,Bhardwaj:2023kri},
with partition function
\begin{align}\label{eq:SPT_4d}
\begin{gathered}
    \mathcal{I}_{5D}(A_B,A_L,A_R) = \\
    \mathcal{J}_{5D}^{N_c}(A_L)\mathcal{J}_{5D}^{-N_c}(A_R)
    \exp\!\left[\frac{i}{8\pi^2} \!\int_{5D}\!\! A_B \tr( F_L^2 - F_R^2)\right]\!,
\end{gathered}
\end{align}
where $F_{L,R}$ is $A_{L,R}$'s field strength. 
For $N_f \ge 3$, $\calJ_{5D}(A)$ is the 5D Chern-Simons partition
function~\cite{Wess:1971yu}, while for $N_f=2$ it is the 5D spin
topological invariant associated with the Witten
anomaly~\cite{Witten:1982fp,Wang:2018qoy}.

As a renormalization group (RG) invariant, the 't Hooft
anomaly~\eqref{eq:SPT_4d} ensures massless
excitations~\cite{Frishman:1980dq,Coleman:1982yg}, but their nature and
interactions depend on $N_c$ and $N_f$. Below the conformal window (very
roughly $N_f\lesssim\frac{8}{3}N_c$~\cite{LatticeStrongDynamics:2018hun,
Kuti:2022ldb,Ingoldby:2023mtf,Hasenfratz:2023wbr,Hasenfratz:2024fad,Bergner:2025yke}),
the continuous symmetry $G$ is spontaneously broken to 
\begin{equation}
    H = \frac{U(1)_B\times SU(N_f)_V}{\Z_{N_f}}\,,
\end{equation}
while the discrete symmetries remain intact. The gapless IR physics is
described via a nonlinear scalar field
\begin{equation}\label{eq:U(x)}
    U(x) \ \in\ G/H\ \simeq\ SU(N_f)\,.
\end{equation}
In this phase, $\mathcal{J}^{N_c}(A_L)\mathcal{J}^{-N_c}(A_R)$ in
Eq.~\eqref{eq:SPT_4d} is matched by the Wess-Zumino-Witten
action~\cite{Witten:1983tw} for $N_f\ge3$, and by the discrete spin
$\theta$-angle~\cite{Witten:1983tw} for $N_f=2$. The $A_B$ term is
matched by identifying the Skyrme current (from $\pi_3(SU(N_f)) = \Z$)
with the baryon current~\cite{Witten:1983tx}.

{\bf Circle compactification.}
We now put massless QCD on $\R^3\times S^1$ with circumference $\beta$. 
An $S^1$-homogeneous gauge field splits into a gauge field on $\R^3$
and a holonomy scalar field on $\R^3$. In particular, the $U(1)_B$
holonomy gives a $2\pi$-periodic scalar field $\theta_B$.%
\footnote{What directly shows up in the QCD Lagrangian is $U(1)_V$
instead of $U(1)_B$. Thus in the path integral, we have to define
$\theta_B$ via $\theta_B =\theta_VN_c$. Then the shift
$\theta_B\sim\theta_B+2\pi$ is generated by an improper $SU(N_c)$ gauge
transformation $g$ that satisfies $g(x,\tau+\beta) = g(x,\tau)\,\e^{2\pi
i/N_c}$ for $\tau\in S^1$~\cite{Roberge:1986mm}.}
For constant $\theta_B$, the circle compactification has two Hamiltonian
interpretations, and all physical observables, including the free energy
density
\begin{equation}
    \label{eq:free_energy}
    f_{\beta}(\theta_B)\equiv-\lim_{V\to\infty}\frac{1}{\beta V}\ln Z_{\beta}(\theta_B),
\end{equation}
must be \emph{simultaneously} interpretable in both pictures:
\begin{itemize}
    \item \textbf{(Thermal)} Hamiltonian evolution along $S^1$ yields a
    grand canonical ensemble on $\R^3$ with temperature $T=1/\beta$ and
    baryonic chemical potential $\mu_B = -iT\theta_B$. Then
    $-f_{\beta}(\theta_B)$ is the thermal pressure.
    \item \textbf{(Quantum)} Hamiltonian evolution along $\R$ yields a
    zero-temperature quantum system on $\R^2\!\times\!S^1$ with
    circumference $\beta$ and a baryonic Aharonov-Bohm phase
    $e^{i\theta_B}$. Then $f_{\beta}(\theta_B)$ is the Casimir energy
    density.
\end{itemize}
Therefore, with real $\theta_B\neq0$, we lose reflection positivity
reflecting $S^1$, but preserve reflection positivity reflecting an
$\R \subset \R^3$.
The imaginary baryon density at zero Matsubara frequency,
\begin{equation}\label{eq:baryon_density}
    \rho_B(x) \equiv \frac{i}{\beta}\int_{0}^{\beta}\!\!\!\d\tau\, j_B^0(x,\tau)\,,
\end{equation}
is anti-hermitian in the thermal picture, but is hermitian in the quantum picture.

The quantum picture is convenient for describing 3D infrared (IR)
effective theories at length scales $\ell \gg\beta$. These theories are
unitary 3D quantum field theories (QFTs) for any real $\theta_B$, and
$\rho_B$ flows to a hermitian operator thereof. When the 3D effective
theory becomes scale-invariant at a critical point, unitarity is
expected to enhance scale invariance to full conformal invariance; see
e.g.~Refs.~\cite{Polchinski:1987dy,El-Showk:2011xbs,Nakayama:2013is,
Delamotte:2015aaa,Paulos:2015jfa,Meneses:2018xpu}.

Upon $S^1$ compactification, the 5D invertible theory~\eqref{eq:SPT_4d}
reduces to the 4D invertible theory
\begin{align}\label{eq:SPT_3D}
    \mathcal{I}_{4D}(\theta_B,A_L,A_R) =
    \exp\!\left[\frac{i}{8\pi^2}\! \int_{4D}\!\! \theta_B \,\tr (F_L^2 -F_R^2)\right]\!.
\end{align}
It is well-defined on closed 4D oriented manifolds, but is not invariant
under $\theta_B\to\theta_B+2\pi$ if the 4D manifold has a 3D boundary.
Hence, with $S^1$-homogeneous $A_{L,R}$, the partition function of
$S^1$-compactified massless QCD
satisfies~\cite{Yonekura:2019vyz,Nishimura:2019umw,Kobayashi:2023ajk}
\begin{align}\label{eq:anomaly_3D}
\begin{gathered}
    Z_{\beta}(\theta_B + 2\pi, A_L, A_R) \\
    = Z_{\beta}(\theta_B, A_L, A_R) \,\calJ_{3D}(A_L) \,\calJ_{3D}^{-1}(A_R)
\end{gathered}
\end{align}
where $\calJ_{3D}(A)$ is the 3D Chern-Simons partition function
\begin{align}
    \calJ_{3D}(A) =  \exp\left[\frac{\i}{4\pi}\int_{3D}
    \mathrm{tr}\left( A \d A - \i\frac{2}{3} A^3 \right)\right] \,.
\end{align}
This is a mixed anomaly between $SU(N_f)_{L,R}$ and a $2\pi$-periodic
coupling~\cite{Kikuchi:2017pcp,Cordova:2019jnf,Cordova:2019uob,Heidenreich:2021xpr,Garcia-Valdecasas:2024cqn,Yu:2024jtk},
or a ``$(-1)$-form $U(1)$ symmetry''~\cite{Najjar:2024vmm,Aloni:2024jpb,
Lin:2025oml,Robbins:2025urk}.

Upon $S^1$ compactification, the discrete symmetry is generated by
$\calC$, the $S^1$-reflection $\calS$, and $\R$-reflections $\calR$ for
$\R\subset\R^3$. They act on $S^1$-homogeneous 3D background fields as
\begin{equation}
\!\!\!\!\begin{gathered}
    \theta_B(x) \xrightarrow{\calC} -\theta_B(x),\quad A_{L,R}(x) \xrightarrow{\calC} -A_{R,L}^*(x),\\
    \theta_B(x) \xrightarrow{\calS} -\theta_B(x),\quad A_{L,R}(x) \xrightarrow{\calS} A_{R,L}(x),\\
    \theta_B(x) \xrightarrow{\calR}\theta_B(\overline{x}),\quad A_{L,R}(x) \xrightarrow{\calR} \overline{A}_{R,L}(\overline{x}),
\end{gathered}
\end{equation}
where the bars indicate an $\R$-reflection.
$\calC$, $\calS$, and $\calR$ exchange $L$ and $R$.
When $N_f=2$, $\calC\calS$ can be undone by $SU(N_f)_{L,R}$.

A constant $\theta_B\neq0,\pi$ preserves only the subgroup generated by
$\calC\calS$ and $\calR$, whereas $\calS$, $\calC$, $\calS\calR$, and
$\calC\calR$ are restored at $\theta_B=\pi$ by the $2\pi$-periodicity
of $\theta_B$. Therefore, at $\theta_B=\pi$,
Eq.~\eqref{eq:anomaly_3D} gives a mixed anomaly between $SU(N_f)_{L,R}$
and $(\Z_2)_{\calS\calR}$~\cite{Yonekura:2019vyz,Nishimura:2019umw,Kobayashi:2023ajk}.
Equivalently, this is an anomaly of
$SU(N_f)_{L,R}\rtimes(\Z_2)_{\calC\calR}$, or of
$[SU(N_f)_L\times SU(N_f)_R]\rtimes(\Z_2)_{\calS}$ or
$\rtimes(\Z_2)_{\calC}$. The 4D invertible theory~\eqref{eq:SPT_3D} captures
this because, at $\theta_B=\pi$, Eq.~\eqref{eq:SPT_3D} can be defined on
unorientable manifolds or with twists that exchange $L$ and $R$.

{\bf Low-$T$ and high-$T$ Phases.}
Although it does not guarantee gapless modes, the 3D
anomaly~\eqref{eq:SPT_3D} excludes $\theta_B$-uniform trivially gapped
phases, as well as a trivially gapped phase at $\theta_B=\pi$. The 3D IR
physics at length scales $\ell\gg\beta$ must see $\theta_B$ and the
symmetry 
\begin{equation}\label{eq:3D_IR_symmetry}
    \frac{SU(N_f)_L\times SU(N_f)_R}{\Z_{N_f}} 
    \begin{cases}
        \!\rtimes (\Z_2)^3_{\calC,\calS,\calR}, & \theta_B=0,\pi \\
        \!\rtimes (\Z_2)^2_{\calC\calS,\calR}, & \theta_B\neq0,\pi
    \end{cases}
\end{equation}
where discrete symmetry needs to quotient out $(\Z_2)_{\calC\calS}$ for
$N_f=2$. The operator $\rho_B$ of Eq.~\eqref{eq:baryon_density} is
neutral under the second line of Eq.~\eqref{eq:3D_IR_symmetry} but
charged under the first. In contrast, the 3D baryon symmetry $U(1)_B$
and chiral holonomies $\theta_{L,R}$ need not appear in the IR.

At sufficiently low $T$, the symmetry $G$ is expected to remain broken
to $H$ just as at $T=0$, and the IR physics is described by the 3D
reduction of the nonlinear scalar field $U(x)$ of Eq.~\eqref{eq:U(x)}.
The 3D symmetry~\eqref{eq:3D_IR_symmetry} acts on it by $U(x)\to L U
R^{-1}$ for $L,R\in SU(N_f)$ and
\begin{equation}
    U(x) \xrightarrow{\calC} U^T(x)\,,\ 
    U(x) \xrightarrow{\calS} U^{\dag}(x)\,,\ 
    U(x) \xrightarrow{\calR} U^{\dag}(\overline{x})\,.
\end{equation}
When $N_f=2$, $\calC\calS$ is absorbed into $SU(2)_{L,R}$.
The low-$T$ 3D IR effective QFT is given by~\cite{Xu:2011sj,Yonekura:2019vyz}
\begin{equation}\label{eq:lowT_EFT}
\begin{gathered}
    \!\!\!\!\int\!\!\calD U\exp\!\left\{\!-\frac{1}{g^2}\!\!\int\!\!\d^3\!x\,\tr \partial_{a}U\, \partial^{a}\!U^{\dagger}\! + i\!\!\int\!\!\theta_B\frac{\tr(U^{\dagger}\d U)^3}{24\pi^2} \!\right\}\!.\!\!
\end{gathered}
\end{equation}
To verify anomaly matching in Eq.~\eqref{eq:lowT_EFT}, one can turn on
background gauge fields $A_{L,R}$ and study their interplay with the
Skyrme 3-form~\cite{Witten:1983tw,Xu:2011sj,Yonekura:2019vyz}. In this
low-$T$ gapless phase, varying $\theta_B$ does not induce phase
transitions since baryons are heavy.

At sufficiently high $T$, perturbative evaluations of some quantities
are trustworthy; see e.g.~Refs.~\cite{Gross:1980br,KapustaGale201102}.
They imply that chiral symmetry is restored and there is a mass gap for
all $\theta_B$~\cite{Roberge:1986mm}. Roberge and
Weiss~\cite{Roberge:1986mm} pointed out a first-order phase transition
at $\theta_B=\pi$. This is associated with discrete symmetry breaking at
$\theta_B=\pi$,
\begin{equation}
\label{eq:broken_Z2}
    (\Z_2)^3_{\calC,\calS,\calR} \ \ \to\ \ (\Z_2)^2_{\calC\calS,\calR},
\end{equation} 
which needs to quotient out $(\Z_2)_{\calC\calS}$ for $N_f=2$.
The order operator is $\rho_B$ with 
\begin{equation}
    \langle\rho_B\rangle = \frac{\partial f_{\beta}(\theta_B)}{\partial\theta_B}\,.
\end{equation}
In this high-$T$ gapped phase, continuous chiral symmetry flows to an
emergent $2$-form $U(1)$ symmetry~\cite{Gaiotto:2014kfa} and the 3D
anomaly~\eqref{eq:SPT_3D} is matched via symmetry
transmutation~\cite{Seiberg:2025bqy}; we explain this high-$T$ IR
effective theory in the Supplemental Material~\cite[Sec.~S1]{supp}.

In summary, chiral symmetry is spontaneously broken in the low-$T$ phase and restored
in the high-$T$ phase, while the discrete symmetry at $\theta_B=\pi$ is spontaneously
broken%
\footnote{ If a CFT matching the symmetry and anomaly existed, a
second-order transition at $\theta_B=\pi$ without symmetry breaking
could occur at lower $T$ in the high-$T$
phase~\cite{Xu:2011sj,Yonekura:2019vyz,Nishimura:2019umw,Kobayashi:2023ajk};
see the Supplemental Material~\cite[Sec.~S2]{supp} for comments.}
in the high-$T$ phase and restored in the low-$T$ phase. This
complementarity of symmetry realizations is a fingerprint of mixed
anomalies.

{\bf Consistent phase diagrams.}
The low-$T$ and high-$T$ phases are separated by a critical line
$\theta_B \mapsto T_c(\theta_B)$. Motivated by recent lattice
results~\cite{Cuteri:2021ikv,Dini:2021hug,Nakamura:2022abk,Zhang:2022kzb,Zhang:2024ldl,Zhang:2025vns,Klinger:2025xxb,DAmbrosio:2025ldv,Klinger:2026pbe},
we postulate that this line generically consists of natural second-order
points with emergent scale invariance.
The IR physics of such a natural critical point is a 3D unitary CFT whose neutral
sector under the symmetry~\eqref{eq:3D_IR_symmetry} has precisely one
relevant primary scalar operator $\calO_{E}$, to which the $T$
perturbation flows. The operator $\rho_B$ also flows to
$\calO_{E}$ at $\theta_B \neq 0,\pi$, or at $\theta_B = 0,\pi$ if
spontaneous symmetry breaking~\eqref{eq:broken_Z2} occurs. 

The 't Hooft anomaly~\eqref{eq:SPT_3D} prohibits the critical line from
flowing to a single CFT for all $\theta_B$. First, to produce
the behavior~\eqref{eq:anomaly_3D}, the IR physics must change as we
smoothly vary $\theta_B$ to $\theta_B+2\pi$. Second, it must differ at
$\theta_B=0$ and $\pi$, since $\theta_B=\pi$ has an additional mixed anomaly
between $SU(N_f)_{L,R}$ and discrete symmetries, which is absent at
$\theta_B=0$. Therefore, the IR physics must change from
$\theta_B=0$ to $\theta_B=\pi$, either abruptly or gradually. Abrupt
changes correspond to phase transitions at special $\theta_B$, like in
the high-$T$ phase. Gradual changes, as in the low-$T$ phase,
correspond to smooth variations of the CFT data along the critical line.

If the changes in the IR physics are abrupt, a neighborhood of
$\theta_B=0$ on the critical line only cares about the breaking
$G \to H$, and should fit the Landau paradigm, i.e.~be described by a
natural fixed point of the Ginzburg-Landau model of
Ref.~\cite{Pisarski:1983ms}. Minimally, it suffices to have one
phase transition at $\theta_B = \pi$ and a
Ginzburg-Landau description for the rest of the critical line.
If the transition at $\theta_B = \pi$ is first-order ({\bf Landau
scenario}, Fig.~\ref{fig:allowed_scenarios} left), discrete
symmetry breaking~\eqref{eq:broken_Z2} with $\langle\rho_B\rangle\neq0$ occurs at $\theta_B = \pi$. There must then be a
window at $\theta_B=\pi$ where chiral and discrete symmetries are
broken, with an Ising transition at its lower edge.%
\footnote{Such a window is not supported by current lattice evidence
with $2+1$ flavors of massive quarks in QCD, see Fig.~9 of
Ref.~\cite{Bonati:2018fvg} and Fig.~1 of Ref.~\cite{Cuteri:2022vwk}.}
If instead the
transition at $\theta_B = \pi$ is second order ({\bf Landau-DQCP
scenario}, Fig.~\ref{fig:allowed_scenarios} middle), it must be a
beyond-Landau critical point analogous to condensed-matter
DQCPs~\cite{Senthil:2003eed,Senthil:2004fuw,Senthil:2023vqd}. The
operator $\rho_B$ flows to a relevant (or marginally relevant) operator
at this DQCP.

If the changes in the IR physics are gradual, the CFT data instead vary
smoothly with $\theta_B$ on the critical line ({\bf conformal-manifold
scenario}, Fig.~\ref{fig:allowed_scenarios} right). This scenario
requires a conformal manifold --- a continuum of distinct CFTs ---
generated by an exactly marginal operator $\calO_{B}$.%
\footnote{It is natural to expect that a conformal manifold is always generated by a marginal operator.
See e.g.~Ref.~\cite{Komatsu:2025cai} for recent progress toward a proof of this relation.}
The operator
$\rho_B$ flows to $\calO_{B}$ at $\theta_B=0,\pi$, and to a linear
combination of $\calO_{B}$ and $\calO_{E}$ at $\theta_B\neq0,\pi$. The
presence of $\calO_{B}$ makes this scenario intrinsically beyond
Ginzburg-Landau.

One can imagine more complicated scenarios with phase transitions
away from $\theta_B=\pi$, but they generically require unnatural
multicritical points. In any case, the three scenarios above are the
minimal ones under the assumption of a second-order critical line.

{\bf Landau scenarios.}
The Landau and Landau-DQCP scenarios posit that the critical line at
$\theta_B \neq \pi$ is described%
\footnote{ For $\theta_B\neq0$, we should allow chiral-symmetric
$(\Z_2)^2_{\calC\calS,\calR}$-neutral
$(\Z_2)^3_{\calC,\calS,\calR}$-charged deformations, such as $i\lambda_4
\int[\tr (\Phi^\dagger d \Phi)^3 + \text{c.c.}]$. However,
operators in this sector are expected to be irrelevant.}
by a natural fixed point of the 3D linear sigma model of
Ref.~\cite{Pisarski:1983ms}:
\begin{align}\label{eq:Landau-Ginzburg}
\begin{gathered}
    \int\!\calD\Phi\,\exp\Big\{\!-\!\int\!\d^3x\Big[\tr \partial_{a}\Phi\,\partial^{a}\Phi^{\dag} + \mu\,\tr\Phi\Phi^{\dag} \\
    +\, \lambda_1\!\left(\tr\Phi\Phi^{\dag}\right)^{\!2} \!+\lambda_2\tr\!\left(\Phi\Phi^{\dag}\right)^{\!2} \!- \!\lambda_3\left(\det\!\Phi + \text{c.c.}\right)\!\Big]\!\Big\},
\end{gathered}
\end{align}
where $\Phi(x) \sim \bar{\psi}_L \psi_R$ is an $N_f$-by-$N_f$
complex matrix field, with the symmetry~\eqref{eq:3D_IR_symmetry} acting
on $\Phi(x)$ in the same way as $U(x)$. The question is whether
this model hits a natural critical point as $\mu$ varies.%
\footnote{One may speculate that the model~\eqref{eq:Landau-Ginzburg}
flows to a fixed point with emergent $U(1)_A$ symmetry, with the
$\lambda_3$ term
irrelevant~\cite{Pisarski:1983ms,Cohen:1996ng,Lee:1996zy,Cohen:2002st,Fukushima:2008wg,Aoki:2012yj,Pelissetto:2013hqa,Cossu:2013uua}.
However, as discussed in the Supplemental Material~\cite[Sec.~S3]{supp},
no convincing evidence supports this possibility in the
model~\eqref{eq:Landau-Ginzburg}. }

When $N_f=2$, the model~\eqref{eq:Landau-Ginzburg} at mean-field level
undergoes an $O(4)$ second-order transition at $\mu=|\lambda_3|$. This
critical point survives fluctuations, and flows%
\footnote{The symmetry~\eqref{eq:3D_IR_symmetry} is
$O(4)\rtimes(\Z_2)_{\calR}$ at $\theta_B=0,\pi$, and
$SO(4)\rtimes(\Z_2)_{\calR}$ for $\theta_B\neq0,\pi$.}%
to the $O(4)$ Wilson-Fisher universality
class~\cite{Pisarski:1983ms,Butti:2003nu}. Hence the Landau and
Landau-DQCP scenarios are both in principle plausible for $N_f=2$.

At mean-field level, the $\lambda_3$ term drives the transition
first-order for $N_f=3$, but allows an $O(2N_f^2)$ second-order
transition at $\mu=0$ for $N_f\ge4$. Unfortunately, beyond mean-field
level, extensive searches have found no stable fixed points in the
model~\eqref{eq:Landau-Ginzburg} for all $N_f\ge
3$~\cite{Pisarski:1983ms,Butti:2003nu,Calabrese:2004uk,Adzhemyan:2021sug,Fejos:2014qga,Resch:2017vjs,Fejos:2022mso,Fejos:2024bgl}.
We thus consider the Landau and Landau-DQCP scenarios to be disfavored
for $N_f\geq 3$.

{\bf Conformal manifold.}
The only option not yet disfavored for $N_f\ge3$ is the
conformal-manifold scenario, which might work for $N_f=2$ as well. We
denote the CFT at $\theta_B$ on the critical line by
$SU(N_f)_{\theta_B}$. We obtain a non-conformal field theory by adding
the deformation
\begin{equation}
   \mu \int\d^3x\ \calO_E\,,
\end{equation}
to the $SU(N_f)_{\theta_{B}}$ CFT. This $SU(N_f)_{\theta_B}$ field theory must
describe the 3D IR physics of finite-$T$ massless QCD for all $T$. It should
flow to a gapped phase for $\mu>0$ and a Nambu-Goldstone
phase for $\mu<0$. 

Whether the $SU(N_f)_{\theta_B}$ CFT exists is a nontrivial question, since
non-supersymmetric conformal manifolds are rare in
3D~\cite{Gaberdiel:2008fn,Bashmakov:2017rko,Behan:2017mwi,Hollands:2017chb,Sen:2017gfr,Giambrone:2021wsm}.
We now argue for the existence of the $SU(N_f)_0$ CFT that generates the
conformal manifold via $\calO_B$ deformations.

If we can arbitrarily ``contract'' the $\mu<0$ RG flow into its IR end by smoothly
deforming the $SU(N_f)_0$ field theory, we can study a deformed version of the $SU(N_f)_0$ CFT via
loop expansions around the Gaussian IR fixed point. The IR tail of
the $\mu<0$ RG flow is described by the 3D $SU(N_f)$ principal chiral model
(PCM) of Eq.~\eqref{eq:lowT_EFT}.
Conveniently, loop expansions in the $(2+\epsilon)$D $SU(N_f)$ PCM with
arbitrarily small $\epsilon>0$ are known to exhibit an arbitrarily short
RG flow from an interacting ultraviolet (UV) fixed point to the Gaussian
IR fixed point~\cite{ZinnJustin_book}. It is tempting to view this UV
fixed point as a weakly-coupled avatar of the $SU(N_f)_0$ CFT.

The $(2+\epsilon)$D UV fixed point should contain the avatar of the exactly marginal operator
$\calO_B$. The key
observation~\cite{Nahum:2015jya,Jones:2024ept,DeCesare:2025ukl,DeCesare:2026dwm}
is that a $d$-dimensional QFT with $d\in\R_+$ is like an
infinite-dimensional QFT, since it is formally an analytic continuation
of a family of $n$-dimensional QFTs for all $n\in\Z_+$. The
$(2+\epsilon)$D $SU(N_f)$ PCM thus sees the topology of the $SU(N_f)$
target space in all finite dimensions, including $\pi_3(SU(N_f))=\Z$. As
a result, at the $(2+\epsilon)$D UV fixed point, the closed Skyrme
3-form has protected scaling dimension exactly $3$ and generates a short
conformal multiplet~\cite{DeCesare:2025ukl}. The $(2+\epsilon)$D Skyrme
3-form is thus a promising avatar for $\calO_B$.

A large-$N$ analysis~\cite{DeCesare:2025ukl} indicates that if the 3D
$SU(2)_0$ CFT exists, it might be close to the 3D $O(4)$ Wilson-Fisher
CFT, so existing lattice measurements~\cite{Philipsen:2021qji} might not
distinguish them. The idea that fixed points in $2+\epsilon$ and $4-\epsilon$
dimensions could have different extrapolations to 3D has recently been
highlighted by De Cesare and Rychkov~\cite{DeCesare:2025ukl,DeCesare:2026dwm}.

To see whether the $(2+\epsilon)$D UV fixed point extrapolates to
$\epsilon=1$, we study the scaling dimensions of $\calO_E$ and the lowest-dimension
primary scalar $\calO_{U}$ in the
$\boldsymbol{N_f}\otimes\overline{\boldsymbol{N_f}}$ representation using
the three-loop results of
Refs.~\cite{McKane:1979cm,Hikami:1980hi,Hikami:1982ak} and the four-loop results of Ref.~\cite{Wegner:1989ss}:
\begin{subequations}\label{eq:epsilon_expansion_main}
\begin{align}
    \Delta_E &= 2 - \frac{1}{2}\epsilon^2 - \frac{1}{4}\epsilon^3 
    - \left( \frac{5}{16} + \frac{9\zeta_3}{4N_f^2} \right)\epsilon^4
    + \calO(\epsilon^5),\\
    \Delta_{U} &= \left(1-\frac{1}{N_f^2}\right)\left[\epsilon 
    - \frac{1}{2}  \epsilon^2 +  \frac{1}{2} \epsilon^3\right] 
    +\mathcal{O}\left(\epsilon ^4\right),
\end{align}
\end{subequations}
where $\zeta_3$ is Ap\'ery's constant.
We estimate $\Delta_E$ and $\Delta_{U}$ in the 3D $SU(N_f)_0$ CFT using
Pad\'e-Borel resummation.  We
summarize our 3D estimates with small $N_f$ in
Table~\ref{tab:Delta_phi_estimates}, which are consistent with unitarity
$\Delta_E,\Delta_{U} \ge \frac{1}{2}$ and relevance $\Delta_E,\Delta_{U}
< 3$. 

We present our resummation procedure in Supplemental Material~\cite[Sec.~S4]{supp}.
We estimate the resummation uncertainties by imaginary parts of the Borel singularities and variations under different Pad\`e approximants. 
The alternating truncated series $\Delta_U$ has a sizably smaller resummation error bar than $\Delta_E$, because the Borel singularities in the Pad\`e approximants associated with $\Delta_U$ are away from the Borel integral contour.
In any $\epsilon$ expansion, there is no reliable way to estimate the systematic errors in an extrapolation to $\epsilon=1$ from a realistic number of expansion terms, so our results should only be viewed as suggestive.

\begin{table}[h]
    \centering
    \renewcommand{\arraystretch}{1.5}
    \begin{tabular}{|c|c|c|c|c|}
        \hline
        $N_f$ & $2$ & $3$ & $4$ & $5$ \\
        \hline
        $\Delta_E$ & $1.94\pm 0.56$ & $1.76\pm 0.67$ & $1.64\pm 0.71$ & $1.56\pm 0.72$ \\
        \hline
        $\Delta_{U}$ & $0.54 \pm 0.02$ & $0.64 \pm 0.02$ & $0.67 \pm
        0.02$ & $0.69 \pm
        0.02$ \\
        \hline
    \end{tabular}
    \caption{Estimates for $\Delta_E$ and $\Delta_{U}$ in the 3D
    $SU(N_f)_0$ CFT obtained from Pad\'e-Borel resummations of the known
    $(2+\epsilon)$D expansions.}
    \label{tab:Delta_phi_estimates}
\end{table}

There are also some conformal bootstrap~\cite{Poland:2018epd}
studies for 3D CFTs with symmetry $[SU(N_f)_L\times
SU(N_f)_R]/\Z_{N_f}$. For $N_f = 2$, Ref.~\cite{Kos:2015mba} found an
island with $1.64\lesssim \Delta_E\lesssim 1.68$ and $0.517 \lesssim
\Delta_{U} \lesssim 0.521$. For $N_f=3$, Ref.~\cite{Kousvos:2022ewl}
found a rough kink at $\Delta_{U} = 0.62 \pm 0.01$. We view the
apparent agreement at $N_f=2$ and $N_f = 3$ between our $(2 + \epsilon)$D
estimates and the bootstrap estimates as encouraging.

Gathering all the evidence, we consider it promising to conjecture%
\footnote{ The most ambitious conjecture would be that the 3D $SU(N_f)$
PCM~\eqref{eq:lowT_EFT} is asymptotically
safe~\cite{Wilson:1973jj,Weinberg:1976xy} and gives a path-integral
description of the 3D $SU(N_f)_{\theta_B}$ field theory, such that
(schematically) $g_c^{-2}-g^{-2}\sim\mu$ for a critical coupling $g_c$.
}
the existence of 3D $SU(N_f)_{\theta_B}$ CFTs, though further tests are
needed. Assuming that these CFTs exist, it is natural to further
conjecture that QCD flows to these CFTs at the chiral transition, and
thus realizes the conformal-manifold scenario in the chiral limit from
$N_f=2$ to the lower edge of the conformal window.

{\bf Outlook.}
Recent lattice
results~\cite{Cuteri:2021ikv,Dini:2021hug,Nakamura:2022abk,Zhang:2022kzb,
Zhang:2024ldl,Zhang:2025vns,Klinger:2025xxb,DAmbrosio:2025ldv,Klinger:2026pbe}
on the QCD chiral transition motivated us to consider that QCD flows to a
natural second-order critical line for all $N_f$ below the conformal
window.  We found that only three scenarios
(Fig.~\ref{fig:allowed_scenarios}) are consistent with this possibility.
Lattice QCD simulations should distinguish them by measuring the leading
IR behaviors of $\rho_B$ at $\theta_B =0,\pi$ on the critical line, in
Table~\ref{tab:fingerprints}.
\begin{table}[ht]
    \centering
    \renewcommand{\arraystretch}{1.5}
    \begin{tabular}{|c|c|c|c|}
        \hline
         $\theta_B$ & \textbf{Landau} & \textbf{Landau-DQCP} & \textbf{Conformal-manifold}  \\
        \hline
        $0$ & irrelevant & irrelevant & exactly marginal \\ \hline
        $\pi$ & $\langle\rho_B\rangle\neq0$ & relevant & exactly marginal \\ \hline
    \end{tabular}
    \caption{The leading IR behaviors of the operator $\rho_B$ at
    $\theta_B =0,\pi$ on the critical line.}
    \label{tab:fingerprints}
\end{table}

The conformal-manifold scenario is the only option not disfavored for
$N_f\ge3$ by current evidence. However, it requires the existence of the
$SU(N_f)_{\theta_B}$ universality class, which is an open question of
independent interest. Demonstrating it will require improved conformal
bootstrap bounds on 3D CFTs with the symmetry~\eqref{eq:3D_IR_symmetry}
and further development of the $(2+\epsilon)$D expansion. Lattice Monte
Carlo simulations of the 3D $SU(N_f)$ PCM using L\"uscher's
admissibility
method~\cite{Luscher:1981zq,Schramm:2000cc,Shiozaki:2024yrm,Lopez-Contreras:2025iik}
(see also Ref.~\cite{Chen:2024ddr}) might be valuable.

While we assume a second-order critical line, a \emph{weakly}
first-order transition is not excluded by
Refs.~\cite{Cuteri:2021ikv,Dini:2021hug,Nakamura:2022abk,
Zhang:2022kzb,Zhang:2024ldl,Zhang:2025vns,Klinger:2025xxb,
DAmbrosio:2025ldv,Klinger:2026pbe}. If such transitions are controlled
by nearby non-unitary complex
CFTs~\cite{Gorbenko:2018ncu,Gorbenko:2018dtm}, such as the
$(2+\epsilon)$D fixed points, much of our discussion would carry over.


{\bf Acknowledgments.}
We are very grateful to Jorge DaSilva Golan, Thomas Dumitrescu, Amey
Gaikwad, Evgenii Ievlev, Theodore Jacobson, David B. Kaplan, Joseph
Kapusta, Zohar Komargodski, Yacine Mehtar-Tani, Swagato Mukherjee, Peter
Petreczky, Owe Philipsen, Jiaxin Qiao, Srimoyee Sen, Laurence Yaffe, and
Yuan Xin for very helpful discussions. OpenAI's Codex 5.4, Anthropic's
Claude Opus 4.6, and Refine.ink were helpful in proofreading the
manuscript. A.C. and S.C. are supported by Simons Foundation award
number 994302, as part of the Simons Collaboration on Confinement and
QCD Strings. R.D.P. is supported by the U.S. Department of Energy under
contract DE-SC0012704, and by the Alexander von Humboldt Foundation.
Our collaboration began at the meeting for the Simons Collaboration on
"Confinement and QCD Strings", Nov. 10-12, 2025.

\bibliography{non_inv}

\begin{thebibliography}{126}%
\makeatletter
\providecommand \@ifxundefined [1]{%
 \@ifx{#1\undefined}
}%
\providecommand \@ifnum [1]{%
 \ifnum #1\expandafter \@firstoftwo
 \else \expandafter \@secondoftwo
 \fi
}%
\providecommand \@ifx [1]{%
 \ifx #1\expandafter \@firstoftwo
 \else \expandafter \@secondoftwo
 \fi
}%
\providecommand \natexlab [1]{#1}%
\providecommand \enquote  [1]{``#1''}%
\providecommand \bibnamefont  [1]{#1}%
\providecommand \bibfnamefont [1]{#1}%
\providecommand \citenamefont [1]{#1}%
\providecommand \href@noop [0]{\@secondoftwo}%
\providecommand \href [0]{\begingroup \@sanitize@url \@href}%
\providecommand \@href[1]{\@@startlink{#1}\@@href}%
\providecommand \@@href[1]{\endgroup#1\@@endlink}%
\providecommand \@sanitize@url [0]{\catcode `\\12\catcode `\$12\catcode
  `\&12\catcode `\#12\catcode `\^12\catcode `\_12\catcode `\%12\relax}%
\providecommand \@@startlink[1]{}%
\providecommand \@@endlink[0]{}%
\providecommand \url  [0]{\begingroup\@sanitize@url \@url }%
\providecommand \@url [1]{\endgroup\@href {#1}{\urlprefix }}%
\providecommand \urlprefix  [0]{URL }%
\providecommand \Eprint [0]{\href }%
\providecommand \doibase [0]{https://doi.org/}%
\providecommand \selectlanguage [0]{\@gobble}%
\providecommand \bibinfo  [0]{\@secondoftwo}%
\providecommand \bibfield  [0]{\@secondoftwo}%
\providecommand \translation [1]{[#1]}%
\providecommand \BibitemOpen [0]{}%
\providecommand \bibitemStop [0]{}%
\providecommand \bibitemNoStop [0]{.\EOS\space}%
\providecommand \EOS [0]{\spacefactor3000\relax}%
\providecommand \BibitemShut  [1]{\csname bibitem#1\endcsname}%
\let\auto@bib@innerbib\@empty
\bibitem [{\citenamefont {Bazavov}\ \emph {et~al.}(2019)\citenamefont {Bazavov}
  \emph {et~al.}}]{HotQCD:2018pds}%
  \BibitemOpen
  \bibfield  {author} {\bibinfo {author} {\bibfnamefont {A.}~\bibnamefont
  {Bazavov}} \emph {et~al.} (\bibinfo {collaboration} {HotQCD}),\ }\bibfield
  {title} {\bibinfo {title} {{Chiral crossover in QCD at zero and non-zero
  chemical potentials}},\ }\href
  {https://doi.org/10.1016/j.physletb.2019.05.013} {\bibfield  {journal}
  {\bibinfo  {journal} {Phys. Lett. B}\ }\textbf {\bibinfo {volume} {795}},\
  \bibinfo {pages} {15} (\bibinfo {year} {2019})},\ \Eprint
  {https://arxiv.org/abs/1812.08235} {arXiv:1812.08235 [hep-lat]} \BibitemShut
  {NoStop}%
\bibitem [{\citenamefont {Borsanyi}\ \emph {et~al.}(2020)\citenamefont
  {Borsanyi}, \citenamefont {Fodor}, \citenamefont {Guenther}, \citenamefont
  {Kara}, \citenamefont {Katz}, \citenamefont {Parotto}, \citenamefont
  {Pasztor}, \citenamefont {Ratti},\ and\ \citenamefont
  {Szabo}}]{Borsanyi:2020fev}%
  \BibitemOpen
  \bibfield  {author} {\bibinfo {author} {\bibfnamefont {S.}~\bibnamefont
  {Borsanyi}}, \bibinfo {author} {\bibfnamefont {Z.}~\bibnamefont {Fodor}},
  \bibinfo {author} {\bibfnamefont {J.~N.}\ \bibnamefont {Guenther}}, \bibinfo
  {author} {\bibfnamefont {R.}~\bibnamefont {Kara}}, \bibinfo {author}
  {\bibfnamefont {S.~D.}\ \bibnamefont {Katz}}, \bibinfo {author}
  {\bibfnamefont {P.}~\bibnamefont {Parotto}}, \bibinfo {author} {\bibfnamefont
  {A.}~\bibnamefont {Pasztor}}, \bibinfo {author} {\bibfnamefont
  {C.}~\bibnamefont {Ratti}},\ and\ \bibinfo {author} {\bibfnamefont {K.~K.}\
  \bibnamefont {Szabo}},\ }\bibfield  {title} {\bibinfo {title} {{QCD Crossover
  at Finite Chemical Potential from Lattice Simulations}},\ }\href
  {https://doi.org/10.1103/PhysRevLett.125.052001} {\bibfield  {journal}
  {\bibinfo  {journal} {Phys. Rev. Lett.}\ }\textbf {\bibinfo {volume} {125}},\
  \bibinfo {pages} {052001} (\bibinfo {year} {2020})},\ \Eprint
  {https://arxiv.org/abs/2002.02821} {arXiv:2002.02821 [hep-lat]} \BibitemShut
  {NoStop}%
\bibitem [{\citenamefont {Pisarski}\ and\ \citenamefont
  {Wilczek}(1984)}]{Pisarski:1983ms}%
  \BibitemOpen
  \bibfield  {author} {\bibinfo {author} {\bibfnamefont {R.~D.}\ \bibnamefont
  {Pisarski}}\ and\ \bibinfo {author} {\bibfnamefont {F.}~\bibnamefont
  {Wilczek}},\ }\bibfield  {title} {\bibinfo {title} {{Remarks on the Chiral
  Phase Transition in Chromodynamics}},\ }\href
  {https://doi.org/10.1103/PhysRevD.29.338} {\bibfield  {journal} {\bibinfo
  {journal} {Phys. Rev. D}\ }\textbf {\bibinfo {volume} {29}},\ \bibinfo
  {pages} {338} (\bibinfo {year} {1984})}\BibitemShut {NoStop}%
\bibitem [{\citenamefont {Brown}\ \emph {et~al.}(1990)\citenamefont {Brown},
  \citenamefont {Butler}, \citenamefont {Chen}, \citenamefont {Christ},
  \citenamefont {Dong}, \citenamefont {Schaffer}, \citenamefont {Unger},\ and\
  \citenamefont {Vaccarino}}]{Brown:1990ev}%
  \BibitemOpen
  \bibfield  {author} {\bibinfo {author} {\bibfnamefont {F.~R.}\ \bibnamefont
  {Brown}}, \bibinfo {author} {\bibfnamefont {F.~P.}\ \bibnamefont {Butler}},
  \bibinfo {author} {\bibfnamefont {H.}~\bibnamefont {Chen}}, \bibinfo {author}
  {\bibfnamefont {N.~H.}\ \bibnamefont {Christ}}, \bibinfo {author}
  {\bibfnamefont {Z.-h.}\ \bibnamefont {Dong}}, \bibinfo {author}
  {\bibfnamefont {W.}~\bibnamefont {Schaffer}}, \bibinfo {author}
  {\bibfnamefont {L.~I.}\ \bibnamefont {Unger}},\ and\ \bibinfo {author}
  {\bibfnamefont {A.}~\bibnamefont {Vaccarino}},\ }\bibfield  {title} {\bibinfo
  {title} {{On the existence of a phase transition for QCD with three light
  quarks}},\ }\href {https://doi.org/10.1103/PhysRevLett.65.2491} {\bibfield
  {journal} {\bibinfo  {journal} {Phys. Rev. Lett.}\ }\textbf {\bibinfo
  {volume} {65}},\ \bibinfo {pages} {2491} (\bibinfo {year}
  {1990})}\BibitemShut {NoStop}%
\bibitem [{\citenamefont {Karsch}\ and\ \citenamefont
  {Laermann}(1994)}]{Karsch:1994hm}%
  \BibitemOpen
  \bibfield  {author} {\bibinfo {author} {\bibfnamefont {F.}~\bibnamefont
  {Karsch}}\ and\ \bibinfo {author} {\bibfnamefont {E.}~\bibnamefont
  {Laermann}},\ }\bibfield  {title} {\bibinfo {title} {{Susceptibilities, the
  specific heat and a cumulant in two flavor QCD}},\ }\href
  {https://doi.org/10.1103/PhysRevD.50.6954} {\bibfield  {journal} {\bibinfo
  {journal} {Phys. Rev. D}\ }\textbf {\bibinfo {volume} {50}},\ \bibinfo
  {pages} {6954} (\bibinfo {year} {1994})},\ \Eprint
  {https://arxiv.org/abs/hep-lat/9406008} {arXiv:hep-lat/9406008} \BibitemShut
  {NoStop}%
\bibitem [{\citenamefont {Iwasaki}\ \emph {et~al.}(1996)\citenamefont
  {Iwasaki}, \citenamefont {Kanaya}, \citenamefont {Sakai},\ and\ \citenamefont
  {Yoshie}}]{Iwasaki:1995ij}%
  \BibitemOpen
  \bibfield  {author} {\bibinfo {author} {\bibfnamefont {Y.}~\bibnamefont
  {Iwasaki}}, \bibinfo {author} {\bibfnamefont {K.}~\bibnamefont {Kanaya}},
  \bibinfo {author} {\bibfnamefont {S.}~\bibnamefont {Sakai}},\ and\ \bibinfo
  {author} {\bibfnamefont {T.}~\bibnamefont {Yoshie}},\ }\bibfield  {title}
  {\bibinfo {title} {{Chiral phase transition in lattice QCD with Wilson
  quarks}},\ }\href {https://doi.org/10.1007/s002880050179} {\bibfield
  {journal} {\bibinfo  {journal} {Z. Phys. C}\ }\textbf {\bibinfo {volume}
  {71}},\ \bibinfo {pages} {337} (\bibinfo {year} {1996})},\ \Eprint
  {https://arxiv.org/abs/hep-lat/9504019} {arXiv:hep-lat/9504019} \BibitemShut
  {NoStop}%
\bibitem [{\citenamefont {Aoki}\ \emph {et~al.}(1998)\citenamefont {Aoki} \emph
  {et~al.}}]{JLQCD:1998qth}%
  \BibitemOpen
  \bibfield  {author} {\bibinfo {author} {\bibfnamefont {S.}~\bibnamefont
  {Aoki}} \emph {et~al.} (\bibinfo {collaboration} {JLQCD}),\ }\bibfield
  {title} {\bibinfo {title} {{Scaling study of the two flavor chiral phase
  transition with the Kogut-Susskind quark action in lattice QCD}},\ }\href
  {https://doi.org/10.1103/PhysRevD.57.3910} {\bibfield  {journal} {\bibinfo
  {journal} {Phys. Rev. D}\ }\textbf {\bibinfo {volume} {57}},\ \bibinfo
  {pages} {3910} (\bibinfo {year} {1998})},\ \Eprint
  {https://arxiv.org/abs/hep-lat/9710048} {arXiv:hep-lat/9710048} \BibitemShut
  {NoStop}%
\bibitem [{\citenamefont {Karsch}\ \emph {et~al.}(2001)\citenamefont {Karsch},
  \citenamefont {Laermann},\ and\ \citenamefont {Schmidt}}]{Karsch:2001nf}%
  \BibitemOpen
  \bibfield  {author} {\bibinfo {author} {\bibfnamefont {F.}~\bibnamefont
  {Karsch}}, \bibinfo {author} {\bibfnamefont {E.}~\bibnamefont {Laermann}},\
  and\ \bibinfo {author} {\bibfnamefont {C.}~\bibnamefont {Schmidt}},\
  }\bibfield  {title} {\bibinfo {title} {{The Chiral critical point in
  three-flavor QCD}},\ }\href {https://doi.org/10.1016/S0370-2693(01)01114-5}
  {\bibfield  {journal} {\bibinfo  {journal} {Phys. Lett. B}\ }\textbf
  {\bibinfo {volume} {520}},\ \bibinfo {pages} {41} (\bibinfo {year} {2001})},\
  \Eprint {https://arxiv.org/abs/hep-lat/0107020} {arXiv:hep-lat/0107020}
  \BibitemShut {NoStop}%
\bibitem [{\citenamefont {Liao}(2002)}]{Liao:2001en}%
  \BibitemOpen
  \bibfield  {author} {\bibinfo {author} {\bibfnamefont {X.}~\bibnamefont
  {Liao}},\ }\bibfield  {title} {\bibinfo {title} {{Study of 3 flavor QCD
  finite temperature phase transition with staggered fermions}},\ }\href
  {https://doi.org/10.1016/S0920-5632(01)01735-2} {\bibfield  {journal}
  {\bibinfo  {journal} {Nucl. Phys. B Proc. Suppl.}\ }\textbf {\bibinfo
  {volume} {106}},\ \bibinfo {pages} {426} (\bibinfo {year} {2002})},\ \Eprint
  {https://arxiv.org/abs/hep-lat/0111013} {arXiv:hep-lat/0111013} \BibitemShut
  {NoStop}%
\bibitem [{\citenamefont {de~Forcrand}\ and\ \citenamefont
  {Philipsen}(2003)}]{deForcrand:2003vyj}%
  \BibitemOpen
  \bibfield  {author} {\bibinfo {author} {\bibfnamefont {P.}~\bibnamefont
  {de~Forcrand}}\ and\ \bibinfo {author} {\bibfnamefont {O.}~\bibnamefont
  {Philipsen}},\ }\bibfield  {title} {\bibinfo {title} {{The QCD phase diagram
  for three degenerate flavors and small baryon density}},\ }\href
  {https://doi.org/10.1016/j.nuclphysb.2003.09.005} {\bibfield  {journal}
  {\bibinfo  {journal} {Nucl. Phys. B}\ }\textbf {\bibinfo {volume} {673}},\
  \bibinfo {pages} {170} (\bibinfo {year} {2003})},\ \Eprint
  {https://arxiv.org/abs/hep-lat/0307020} {arXiv:hep-lat/0307020} \BibitemShut
  {NoStop}%
\bibitem [{\citenamefont {Jin}\ \emph {et~al.}(2015)\citenamefont {Jin},
  \citenamefont {Kuramashi}, \citenamefont {Nakamura}, \citenamefont {Takeda},\
  and\ \citenamefont {Ukawa}}]{Jin:2014hea}%
  \BibitemOpen
  \bibfield  {author} {\bibinfo {author} {\bibfnamefont {X.-Y.}\ \bibnamefont
  {Jin}}, \bibinfo {author} {\bibfnamefont {Y.}~\bibnamefont {Kuramashi}},
  \bibinfo {author} {\bibfnamefont {Y.}~\bibnamefont {Nakamura}}, \bibinfo
  {author} {\bibfnamefont {S.}~\bibnamefont {Takeda}},\ and\ \bibinfo {author}
  {\bibfnamefont {A.}~\bibnamefont {Ukawa}},\ }\bibfield  {title} {\bibinfo
  {title} {{Critical endpoint of the finite temperature phase transition for
  three flavor QCD}},\ }\href {https://doi.org/10.1103/PhysRevD.91.014508}
  {\bibfield  {journal} {\bibinfo  {journal} {Phys. Rev. D}\ }\textbf {\bibinfo
  {volume} {91}},\ \bibinfo {pages} {014508} (\bibinfo {year} {2015})},\
  \Eprint {https://arxiv.org/abs/1411.7461} {arXiv:1411.7461 [hep-lat]}
  \BibitemShut {NoStop}%
\bibitem [{\citenamefont {Kuramashi}\ \emph {et~al.}(2020)\citenamefont
  {Kuramashi}, \citenamefont {Nakamura}, \citenamefont {Ohno},\ and\
  \citenamefont {Takeda}}]{Kuramashi:2020meg}%
  \BibitemOpen
  \bibfield  {author} {\bibinfo {author} {\bibfnamefont {Y.}~\bibnamefont
  {Kuramashi}}, \bibinfo {author} {\bibfnamefont {Y.}~\bibnamefont {Nakamura}},
  \bibinfo {author} {\bibfnamefont {H.}~\bibnamefont {Ohno}},\ and\ \bibinfo
  {author} {\bibfnamefont {S.}~\bibnamefont {Takeda}},\ }\bibfield  {title}
  {\bibinfo {title} {{Nature of the phase transition for finite temperature
  $N_{\rm f}=3$ QCD with nonperturbatively O($a$) improved Wilson fermions at
  $N_{\rm t}=12$}},\ }\href {https://doi.org/10.1103/PhysRevD.101.054509}
  {\bibfield  {journal} {\bibinfo  {journal} {Phys. Rev. D}\ }\textbf {\bibinfo
  {volume} {101}},\ \bibinfo {pages} {054509} (\bibinfo {year} {2020})},\
  \Eprint {https://arxiv.org/abs/2001.04398} {arXiv:2001.04398 [hep-lat]}
  \BibitemShut {NoStop}%
\bibitem [{\citenamefont {Cuteri}\ \emph {et~al.}(2021)\citenamefont {Cuteri},
  \citenamefont {Philipsen},\ and\ \citenamefont {Sciarra}}]{Cuteri:2021ikv}%
  \BibitemOpen
  \bibfield  {author} {\bibinfo {author} {\bibfnamefont {F.}~\bibnamefont
  {Cuteri}}, \bibinfo {author} {\bibfnamefont {O.}~\bibnamefont {Philipsen}},\
  and\ \bibinfo {author} {\bibfnamefont {A.}~\bibnamefont {Sciarra}},\
  }\bibfield  {title} {\bibinfo {title} {{On the order of the QCD chiral phase
  transition for different numbers of quark flavours}},\ }\href
  {https://doi.org/10.1007/JHEP11(2021)141} {\bibfield  {journal} {\bibinfo
  {journal} {JHEP}\ }\textbf {\bibinfo {volume} {11}},\ \bibinfo {pages}
  {141}},\ \Eprint {https://arxiv.org/abs/2107.12739} {arXiv:2107.12739
  [hep-lat]} \BibitemShut {NoStop}%
\bibitem [{\citenamefont {Dini}\ \emph {et~al.}(2022)\citenamefont {Dini},
  \citenamefont {Hegde}, \citenamefont {Karsch}, \citenamefont {Lahiri},
  \citenamefont {Schmidt},\ and\ \citenamefont {Sharma}}]{Dini:2021hug}%
  \BibitemOpen
  \bibfield  {author} {\bibinfo {author} {\bibfnamefont {L.}~\bibnamefont
  {Dini}}, \bibinfo {author} {\bibfnamefont {P.}~\bibnamefont {Hegde}},
  \bibinfo {author} {\bibfnamefont {F.}~\bibnamefont {Karsch}}, \bibinfo
  {author} {\bibfnamefont {A.}~\bibnamefont {Lahiri}}, \bibinfo {author}
  {\bibfnamefont {C.}~\bibnamefont {Schmidt}},\ and\ \bibinfo {author}
  {\bibfnamefont {S.}~\bibnamefont {Sharma}},\ }\bibfield  {title} {\bibinfo
  {title} {{Chiral phase transition in three-flavor QCD from lattice QCD}},\
  }\href {https://doi.org/10.1103/PhysRevD.105.034510} {\bibfield  {journal}
  {\bibinfo  {journal} {Phys. Rev. D}\ }\textbf {\bibinfo {volume} {105}},\
  \bibinfo {pages} {034510} (\bibinfo {year} {2022})},\ \Eprint
  {https://arxiv.org/abs/2111.12599} {arXiv:2111.12599 [hep-lat]} \BibitemShut
  {NoStop}%
\bibitem [{\citenamefont {Nakamura}\ \emph {et~al.}(2022)\citenamefont
  {Nakamura}, \citenamefont {Aoki}, \citenamefont {Hashimoto}, \citenamefont
  {Kanamori}, \citenamefont {Kaneko},\ and\ \citenamefont
  {Zhang}}]{Nakamura:2022abk}%
  \BibitemOpen
  \bibfield  {author} {\bibinfo {author} {\bibfnamefont {Y.}~\bibnamefont
  {Nakamura}}, \bibinfo {author} {\bibfnamefont {Y.}~\bibnamefont {Aoki}},
  \bibinfo {author} {\bibfnamefont {S.}~\bibnamefont {Hashimoto}}, \bibinfo
  {author} {\bibfnamefont {I.}~\bibnamefont {Kanamori}}, \bibinfo {author}
  {\bibfnamefont {T.}~\bibnamefont {Kaneko}},\ and\ \bibinfo {author}
  {\bibfnamefont {Y.}~\bibnamefont {Zhang}},\ }\bibfield  {title} {\bibinfo
  {title} {{Finite temperature phase transition for three flavor QCD with
  M{\"o}bius-domain wall fermions}},\ }\href
  {https://doi.org/10.22323/1.396.0080} {\bibfield  {journal} {\bibinfo
  {journal} {PoS}\ }\textbf {\bibinfo {volume} {LATTICE2021}},\ \bibinfo
  {pages} {080} (\bibinfo {year} {2022})}\BibitemShut {NoStop}%
\bibitem [{\citenamefont {Zhang}\ \emph {et~al.}(2023)\citenamefont {Zhang},
  \citenamefont {Aoki}, \citenamefont {Hashimoto}, \citenamefont {Kanamori},
  \citenamefont {Kaneko},\ and\ \citenamefont {Nakamura}}]{Zhang:2022kzb}%
  \BibitemOpen
  \bibfield  {author} {\bibinfo {author} {\bibfnamefont {Y.}~\bibnamefont
  {Zhang}}, \bibinfo {author} {\bibfnamefont {Y.}~\bibnamefont {Aoki}},
  \bibinfo {author} {\bibfnamefont {S.}~\bibnamefont {Hashimoto}}, \bibinfo
  {author} {\bibfnamefont {I.}~\bibnamefont {Kanamori}}, \bibinfo {author}
  {\bibfnamefont {T.}~\bibnamefont {Kaneko}},\ and\ \bibinfo {author}
  {\bibfnamefont {Y.}~\bibnamefont {Nakamura}},\ }\bibfield  {title} {\bibinfo
  {title} {{Finite temperature QCD phase transition with 3 flavors of Mobius
  domain wall fermions}},\ }\href {https://doi.org/10.22323/1.430.0197}
  {\bibfield  {journal} {\bibinfo  {journal} {PoS}\ }\textbf {\bibinfo {volume}
  {LATTICE2022}},\ \bibinfo {pages} {197} (\bibinfo {year} {2023})},\ \Eprint
  {https://arxiv.org/abs/2212.10021} {arXiv:2212.10021 [hep-lat]} \BibitemShut
  {NoStop}%
\bibitem [{\citenamefont {Zhang}\ \emph {et~al.}(2024)\citenamefont {Zhang},
  \citenamefont {Aoki}, \citenamefont {Hashimoto}, \citenamefont {Kanamori},
  \citenamefont {Kaneko},\ and\ \citenamefont {Nakamura}}]{Zhang:2024ldl}%
  \BibitemOpen
  \bibfield  {author} {\bibinfo {author} {\bibfnamefont {Y.}~\bibnamefont
  {Zhang}}, \bibinfo {author} {\bibfnamefont {Y.}~\bibnamefont {Aoki}},
  \bibinfo {author} {\bibfnamefont {S.}~\bibnamefont {Hashimoto}}, \bibinfo
  {author} {\bibfnamefont {I.}~\bibnamefont {Kanamori}}, \bibinfo {author}
  {\bibfnamefont {T.}~\bibnamefont {Kaneko}},\ and\ \bibinfo {author}
  {\bibfnamefont {Y.}~\bibnamefont {Nakamura}},\ }\bibfield  {title} {\bibinfo
  {title} {{Exploring the QCD phase diagram with three flavors of M{\"o}bius
  domain wall fermions}},\ }\href {https://doi.org/10.22323/1.453.0203}
  {\bibfield  {journal} {\bibinfo  {journal} {PoS}\ }\textbf {\bibinfo {volume}
  {LATTICE2023}},\ \bibinfo {pages} {203} (\bibinfo {year} {2024})},\ \Eprint
  {https://arxiv.org/abs/2401.05066} {arXiv:2401.05066 [hep-lat]} \BibitemShut
  {NoStop}%
\bibitem [{\citenamefont {Zhang}\ \emph {et~al.}(2025)\citenamefont {Zhang},
  \citenamefont {Aoki}, \citenamefont {Hashimoto}, \citenamefont {Kanamori},
  \citenamefont {Kaneko},\ and\ \citenamefont {Nakamura}}]{Zhang:2025vns}%
  \BibitemOpen
  \bibfield  {author} {\bibinfo {author} {\bibfnamefont {Y.}~\bibnamefont
  {Zhang}}, \bibinfo {author} {\bibfnamefont {Y.}~\bibnamefont {Aoki}},
  \bibinfo {author} {\bibfnamefont {S.}~\bibnamefont {Hashimoto}}, \bibinfo
  {author} {\bibfnamefont {I.}~\bibnamefont {Kanamori}}, \bibinfo {author}
  {\bibfnamefont {T.}~\bibnamefont {Kaneko}},\ and\ \bibinfo {author}
  {\bibfnamefont {Y.}~\bibnamefont {Nakamura}},\ }\bibfield  {title} {\bibinfo
  {title} {{Three flavor QCD phase transition with Mobius domain wall
  fermions}},\ }\href {https://doi.org/10.22323/1.466.0193} {\bibfield
  {journal} {\bibinfo  {journal} {PoS}\ }\textbf {\bibinfo {volume}
  {LATTICE2024}},\ \bibinfo {pages} {193} (\bibinfo {year} {2025})},\ \Eprint
  {https://arxiv.org/abs/2501.15494} {arXiv:2501.15494 [hep-lat]} \BibitemShut
  {NoStop}%
\bibitem [{\citenamefont {Klinger}\ \emph {et~al.}(2025)\citenamefont
  {Klinger}, \citenamefont {Kaiser},\ and\ \citenamefont
  {Philipsen}}]{Klinger:2025xxb}%
  \BibitemOpen
  \bibfield  {author} {\bibinfo {author} {\bibfnamefont {J.~P.}\ \bibnamefont
  {Klinger}}, \bibinfo {author} {\bibfnamefont {R.}~\bibnamefont {Kaiser}},\
  and\ \bibinfo {author} {\bibfnamefont {O.}~\bibnamefont {Philipsen}},\
  }\bibfield  {title} {\bibinfo {title} {{The order of the chiral phase
  transition in massless many-flavour lattice QCD}},\ }\href
  {https://doi.org/10.22323/1.466.0172} {\bibfield  {journal} {\bibinfo
  {journal} {PoS}\ }\textbf {\bibinfo {volume} {LATTICE2024}},\ \bibinfo
  {pages} {172} (\bibinfo {year} {2025})},\ \Eprint
  {https://arxiv.org/abs/2501.19251} {arXiv:2501.19251 [hep-lat]} \BibitemShut
  {NoStop}%
\bibitem [{\citenamefont {D'Ambrosio}\ \emph {et~al.}(2025)\citenamefont
  {D'Ambrosio}, \citenamefont {Fromm}, \citenamefont {Kaiser},\ and\
  \citenamefont {Philipsen}}]{DAmbrosio:2025ldv}%
  \BibitemOpen
  \bibfield  {author} {\bibinfo {author} {\bibfnamefont {A.}~\bibnamefont
  {D'Ambrosio}}, \bibinfo {author} {\bibfnamefont {M.}~\bibnamefont {Fromm}},
  \bibinfo {author} {\bibfnamefont {R.}~\bibnamefont {Kaiser}},\ and\ \bibinfo
  {author} {\bibfnamefont {O.}~\bibnamefont {Philipsen}},\ }\bibfield  {title}
  {\bibinfo {title} {{On the nature of the QCD chiral phase transition with
  imaginary chemical potential}},\ }\Eprint {https://arxiv.org/abs/2512.15418}
  {arXiv:2512.15418 [hep-lat]}  (\bibinfo {year} {2025}),\ \bibinfo {note}
  {preprint}\BibitemShut {NoStop}%
\bibitem [{\citenamefont {Klinger}\ \emph {et~al.}(2026)\citenamefont
  {Klinger}, \citenamefont {Kaiser}, \citenamefont {Philipsen},\ and\
  \citenamefont {Schaible}}]{Klinger:2026pbe}%
  \BibitemOpen
  \bibfield  {author} {\bibinfo {author} {\bibfnamefont {J.~P.}\ \bibnamefont
  {Klinger}}, \bibinfo {author} {\bibfnamefont {R.}~\bibnamefont {Kaiser}},
  \bibinfo {author} {\bibfnamefont {O.}~\bibnamefont {Philipsen}},\ and\
  \bibinfo {author} {\bibfnamefont {J.}~\bibnamefont {Schaible}},\ }\bibfield
  {title} {\bibinfo {title} {{On the phase structure of massless many-flavour
  QCD with staggered fermions}}\ }(\bibinfo {year} {2026})\ \Eprint
  {https://arxiv.org/abs/2603.20099} {arXiv:2603.20099 [hep-lat]} \BibitemShut
  {NoStop}%
\bibitem [{\citenamefont {Fejos}(2022)}]{Fejos:2022mso}%
  \BibitemOpen
  \bibfield  {author} {\bibinfo {author} {\bibfnamefont {G.}~\bibnamefont
  {Fejos}},\ }\bibfield  {title} {\bibinfo {title} {{Second-order chiral phase
  transition in three-flavor quantum chromodynamics?}},\ }\href
  {https://doi.org/10.1103/PhysRevD.105.L071506} {\bibfield  {journal}
  {\bibinfo  {journal} {Phys. Rev. D}\ }\textbf {\bibinfo {volume} {105}},\
  \bibinfo {pages} {L071506} (\bibinfo {year} {2022})},\ \Eprint
  {https://arxiv.org/abs/2201.07909} {arXiv:2201.07909 [hep-ph]} \BibitemShut
  {NoStop}%
\bibitem [{\citenamefont {Bernhardt}\ and\ \citenamefont
  {Fischer}(2023)}]{Bernhardt:2023hpr}%
  \BibitemOpen
  \bibfield  {author} {\bibinfo {author} {\bibfnamefont {J.}~\bibnamefont
  {Bernhardt}}\ and\ \bibinfo {author} {\bibfnamefont {C.~S.}\ \bibnamefont
  {Fischer}},\ }\bibfield  {title} {\bibinfo {title} {{QCD phase transitions in
  the light quark chiral limit}},\ }\href
  {https://doi.org/10.1103/PhysRevD.108.114018} {\bibfield  {journal} {\bibinfo
   {journal} {Phys. Rev. D}\ }\textbf {\bibinfo {volume} {108}},\ \bibinfo
  {pages} {114018} (\bibinfo {year} {2023})},\ \Eprint
  {https://arxiv.org/abs/2309.06737} {arXiv:2309.06737 [hep-ph]} \BibitemShut
  {NoStop}%
\bibitem [{\citenamefont {Fejos}\ and\ \citenamefont
  {Hatsuda}(2024)}]{Fejos:2024bgl}%
  \BibitemOpen
  \bibfield  {author} {\bibinfo {author} {\bibfnamefont {G.}~\bibnamefont
  {Fejos}}\ and\ \bibinfo {author} {\bibfnamefont {T.}~\bibnamefont
  {Hatsuda}},\ }\bibfield  {title} {\bibinfo {title} {{Order of the
  SU(Nf){\texttimes}SU(Nf) chiral transition via the functional renormalization
  group}},\ }\href {https://doi.org/10.1103/PhysRevD.110.016021} {\bibfield
  {journal} {\bibinfo  {journal} {Phys. Rev. D}\ }\textbf {\bibinfo {volume}
  {110}},\ \bibinfo {pages} {016021} (\bibinfo {year} {2024})},\ \Eprint
  {https://arxiv.org/abs/2404.00554} {arXiv:2404.00554 [hep-ph]} \BibitemShut
  {NoStop}%
\bibitem [{\citenamefont {Giacosa}\ \emph {et~al.}(2025)\citenamefont
  {Giacosa}, \citenamefont {Kov{\'a}cs}, \citenamefont {Kov{\'a}cs},
  \citenamefont {Pisarski},\ and\ \citenamefont {Rennecke}}]{Giacosa:2024orp}%
  \BibitemOpen
  \bibfield  {author} {\bibinfo {author} {\bibfnamefont {F.}~\bibnamefont
  {Giacosa}}, \bibinfo {author} {\bibfnamefont {G.}~\bibnamefont {Kov{\'a}cs}},
  \bibinfo {author} {\bibfnamefont {P.}~\bibnamefont {Kov{\'a}cs}}, \bibinfo
  {author} {\bibfnamefont {R.~D.}\ \bibnamefont {Pisarski}},\ and\ \bibinfo
  {author} {\bibfnamefont {F.}~\bibnamefont {Rennecke}},\ }\bibfield  {title}
  {\bibinfo {title} {{Anomalous U(1)A couplings and the Columbia plot}},\
  }\href {https://doi.org/10.1103/PhysRevD.111.016014} {\bibfield  {journal}
  {\bibinfo  {journal} {Phys. Rev. D}\ }\textbf {\bibinfo {volume} {111}},\
  \bibinfo {pages} {016014} (\bibinfo {year} {2025})},\ \Eprint
  {https://arxiv.org/abs/2410.08185} {arXiv:2410.08185 [hep-ph]} \BibitemShut
  {NoStop}%
\bibitem [{\citenamefont {Pisarski}\ and\ \citenamefont
  {Rennecke}(2024)}]{Pisarski:2024esv}%
  \BibitemOpen
  \bibfield  {author} {\bibinfo {author} {\bibfnamefont {R.~D.}\ \bibnamefont
  {Pisarski}}\ and\ \bibinfo {author} {\bibfnamefont {F.}~\bibnamefont
  {Rennecke}},\ }\bibfield  {title} {\bibinfo {title} {{Conjectures about the
  Chiral Phase Transition in QCD from Anomalous Multi-Instanton
  Interactions}},\ }\href {https://doi.org/10.1103/PhysRevLett.132.251903}
  {\bibfield  {journal} {\bibinfo  {journal} {Phys. Rev. Lett.}\ }\textbf
  {\bibinfo {volume} {132}},\ \bibinfo {pages} {251903} (\bibinfo {year}
  {2024})},\ \Eprint {https://arxiv.org/abs/2401.06130} {arXiv:2401.06130
  [hep-ph]} \BibitemShut {NoStop}%
\bibitem [{\citenamefont {Yonekura}(2019)}]{Yonekura:2019vyz}%
  \BibitemOpen
  \bibfield  {author} {\bibinfo {author} {\bibfnamefont {K.}~\bibnamefont
  {Yonekura}},\ }\bibfield  {title} {\bibinfo {title} {{Anomaly matching in QCD
  thermal phase transition}},\ }\href {https://doi.org/10.1007/JHEP05(2019)062}
  {\bibfield  {journal} {\bibinfo  {journal} {JHEP}\ }\textbf {\bibinfo
  {volume} {05}},\ \bibinfo {pages} {062}},\ \Eprint
  {https://arxiv.org/abs/1901.08188} {arXiv:1901.08188 [hep-th]} \BibitemShut
  {NoStop}%
\bibitem [{\citenamefont {Nishimura}\ and\ \citenamefont
  {Tanizaki}(2019)}]{Nishimura:2019umw}%
  \BibitemOpen
  \bibfield  {author} {\bibinfo {author} {\bibfnamefont {H.}~\bibnamefont
  {Nishimura}}\ and\ \bibinfo {author} {\bibfnamefont {Y.}~\bibnamefont
  {Tanizaki}},\ }\bibfield  {title} {\bibinfo {title} {{High-temperature domain
  walls of QCD with imaginary chemical potentials}},\ }\href
  {https://doi.org/10.1007/JHEP06(2019)040} {\bibfield  {journal} {\bibinfo
  {journal} {JHEP}\ }\textbf {\bibinfo {volume} {06}},\ \bibinfo {pages}
  {040}},\ \Eprint {https://arxiv.org/abs/1903.04014} {arXiv:1903.04014
  [hep-th]} \BibitemShut {NoStop}%
\bibitem [{\citenamefont {Kobayashi}\ \emph {et~al.}(2023)\citenamefont
  {Kobayashi}, \citenamefont {Yokokura},\ and\ \citenamefont
  {Yonekura}}]{Kobayashi:2023ajk}%
  \BibitemOpen
  \bibfield  {author} {\bibinfo {author} {\bibfnamefont {S.~K.}\ \bibnamefont
  {Kobayashi}}, \bibinfo {author} {\bibfnamefont {T.}~\bibnamefont
  {Yokokura}},\ and\ \bibinfo {author} {\bibfnamefont {K.}~\bibnamefont
  {Yonekura}},\ }\bibfield  {title} {\bibinfo {title} {{The QCD phase diagram
  in the space of imaginary chemical potential via {\textquoteright}t Hooft
  anomalies}},\ }\href {https://doi.org/10.1007/JHEP08(2023)132} {\bibfield
  {journal} {\bibinfo  {journal} {JHEP}\ }\textbf {\bibinfo {volume} {08}},\
  \bibinfo {pages} {132}},\ \Eprint {https://arxiv.org/abs/2305.01217}
  {arXiv:2305.01217 [hep-th]} \BibitemShut {NoStop}%
\bibitem [{\citenamefont {Roberge}\ and\ \citenamefont
  {Weiss}(1986)}]{Roberge:1986mm}%
  \BibitemOpen
  \bibfield  {author} {\bibinfo {author} {\bibfnamefont {A.}~\bibnamefont
  {Roberge}}\ and\ \bibinfo {author} {\bibfnamefont {N.}~\bibnamefont
  {Weiss}},\ }\bibfield  {title} {\bibinfo {title} {{Gauge Theories With
  Imaginary Chemical Potential and the Phases of {QCD}}},\ }\href
  {https://doi.org/10.1016/0550-3213(86)90582-1} {\bibfield  {journal}
  {\bibinfo  {journal} {Nucl. Phys.}\ }\textbf {\bibinfo {volume} {B275}},\
  \bibinfo {pages} {734} (\bibinfo {year} {1986})}\BibitemShut {NoStop}%
\bibitem [{\citenamefont {Senthil}\ \emph
  {et~al.}(2004{\natexlab{a}})\citenamefont {Senthil}, \citenamefont
  {Vishwanath}, \citenamefont {Balents}, \citenamefont {Sachdev},\ and\
  \citenamefont {Fisher}}]{Senthil:2003eed}%
  \BibitemOpen
  \bibfield  {author} {\bibinfo {author} {\bibfnamefont {T.}~\bibnamefont
  {Senthil}}, \bibinfo {author} {\bibfnamefont {A.}~\bibnamefont {Vishwanath}},
  \bibinfo {author} {\bibfnamefont {L.}~\bibnamefont {Balents}}, \bibinfo
  {author} {\bibfnamefont {S.}~\bibnamefont {Sachdev}},\ and\ \bibinfo {author}
  {\bibfnamefont {M.~P.~A.}\ \bibnamefont {Fisher}},\ }\bibfield  {title}
  {\bibinfo {title} {{Deconfined Quantum Critical Points}},\ }\href
  {https://doi.org/10.1126/science.1091806} {\bibfield  {journal} {\bibinfo
  {journal} {Science}\ }\textbf {\bibinfo {volume} {303}},\ \bibinfo {pages}
  {1490} (\bibinfo {year} {2004}{\natexlab{a}})},\ \Eprint
  {https://arxiv.org/abs/cond-mat/0311326} {arXiv:cond-mat/0311326}
  \BibitemShut {NoStop}%
\bibitem [{\citenamefont {Senthil}\ \emph
  {et~al.}(2004{\natexlab{b}})\citenamefont {Senthil}, \citenamefont {Balents},
  \citenamefont {Sachdev}, \citenamefont {Vishwanath},\ and\ \citenamefont
  {Fisher}}]{Senthil:2004fuw}%
  \BibitemOpen
  \bibfield  {author} {\bibinfo {author} {\bibfnamefont {T.}~\bibnamefont
  {Senthil}}, \bibinfo {author} {\bibfnamefont {L.}~\bibnamefont {Balents}},
  \bibinfo {author} {\bibfnamefont {S.}~\bibnamefont {Sachdev}}, \bibinfo
  {author} {\bibfnamefont {A.}~\bibnamefont {Vishwanath}},\ and\ \bibinfo
  {author} {\bibfnamefont {M.~P.~A.}\ \bibnamefont {Fisher}},\ }\bibfield
  {title} {\bibinfo {title} {{Quantum criticality beyond the
  Landau-Ginzburg-Wilson paradigm}},\ }\href
  {https://doi.org/10.1103/PhysRevB.70.144407} {\bibfield  {journal} {\bibinfo
  {journal} {Phys. Rev. B}\ }\textbf {\bibinfo {volume} {70}},\ \bibinfo
  {pages} {144407} (\bibinfo {year} {2004}{\natexlab{b}})}\BibitemShut
  {NoStop}%
\bibitem [{\citenamefont {Senthil}(2023)}]{Senthil:2023vqd}%
  \BibitemOpen
  \bibfield  {author} {\bibinfo {author} {\bibfnamefont {T.}~\bibnamefont
  {Senthil}},\ }\bibfield  {title} {\bibinfo {title} {{Deconfined quantum
  critical points: a review}},\ }\Eprint {https://arxiv.org/abs/2306.12638}
  {arXiv:2306.12638 [cond-mat.str-el]}  (\bibinfo {year} {2023}),\ \bibinfo
  {note} {preprint}\BibitemShut {NoStop}%
\bibitem [{\citenamefont {'t~Hooft}(1980)}]{tHooft:1979rat}%
  \BibitemOpen
  \bibfield  {author} {\bibinfo {author} {\bibfnamefont {G.}~\bibnamefont
  {'t~Hooft}},\ }\bibfield  {title} {\bibinfo {title} {{Naturalness, chiral
  symmetry, and spontaneous chiral symmetry breaking}},\ }in\ \href
  {https://doi.org/10.1007/978-1-4684-7571-5_9} {\emph {\bibinfo {booktitle}
  {{Recent Developments in Gauge Theories. Proceedings, Nato Advanced Study
  Institute, Cargese, France, August 26 - September 8, 1979}}}},\ Vol.~\bibinfo
  {volume} {59}\ (\bibinfo {year} {1980})\ pp.\ \bibinfo {pages}
  {135--157}\BibitemShut {NoStop}%
\bibitem [{Note1()}]{Note1}%
  \BibitemOpen
  \bibinfo {note} {An invertible field theory evaluates to a pure phase on
  closed manifolds and has an anomalous variation on boundaries, encoding the
  't~Hooft anomaly of the theory living there.}\BibitemShut {Stop}%
\bibitem [{\citenamefont {Freed}(2014)}]{Freed:2014iua}%
  \BibitemOpen
  \bibfield  {author} {\bibinfo {author} {\bibfnamefont {D.~S.}\ \bibnamefont
  {Freed}},\ }\bibfield  {title} {\bibinfo {title} {{Anomalies and Invertible
  Field Theories}},\ }\href {https://doi.org/10.1090/pspum/088/01462}
  {\bibfield  {journal} {\bibinfo  {journal} {Proc. Symp. Pure Math.}\ }\textbf
  {\bibinfo {volume} {88}},\ \bibinfo {pages} {25} (\bibinfo {year} {2014})},\
  \Eprint {https://arxiv.org/abs/1404.7224} {arXiv:1404.7224 [hep-th]}
  \BibitemShut {NoStop}%
\bibitem [{\citenamefont {Kapustin}\ and\ \citenamefont
  {Thorngren}(2014)}]{Kapustin:2014lwa}%
  \BibitemOpen
  \bibfield  {author} {\bibinfo {author} {\bibfnamefont {A.}~\bibnamefont
  {Kapustin}}\ and\ \bibinfo {author} {\bibfnamefont {R.}~\bibnamefont
  {Thorngren}},\ }\bibfield  {title} {\bibinfo {title} {{Anomalies of discrete
  symmetries in three dimensions and group cohomology}},\ }\href
  {https://doi.org/10.1103/PhysRevLett.112.231602} {\bibfield  {journal}
  {\bibinfo  {journal} {Phys. Rev. Lett.}\ }\textbf {\bibinfo {volume} {112}},\
  \bibinfo {pages} {231602} (\bibinfo {year} {2014})},\ \Eprint
  {https://arxiv.org/abs/1403.0617} {arXiv:1403.0617 [hep-th]} \BibitemShut
  {NoStop}%
\bibitem [{\citenamefont {Witten}(2016)}]{Witten:2015aba}%
  \BibitemOpen
  \bibfield  {author} {\bibinfo {author} {\bibfnamefont {E.}~\bibnamefont
  {Witten}},\ }\bibfield  {title} {\bibinfo {title} {{Fermion Path Integrals
  And Topological Phases}},\ }\href
  {https://doi.org/10.1103/RevModPhys.88.035001} {\bibfield  {journal}
  {\bibinfo  {journal} {Rev. Mod. Phys.}\ }\textbf {\bibinfo {volume} {88}},\
  \bibinfo {pages} {035001} (\bibinfo {year} {2016})},\ \Eprint
  {https://arxiv.org/abs/1508.04715} {arXiv:1508.04715 [cond-mat.mes-hall]}
  \BibitemShut {NoStop}%
\bibitem [{\citenamefont {Yonekura}(2016)}]{Yonekura:2016wuc}%
  \BibitemOpen
  \bibfield  {author} {\bibinfo {author} {\bibfnamefont {K.}~\bibnamefont
  {Yonekura}},\ }\bibfield  {title} {\bibinfo {title} {{Dai-Freed theorem and
  topological phases of matter}},\ }\href
  {https://doi.org/10.1007/JHEP09(2016)022} {\bibfield  {journal} {\bibinfo
  {journal} {JHEP}\ }\textbf {\bibinfo {volume} {09}},\ \bibinfo {pages}
  {022}},\ \Eprint {https://arxiv.org/abs/1607.01873} {arXiv:1607.01873
  [hep-th]} \BibitemShut {NoStop}%
\bibitem [{\citenamefont {Bhardwaj}\ \emph {et~al.}(2024)\citenamefont
  {Bhardwaj}, \citenamefont {Bottini}, \citenamefont {Fraser-Taliente},
  \citenamefont {Gladden}, \citenamefont {Gould}, \citenamefont {Platschorre},\
  and\ \citenamefont {Tillim}}]{Bhardwaj:2023kri}%
  \BibitemOpen
  \bibfield  {author} {\bibinfo {author} {\bibfnamefont {L.}~\bibnamefont
  {Bhardwaj}}, \bibinfo {author} {\bibfnamefont {L.~E.}\ \bibnamefont
  {Bottini}}, \bibinfo {author} {\bibfnamefont {L.}~\bibnamefont
  {Fraser-Taliente}}, \bibinfo {author} {\bibfnamefont {L.}~\bibnamefont
  {Gladden}}, \bibinfo {author} {\bibfnamefont {D.~S.~W.}\ \bibnamefont
  {Gould}}, \bibinfo {author} {\bibfnamefont {A.}~\bibnamefont {Platschorre}},\
  and\ \bibinfo {author} {\bibfnamefont {H.}~\bibnamefont {Tillim}},\
  }\bibfield  {title} {\bibinfo {title} {{Lectures on generalized
  symmetries}},\ }\href {https://doi.org/10.1016/j.physrep.2023.11.002}
  {\bibfield  {journal} {\bibinfo  {journal} {Phys. Rept.}\ }\textbf {\bibinfo
  {volume} {1051}},\ \bibinfo {pages} {1} (\bibinfo {year} {2024})},\ \Eprint
  {https://arxiv.org/abs/2307.07547} {arXiv:2307.07547 [hep-th]} \BibitemShut
  {NoStop}%
\bibitem [{\citenamefont {Wess}\ and\ \citenamefont
  {Zumino}(1971)}]{Wess:1971yu}%
  \BibitemOpen
  \bibfield  {author} {\bibinfo {author} {\bibfnamefont {J.}~\bibnamefont
  {Wess}}\ and\ \bibinfo {author} {\bibfnamefont {B.}~\bibnamefont {Zumino}},\
  }\bibfield  {title} {\bibinfo {title} {{Consequences of anomalous Ward
  identities}},\ }\href {https://doi.org/10.1016/0370-2693(71)90582-X}
  {\bibfield  {journal} {\bibinfo  {journal} {Phys. Lett.}\ }\textbf {\bibinfo
  {volume} {37B}},\ \bibinfo {pages} {95} (\bibinfo {year} {1971})}\BibitemShut
  {NoStop}%
\bibitem [{\citenamefont {Witten}(1982)}]{Witten:1982fp}%
  \BibitemOpen
  \bibfield  {author} {\bibinfo {author} {\bibfnamefont {E.}~\bibnamefont
  {Witten}},\ }\bibfield  {title} {\bibinfo {title} {{An SU(2) Anomaly}},\
  }\href {https://doi.org/10.1016/0370-2693(82)90728-6} {\bibfield  {journal}
  {\bibinfo  {journal} {Phys. Lett. B}\ }\textbf {\bibinfo {volume} {117}},\
  \bibinfo {pages} {324} (\bibinfo {year} {1982})}\BibitemShut {NoStop}%
\bibitem [{\citenamefont {Wang}\ \emph {et~al.}(2019)\citenamefont {Wang},
  \citenamefont {Wen},\ and\ \citenamefont {Witten}}]{Wang:2018qoy}%
  \BibitemOpen
  \bibfield  {author} {\bibinfo {author} {\bibfnamefont {J.}~\bibnamefont
  {Wang}}, \bibinfo {author} {\bibfnamefont {X.-G.}\ \bibnamefont {Wen}},\ and\
  \bibinfo {author} {\bibfnamefont {E.}~\bibnamefont {Witten}},\ }\bibfield
  {title} {\bibinfo {title} {{A New SU(2) Anomaly}},\ }\href
  {https://doi.org/10.1063/1.5082852} {\bibfield  {journal} {\bibinfo
  {journal} {J. Math. Phys.}\ }\textbf {\bibinfo {volume} {60}},\ \bibinfo
  {pages} {052301} (\bibinfo {year} {2019})},\ \Eprint
  {https://arxiv.org/abs/1810.00844} {arXiv:1810.00844 [hep-th]} \BibitemShut
  {NoStop}%
\bibitem [{\citenamefont {Frishman}\ \emph {et~al.}(1981)\citenamefont
  {Frishman}, \citenamefont {Schwimmer}, \citenamefont {Banks},\ and\
  \citenamefont {Yankielowicz}}]{Frishman:1980dq}%
  \BibitemOpen
  \bibfield  {author} {\bibinfo {author} {\bibfnamefont {Y.}~\bibnamefont
  {Frishman}}, \bibinfo {author} {\bibfnamefont {A.}~\bibnamefont {Schwimmer}},
  \bibinfo {author} {\bibfnamefont {T.}~\bibnamefont {Banks}},\ and\ \bibinfo
  {author} {\bibfnamefont {S.}~\bibnamefont {Yankielowicz}},\ }\bibfield
  {title} {\bibinfo {title} {{The Axial Anomaly and the Bound State Spectrum in
  Confining Theories}},\ }\href {https://doi.org/10.1016/0550-3213(81)90268-6}
  {\bibfield  {journal} {\bibinfo  {journal} {Nucl. Phys.}\ }\textbf {\bibinfo
  {volume} {B177}},\ \bibinfo {pages} {157} (\bibinfo {year}
  {1981})}\BibitemShut {NoStop}%
\bibitem [{\citenamefont {Coleman}\ and\ \citenamefont
  {Grossman}(1982)}]{Coleman:1982yg}%
  \BibitemOpen
  \bibfield  {author} {\bibinfo {author} {\bibfnamefont {S.~R.}\ \bibnamefont
  {Coleman}}\ and\ \bibinfo {author} {\bibfnamefont {B.}~\bibnamefont
  {Grossman}},\ }\bibfield  {title} {\bibinfo {title} {{'t Hooft's Consistency
  Condition as a Consequence of Analyticity and Unitarity}},\ }\href
  {https://doi.org/10.1016/0550-3213(82)90028-1} {\bibfield  {journal}
  {\bibinfo  {journal} {Nucl. Phys.}\ }\textbf {\bibinfo {volume} {B203}},\
  \bibinfo {pages} {205} (\bibinfo {year} {1982})}\BibitemShut {NoStop}%
\bibitem [{\citenamefont {Appelquist}\ \emph {et~al.}(2019)\citenamefont
  {Appelquist} \emph {et~al.}}]{LatticeStrongDynamics:2018hun}%
  \BibitemOpen
  \bibfield  {author} {\bibinfo {author} {\bibfnamefont {T.}~\bibnamefont
  {Appelquist}} \emph {et~al.} (\bibinfo {collaboration} {Lattice Strong
  Dynamics}),\ }\bibfield  {title} {\bibinfo {title} {{Nonperturbative
  investigations of SU(3) gauge theory with eight dynamical flavors}},\ }\href
  {https://doi.org/10.1103/PhysRevD.99.014509} {\bibfield  {journal} {\bibinfo
  {journal} {Phys. Rev. D}\ }\textbf {\bibinfo {volume} {99}},\ \bibinfo
  {pages} {014509} (\bibinfo {year} {2019})},\ \Eprint
  {https://arxiv.org/abs/1807.08411} {arXiv:1807.08411 [hep-lat]} \BibitemShut
  {NoStop}%
\bibitem [{\citenamefont {Kuti}\ \emph {et~al.}(2022)\citenamefont {Kuti},
  \citenamefont {Fodor}, \citenamefont {Holland},\ and\ \citenamefont
  {Wong}}]{Kuti:2022ldb}%
  \BibitemOpen
  \bibfield  {author} {\bibinfo {author} {\bibfnamefont {J.}~\bibnamefont
  {Kuti}}, \bibinfo {author} {\bibfnamefont {Z.}~\bibnamefont {Fodor}},
  \bibinfo {author} {\bibfnamefont {K.}~\bibnamefont {Holland}},\ and\ \bibinfo
  {author} {\bibfnamefont {C.~H.}\ \bibnamefont {Wong}},\ }\bibfield  {title}
  {\bibinfo {title} {{From ten-flavor tests of the $\beta$-function to
  $\alpha_s$ at the Z-pole}},\ }\href {https://doi.org/10.22323/1.396.0321}
  {\bibfield  {journal} {\bibinfo  {journal} {PoS}\ }\textbf {\bibinfo {volume}
  {LATTICE2021}},\ \bibinfo {pages} {321} (\bibinfo {year} {2022})},\ \Eprint
  {https://arxiv.org/abs/2203.15847} {arXiv:2203.15847 [hep-lat]} \BibitemShut
  {NoStop}%
\bibitem [{\citenamefont {Ingoldby}(2024)}]{Ingoldby:2023mtf}%
  \BibitemOpen
  \bibfield  {author} {\bibinfo {author} {\bibfnamefont {J.}~\bibnamefont
  {Ingoldby}} (\bibinfo {collaboration} {Lattice Strong Dynamics}),\ }\bibfield
   {title} {\bibinfo {title} {{Hidden Conformal Symmetry from Eight Flavors}},\
  }\href {https://doi.org/10.22323/1.453.0091} {\bibfield  {journal} {\bibinfo
  {journal} {PoS}\ }\textbf {\bibinfo {volume} {LATTICE2023}},\ \bibinfo
  {pages} {091} (\bibinfo {year} {2024})},\ \Eprint
  {https://arxiv.org/abs/2401.00267} {arXiv:2401.00267 [hep-lat]} \BibitemShut
  {NoStop}%
\bibitem [{\citenamefont {Hasenfratz}\ \emph {et~al.}(2023)\citenamefont
  {Hasenfratz}, \citenamefont {Neil}, \citenamefont {Shamir}, \citenamefont
  {Svetitsky},\ and\ \citenamefont {Witzel}}]{Hasenfratz:2023wbr}%
  \BibitemOpen
  \bibfield  {author} {\bibinfo {author} {\bibfnamefont {A.}~\bibnamefont
  {Hasenfratz}}, \bibinfo {author} {\bibfnamefont {E.~T.}\ \bibnamefont
  {Neil}}, \bibinfo {author} {\bibfnamefont {Y.}~\bibnamefont {Shamir}},
  \bibinfo {author} {\bibfnamefont {B.}~\bibnamefont {Svetitsky}},\ and\
  \bibinfo {author} {\bibfnamefont {O.}~\bibnamefont {Witzel}},\ }\bibfield
  {title} {\bibinfo {title} {{Infrared fixed point of the SU(3) gauge theory
  with Nf=10 flavors}},\ }\href {https://doi.org/10.1103/PhysRevD.108.L071503}
  {\bibfield  {journal} {\bibinfo  {journal} {Phys. Rev. D}\ }\textbf {\bibinfo
  {volume} {108}},\ \bibinfo {pages} {L071503} (\bibinfo {year} {2023})},\
  \Eprint {https://arxiv.org/abs/2306.07236} {arXiv:2306.07236 [hep-lat]}
  \BibitemShut {NoStop}%
\bibitem [{\citenamefont {Hasenfratz}\ and\ \citenamefont
  {Peterson}(2024)}]{Hasenfratz:2024fad}%
  \BibitemOpen
  \bibfield  {author} {\bibinfo {author} {\bibfnamefont {A.}~\bibnamefont
  {Hasenfratz}}\ and\ \bibinfo {author} {\bibfnamefont {C.~T.}\ \bibnamefont
  {Peterson}},\ }\bibfield  {title} {\bibinfo {title} {{Infrared fixed point in
  the massless twelve-flavor SU(3) gauge-fermion system}},\ }\href
  {https://doi.org/10.1103/PhysRevD.109.114507} {\bibfield  {journal} {\bibinfo
   {journal} {Phys. Rev. D}\ }\textbf {\bibinfo {volume} {109}},\ \bibinfo
  {pages} {114507} (\bibinfo {year} {2024})},\ \Eprint
  {https://arxiv.org/abs/2402.18038} {arXiv:2402.18038 [hep-lat]} \BibitemShut
  {NoStop}%
\bibitem [{\citenamefont {Bergner}(2025)}]{Bergner:2025yke}%
  \BibitemOpen
  \bibfield  {author} {\bibinfo {author} {\bibfnamefont {G.}~\bibnamefont
  {Bergner}},\ }\bibfield  {title} {\bibinfo {title} {{Prospects for lattice
  field theory beyond the Standard Model}},\ }\href
  {https://doi.org/10.22323/1.466.0007} {\bibfield  {journal} {\bibinfo
  {journal} {PoS}\ }\textbf {\bibinfo {volume} {LATTICE2024}},\ \bibinfo
  {pages} {007} (\bibinfo {year} {2025})}\BibitemShut {NoStop}%
\bibitem [{\citenamefont {Witten}(1983{\natexlab{a}})}]{Witten:1983tw}%
  \BibitemOpen
  \bibfield  {author} {\bibinfo {author} {\bibfnamefont {E.}~\bibnamefont
  {Witten}},\ }\bibfield  {title} {\bibinfo {title} {{Global aspects of current
  algebra}},\ }\href {https://doi.org/10.1016/0550-3213(83)90063-9} {\bibfield
  {journal} {\bibinfo  {journal} {Nucl. Phys.}\ }\textbf {\bibinfo {volume}
  {B223}},\ \bibinfo {pages} {422} (\bibinfo {year}
  {1983}{\natexlab{a}})}\BibitemShut {NoStop}%
\bibitem [{\citenamefont {Witten}(1983{\natexlab{b}})}]{Witten:1983tx}%
  \BibitemOpen
  \bibfield  {author} {\bibinfo {author} {\bibfnamefont {E.}~\bibnamefont
  {Witten}},\ }\bibfield  {title} {\bibinfo {title} {{Current Algebra, Baryons,
  and Quark Confinement}},\ }\href
  {https://doi.org/10.1016/0550-3213(83)90064-0} {\bibfield  {journal}
  {\bibinfo  {journal} {Nucl. Phys. B}\ }\textbf {\bibinfo {volume} {223}},\
  \bibinfo {pages} {433} (\bibinfo {year} {1983}{\natexlab{b}})}\BibitemShut
  {NoStop}%
\bibitem [{Note2()}]{Note2}%
  \BibitemOpen
  \bibinfo {note} {What directly shows up in the QCD Lagrangian is $U(1)_V$
  instead of $U(1)_B$. Thus in the path integral, we have to define $\theta _B$
  via $\theta _B =\theta _VN_c$. Then the shift $\theta _B\sim \theta _B+2\pi $
  is generated by an improper $SU(N_c)$ gauge transformation $g$ that satisfies
  $g(x,\tau +\beta ) = g(x,\tau )\protect \,\protect \mathrm {e}^{2\pi i/N_c}$
  for $\tau \in S^1$~\cite {Roberge:1986mm}.}\BibitemShut {Stop}%
\bibitem [{\citenamefont {Polchinski}(1988)}]{Polchinski:1987dy}%
  \BibitemOpen
  \bibfield  {author} {\bibinfo {author} {\bibfnamefont {J.}~\bibnamefont
  {Polchinski}},\ }\bibfield  {title} {\bibinfo {title} {{Scale and Conformal
  Invariance in Quantum Field Theory}},\ }\href
  {https://doi.org/10.1016/0550-3213(88)90179-4} {\bibfield  {journal}
  {\bibinfo  {journal} {Nucl. Phys. B}\ }\textbf {\bibinfo {volume} {303}},\
  \bibinfo {pages} {226} (\bibinfo {year} {1988})}\BibitemShut {NoStop}%
\bibitem [{\citenamefont {El-Showk}\ \emph {et~al.}(2011)\citenamefont
  {El-Showk}, \citenamefont {Nakayama},\ and\ \citenamefont
  {Rychkov}}]{El-Showk:2011xbs}%
  \BibitemOpen
  \bibfield  {author} {\bibinfo {author} {\bibfnamefont {S.}~\bibnamefont
  {El-Showk}}, \bibinfo {author} {\bibfnamefont {Y.}~\bibnamefont {Nakayama}},\
  and\ \bibinfo {author} {\bibfnamefont {S.}~\bibnamefont {Rychkov}},\
  }\bibfield  {title} {\bibinfo {title} {{What Maxwell Theory in
  D{\ensuremath{<}}{\ensuremath{>}}4 teaches us about scale and conformal
  invariance}},\ }\href {https://doi.org/10.1016/j.nuclphysb.2011.03.008}
  {\bibfield  {journal} {\bibinfo  {journal} {Nucl. Phys. B}\ }\textbf
  {\bibinfo {volume} {848}},\ \bibinfo {pages} {578} (\bibinfo {year}
  {2011})},\ \Eprint {https://arxiv.org/abs/1101.5385} {arXiv:1101.5385
  [hep-th]} \BibitemShut {NoStop}%
\bibitem [{\citenamefont {Nakayama}(2015)}]{Nakayama:2013is}%
  \BibitemOpen
  \bibfield  {author} {\bibinfo {author} {\bibfnamefont {Y.}~\bibnamefont
  {Nakayama}},\ }\bibfield  {title} {\bibinfo {title} {{Scale invariance vs
  conformal invariance}},\ }\href
  {https://doi.org/10.1016/j.physrep.2014.12.003} {\bibfield  {journal}
  {\bibinfo  {journal} {Phys. Rept.}\ }\textbf {\bibinfo {volume} {569}},\
  \bibinfo {pages} {1} (\bibinfo {year} {2015})},\ \Eprint
  {https://arxiv.org/abs/1302.0884} {arXiv:1302.0884 [hep-th]} \BibitemShut
  {NoStop}%
\bibitem [{\citenamefont {Delamotte}\ \emph {et~al.}(2016)\citenamefont
  {Delamotte}, \citenamefont {Tissier},\ and\ \citenamefont
  {Wschebor}}]{Delamotte:2015aaa}%
  \BibitemOpen
  \bibfield  {author} {\bibinfo {author} {\bibfnamefont {B.}~\bibnamefont
  {Delamotte}}, \bibinfo {author} {\bibfnamefont {M.}~\bibnamefont {Tissier}},\
  and\ \bibinfo {author} {\bibfnamefont {N.}~\bibnamefont {Wschebor}},\
  }\bibfield  {title} {\bibinfo {title} {{Scale invariance implies conformal
  invariance for the three-dimensional Ising model}},\ }\href
  {https://doi.org/10.1103/PhysRevE.93.012144} {\bibfield  {journal} {\bibinfo
  {journal} {Phys. Rev. E}\ }\textbf {\bibinfo {volume} {93}},\ \bibinfo
  {pages} {012144} (\bibinfo {year} {2016})},\ \Eprint
  {https://arxiv.org/abs/1501.01776} {arXiv:1501.01776 [cond-mat.stat-mech]}
  \BibitemShut {NoStop}%
\bibitem [{\citenamefont {Paulos}\ \emph {et~al.}(2016)\citenamefont {Paulos},
  \citenamefont {Rychkov}, \citenamefont {van Rees},\ and\ \citenamefont
  {Zan}}]{Paulos:2015jfa}%
  \BibitemOpen
  \bibfield  {author} {\bibinfo {author} {\bibfnamefont {M.~F.}\ \bibnamefont
  {Paulos}}, \bibinfo {author} {\bibfnamefont {S.}~\bibnamefont {Rychkov}},
  \bibinfo {author} {\bibfnamefont {B.~C.}\ \bibnamefont {van Rees}},\ and\
  \bibinfo {author} {\bibfnamefont {B.}~\bibnamefont {Zan}},\ }\bibfield
  {title} {\bibinfo {title} {{Conformal Invariance in the Long-Range Ising
  Model}},\ }\href {https://doi.org/10.1016/j.nuclphysb.2015.10.018} {\bibfield
   {journal} {\bibinfo  {journal} {Nucl. Phys. B}\ }\textbf {\bibinfo {volume}
  {902}},\ \bibinfo {pages} {246} (\bibinfo {year} {2016})},\ \Eprint
  {https://arxiv.org/abs/1509.00008} {arXiv:1509.00008 [hep-th]} \BibitemShut
  {NoStop}%
\bibitem [{\citenamefont {Meneses}\ \emph {et~al.}(2019)\citenamefont
  {Meneses}, \citenamefont {Penedones}, \citenamefont {Rychkov}, \citenamefont
  {Viana Parente~Lopes},\ and\ \citenamefont {Yvernay}}]{Meneses:2018xpu}%
  \BibitemOpen
  \bibfield  {author} {\bibinfo {author} {\bibfnamefont {S.}~\bibnamefont
  {Meneses}}, \bibinfo {author} {\bibfnamefont {J.}~\bibnamefont {Penedones}},
  \bibinfo {author} {\bibfnamefont {S.}~\bibnamefont {Rychkov}}, \bibinfo
  {author} {\bibfnamefont {J.~M.}\ \bibnamefont {Viana Parente~Lopes}},\ and\
  \bibinfo {author} {\bibfnamefont {P.}~\bibnamefont {Yvernay}},\ }\bibfield
  {title} {\bibinfo {title} {{A structural test for the conformal invariance of
  the critical 3d Ising model}},\ }\href
  {https://doi.org/10.1007/JHEP04(2019)115} {\bibfield  {journal} {\bibinfo
  {journal} {JHEP}\ }\textbf {\bibinfo {volume} {04}},\ \bibinfo {pages}
  {115}},\ \Eprint {https://arxiv.org/abs/1802.02319} {arXiv:1802.02319
  [hep-th]} \BibitemShut {NoStop}%
\bibitem [{\citenamefont {Kikuchi}\ and\ \citenamefont
  {Tanizaki}(2017)}]{Kikuchi:2017pcp}%
  \BibitemOpen
  \bibfield  {author} {\bibinfo {author} {\bibfnamefont {Y.}~\bibnamefont
  {Kikuchi}}\ and\ \bibinfo {author} {\bibfnamefont {Y.}~\bibnamefont
  {Tanizaki}},\ }\bibfield  {title} {\bibinfo {title} {{Global inconsistency,
  't~Hooft anomaly, and level crossing in quantum mechanics}},\ }\href
  {https://doi.org/10.1093/ptep/ptx148} {\bibfield  {journal} {\bibinfo
  {journal} {Prog. Theor. Exp. Phys.}\ }\textbf {\bibinfo {volume} {2017}},\
  \bibinfo {pages} {113B05} (\bibinfo {year} {2017})},\ \Eprint
  {https://arxiv.org/abs/1708.01962} {arXiv:1708.01962 [hep-th]} \BibitemShut
  {NoStop}%
\bibitem [{\citenamefont {C\'ordova}\ \emph
  {et~al.}(2020{\natexlab{a}})\citenamefont {C\'ordova}, \citenamefont {Freed},
  \citenamefont {Lam},\ and\ \citenamefont {Seiberg}}]{Cordova:2019jnf}%
  \BibitemOpen
  \bibfield  {author} {\bibinfo {author} {\bibfnamefont {C.}~\bibnamefont
  {C\'ordova}}, \bibinfo {author} {\bibfnamefont {D.~S.}\ \bibnamefont
  {Freed}}, \bibinfo {author} {\bibfnamefont {H.~T.}\ \bibnamefont {Lam}},\
  and\ \bibinfo {author} {\bibfnamefont {N.}~\bibnamefont {Seiberg}},\
  }\bibfield  {title} {\bibinfo {title} {{Anomalies in the Space of Coupling
  Constants and Their Dynamical Applications I}},\ }\href
  {https://doi.org/10.21468/SciPostPhys.8.1.001} {\bibfield  {journal}
  {\bibinfo  {journal} {SciPost Phys.}\ }\textbf {\bibinfo {volume} {8}},\
  \bibinfo {pages} {001} (\bibinfo {year} {2020}{\natexlab{a}})},\ \Eprint
  {https://arxiv.org/abs/1905.09315} {arXiv:1905.09315 [hep-th]} \BibitemShut
  {NoStop}%
\bibitem [{\citenamefont {C\'ordova}\ \emph
  {et~al.}(2020{\natexlab{b}})\citenamefont {C\'ordova}, \citenamefont {Freed},
  \citenamefont {Lam},\ and\ \citenamefont {Seiberg}}]{Cordova:2019uob}%
  \BibitemOpen
  \bibfield  {author} {\bibinfo {author} {\bibfnamefont {C.}~\bibnamefont
  {C\'ordova}}, \bibinfo {author} {\bibfnamefont {D.~S.}\ \bibnamefont
  {Freed}}, \bibinfo {author} {\bibfnamefont {H.~T.}\ \bibnamefont {Lam}},\
  and\ \bibinfo {author} {\bibfnamefont {N.}~\bibnamefont {Seiberg}},\
  }\bibfield  {title} {\bibinfo {title} {{Anomalies in the Space of Coupling
  Constants and Their Dynamical Applications II}},\ }\href
  {https://doi.org/10.21468/SciPostPhys.8.1.002} {\bibfield  {journal}
  {\bibinfo  {journal} {SciPost Phys.}\ }\textbf {\bibinfo {volume} {8}},\
  \bibinfo {pages} {002} (\bibinfo {year} {2020}{\natexlab{b}})},\ \Eprint
  {https://arxiv.org/abs/1905.13361} {arXiv:1905.13361 [hep-th]} \BibitemShut
  {NoStop}%
\bibitem [{\citenamefont {Heidenreich}\ \emph {et~al.}(2021)\citenamefont
  {Heidenreich}, \citenamefont {McNamara}, \citenamefont {Montero},
  \citenamefont {Reece}, \citenamefont {Rudelius},\ and\ \citenamefont
  {Valenzuela}}]{Heidenreich:2021xpr}%
  \BibitemOpen
  \bibfield  {author} {\bibinfo {author} {\bibfnamefont {B.}~\bibnamefont
  {Heidenreich}}, \bibinfo {author} {\bibfnamefont {J.}~\bibnamefont
  {McNamara}}, \bibinfo {author} {\bibfnamefont {M.}~\bibnamefont {Montero}},
  \bibinfo {author} {\bibfnamefont {M.}~\bibnamefont {Reece}}, \bibinfo
  {author} {\bibfnamefont {T.}~\bibnamefont {Rudelius}},\ and\ \bibinfo
  {author} {\bibfnamefont {I.}~\bibnamefont {Valenzuela}},\ }\bibfield  {title}
  {\bibinfo {title} {{Non-invertible global symmetries and completeness of the
  spectrum}},\ }\href {https://doi.org/10.1007/JHEP09(2021)203} {\bibfield
  {journal} {\bibinfo  {journal} {JHEP}\ }\textbf {\bibinfo {volume} {09}},\
  \bibinfo {pages} {203}},\ \Eprint {https://arxiv.org/abs/2104.07036}
  {arXiv:2104.07036 [hep-th]} \BibitemShut {NoStop}%
\bibitem [{\citenamefont {Garc{\'\i}a-Valdecasas}\ \emph
  {et~al.}(2025)\citenamefont {Garc{\'\i}a-Valdecasas}, \citenamefont {Reece},\
  and\ \citenamefont {Suzuki}}]{Garcia-Valdecasas:2024cqn}%
  \BibitemOpen
  \bibfield  {author} {\bibinfo {author} {\bibfnamefont {E.}~\bibnamefont
  {Garc{\'\i}a-Valdecasas}}, \bibinfo {author} {\bibfnamefont {M.}~\bibnamefont
  {Reece}},\ and\ \bibinfo {author} {\bibfnamefont {M.}~\bibnamefont
  {Suzuki}},\ }\bibfield  {title} {\bibinfo {title} {{Monopole breaking of
  Chern-Weil symmetries}},\ }\href
  {https://doi.org/10.21468/SciPostPhys.18.5.162} {\bibfield  {journal}
  {\bibinfo  {journal} {SciPost Phys.}\ }\textbf {\bibinfo {volume} {18}},\
  \bibinfo {pages} {162} (\bibinfo {year} {2025})},\ \Eprint
  {https://arxiv.org/abs/2408.00067} {arXiv:2408.00067 [hep-th]} \BibitemShut
  {NoStop}%
\bibitem [{\citenamefont {Yu}(2025)}]{Yu:2024jtk}%
  \BibitemOpen
  \bibfield  {author} {\bibinfo {author} {\bibfnamefont {X.}~\bibnamefont
  {Yu}},\ }\bibfield  {title} {\bibinfo {title} {{Gauging in parameter space: A
  top-down perspective}},\ }\href {https://doi.org/10.1103/638n-qwnm}
  {\bibfield  {journal} {\bibinfo  {journal} {Phys. Rev. D}\ }\textbf {\bibinfo
  {volume} {112}},\ \bibinfo {pages} {025020} (\bibinfo {year} {2025})},\
  \Eprint {https://arxiv.org/abs/2411.14997} {arXiv:2411.14997 [hep-th]}
  \BibitemShut {NoStop}%
\bibitem [{\citenamefont {Najjar}\ \emph {et~al.}(2025)\citenamefont {Najjar},
  \citenamefont {Santilli},\ and\ \citenamefont {Wang}}]{Najjar:2024vmm}%
  \BibitemOpen
  \bibfield  {author} {\bibinfo {author} {\bibfnamefont {M.}~\bibnamefont
  {Najjar}}, \bibinfo {author} {\bibfnamefont {L.}~\bibnamefont {Santilli}},\
  and\ \bibinfo {author} {\bibfnamefont {Y.-N.}\ \bibnamefont {Wang}},\
  }\bibfield  {title} {\bibinfo {title} {{({\ensuremath{-}}1)-form symmetries
  from M-theory and SymTFTs}},\ }\href
  {https://doi.org/10.1007/JHEP03(2025)134} {\bibfield  {journal} {\bibinfo
  {journal} {JHEP}\ }\textbf {\bibinfo {volume} {03}},\ \bibinfo {pages}
  {134}},\ \Eprint {https://arxiv.org/abs/2411.19683} {arXiv:2411.19683
  [hep-th]} \BibitemShut {NoStop}%
\bibitem [{\citenamefont {Aloni}\ \emph {et~al.}(2024)\citenamefont {Aloni},
  \citenamefont {Garcia-Valdecasas}, \citenamefont {Reece},\ and\ \citenamefont
  {Suzuki}}]{Aloni:2024jpb}%
  \BibitemOpen
  \bibfield  {author} {\bibinfo {author} {\bibfnamefont {D.}~\bibnamefont
  {Aloni}}, \bibinfo {author} {\bibfnamefont {E.}~\bibnamefont
  {Garcia-Valdecasas}}, \bibinfo {author} {\bibfnamefont {M.}~\bibnamefont
  {Reece}},\ and\ \bibinfo {author} {\bibfnamefont {M.}~\bibnamefont
  {Suzuki}},\ }\bibfield  {title} {\bibinfo {title} {{Spontaneously broken
  $(-1)$-form {$U(1)$} symmetries}},\ }\href
  {https://doi.org/10.21468/SciPostPhys.17.2.031} {\bibfield  {journal}
  {\bibinfo  {journal} {SciPost Phys.}\ }\textbf {\bibinfo {volume} {17}},\
  \bibinfo {pages} {031} (\bibinfo {year} {2024})},\ \Eprint
  {https://arxiv.org/abs/2402.00117} {arXiv:2402.00117 [hep-th]} \BibitemShut
  {NoStop}%
\bibitem [{\citenamefont {Lin}\ \emph {et~al.}(2025)\citenamefont {Lin},
  \citenamefont {Robbins},\ and\ \citenamefont {Roy}}]{Lin:2025oml}%
  \BibitemOpen
  \bibfield  {author} {\bibinfo {author} {\bibfnamefont {L.}~\bibnamefont
  {Lin}}, \bibinfo {author} {\bibfnamefont {D.}~\bibnamefont {Robbins}},\ and\
  \bibinfo {author} {\bibfnamefont {S.}~\bibnamefont {Roy}},\ }\bibfield
  {title} {\bibinfo {title} {{Decomposition and (non-invertible)
  ({\ensuremath{-}}1)-form symmetries from the symmetry topological field
  theory}},\ }\href {https://doi.org/10.1007/JHEP09(2025)131} {\bibfield
  {journal} {\bibinfo  {journal} {JHEP}\ }\textbf {\bibinfo {volume} {09}},\
  \bibinfo {pages} {131}},\ \Eprint {https://arxiv.org/abs/2503.21862}
  {arXiv:2503.21862 [hep-th]} \BibitemShut {NoStop}%
\bibitem [{\citenamefont {Robbins}\ and\ \citenamefont
  {Roy}(2025)}]{Robbins:2025urk}%
  \BibitemOpen
  \bibfield  {author} {\bibinfo {author} {\bibfnamefont {D.}~\bibnamefont
  {Robbins}}\ and\ \bibinfo {author} {\bibfnamefont {S.}~\bibnamefont {Roy}},\
  }\bibfield  {title} {\bibinfo {title} {{({\ensuremath{-}}1)-form symmetries
  and anomaly shifting from symmetry topological field theory}},\ }\href
  {https://doi.org/10.1103/2zk8-jd5t} {\bibfield  {journal} {\bibinfo
  {journal} {Phys. Rev. D}\ }\textbf {\bibinfo {volume} {112}},\ \bibinfo
  {pages} {105020} (\bibinfo {year} {2025})}\BibitemShut {NoStop}%
\bibitem [{\citenamefont {Xu}\ and\ \citenamefont {Ludwig}(2013)}]{Xu:2011sj}%
  \BibitemOpen
  \bibfield  {author} {\bibinfo {author} {\bibfnamefont {C.}~\bibnamefont
  {Xu}}\ and\ \bibinfo {author} {\bibfnamefont {A.~W.~W.}\ \bibnamefont
  {Ludwig}},\ }\bibfield  {title} {\bibinfo {title} {{Nonperturbative effects
  of Topological $\Theta$-term on Principal Chiral Nonlinear Sigma Models in
  (2+1) dimensions}},\ }\href {https://doi.org/10.1103/PhysRevLett.110.200405}
  {\bibfield  {journal} {\bibinfo  {journal} {Phys. Rev. Lett.}\ }\textbf
  {\bibinfo {volume} {110}},\ \bibinfo {pages} {200405} (\bibinfo {year}
  {2013})},\ \Eprint {https://arxiv.org/abs/1112.5303} {arXiv:1112.5303
  [cond-mat.str-el]} \BibitemShut {NoStop}%
\bibitem [{\citenamefont {Gross}\ \emph {et~al.}(1981)\citenamefont {Gross},
  \citenamefont {Pisarski},\ and\ \citenamefont {Yaffe}}]{Gross:1980br}%
  \BibitemOpen
  \bibfield  {author} {\bibinfo {author} {\bibfnamefont {D.~J.}\ \bibnamefont
  {Gross}}, \bibinfo {author} {\bibfnamefont {R.~D.}\ \bibnamefont
  {Pisarski}},\ and\ \bibinfo {author} {\bibfnamefont {L.~G.}\ \bibnamefont
  {Yaffe}},\ }\bibfield  {title} {\bibinfo {title} {{QCD and Instantons at
  Finite Temperature}},\ }\href {https://doi.org/10.1103/RevModPhys.53.43}
  {\bibfield  {journal} {\bibinfo  {journal} {Rev. Mod. Phys.}\ }\textbf
  {\bibinfo {volume} {53}},\ \bibinfo {pages} {43} (\bibinfo {year}
  {1981})}\BibitemShut {NoStop}%
\bibitem [{\citenamefont {Kapusta}\ and\ \citenamefont
  {Gale}(2011)}]{KapustaGale201102}%
  \BibitemOpen
  \bibfield  {author} {\bibinfo {author} {\bibfnamefont {J.~I.}\ \bibnamefont
  {Kapusta}}\ and\ \bibinfo {author} {\bibfnamefont {C.}~\bibnamefont {Gale}},\
  }\href@noop {} {\emph {\bibinfo {title} {Finite-Temperature Field Theory:
  Principles and Applications (Cambridge Monographs on Mathematical
  Physics)}}},\ \bibinfo {edition} {2nd}\ ed.\ (\bibinfo  {publisher}
  {Cambridge University Press},\ \bibinfo {year} {2011})\BibitemShut {NoStop}%
\bibitem [{\citenamefont {Gaiotto}\ \emph {et~al.}(2015)\citenamefont
  {Gaiotto}, \citenamefont {Kapustin}, \citenamefont {Seiberg},\ and\
  \citenamefont {Willett}}]{Gaiotto:2014kfa}%
  \BibitemOpen
  \bibfield  {author} {\bibinfo {author} {\bibfnamefont {D.}~\bibnamefont
  {Gaiotto}}, \bibinfo {author} {\bibfnamefont {A.}~\bibnamefont {Kapustin}},
  \bibinfo {author} {\bibfnamefont {N.}~\bibnamefont {Seiberg}},\ and\ \bibinfo
  {author} {\bibfnamefont {B.}~\bibnamefont {Willett}},\ }\bibfield  {title}
  {\bibinfo {title} {{Generalized Global Symmetries}},\ }\href
  {https://doi.org/10.1007/JHEP02(2015)172} {\bibfield  {journal} {\bibinfo
  {journal} {JHEP}\ }\textbf {\bibinfo {volume} {02}},\ \bibinfo {pages}
  {172}},\ \Eprint {https://arxiv.org/abs/1412.5148} {arXiv:1412.5148 [hep-th]}
  \BibitemShut {NoStop}%
\bibitem [{\citenamefont {Seiberg}\ and\ \citenamefont
  {Seifnashri}(2025)}]{Seiberg:2025bqy}%
  \BibitemOpen
  \bibfield  {author} {\bibinfo {author} {\bibfnamefont {N.}~\bibnamefont
  {Seiberg}}\ and\ \bibinfo {author} {\bibfnamefont {S.}~\bibnamefont
  {Seifnashri}},\ }\bibfield  {title} {\bibinfo {title} {{Symmetry
  transmutation and anomaly matching}},\ }\href
  {https://doi.org/10.1007/JHEP09(2025)014} {\bibfield  {journal} {\bibinfo
  {journal} {JHEP}\ }\textbf {\bibinfo {volume} {09}},\ \bibinfo {pages}
  {014}},\ \Eprint {https://arxiv.org/abs/2505.08618} {arXiv:2505.08618
  [hep-th]} \BibitemShut {NoStop}%
\bibitem [{sup()}]{supp}%
  \BibitemOpen
  \href@noop {} {}\bibinfo {note} {See Supplemental Material for the discussion on the 3D IR effective theory for
  the high-$T$ gapped phase (Sec.~S1), the comments on the possibility of a
  second-order phase transitions at $\theta=\pi$ (Sec.~S2), the comments on the
  possibility of emergent $U(1)_A$ symmetry at the phase transition (Sec.~S3),
  and the calculation details of the Pad\'e-Borel resummations for the
  $(2+\epsilon)$D perturbative series (Sec.~S4).}\BibitemShut {Stop}%
\bibitem [{Note3()}]{Note3}%
  \BibitemOpen
  \bibinfo {note} {If a CFT matching the symmetry and anomaly existed, a
  second-order transition at $\theta _B=\pi $ without symmetry breaking could
  occur at lower $T$ in the high-$T$ phase~\cite
  {Xu:2011sj,Yonekura:2019vyz,Nishimura:2019umw,Kobayashi:2023ajk}; see the
  Supplemental Material~\cite [Sec.~S2]{supp} for comments.}\BibitemShut
  {Stop}%
\bibitem [{Note4()}]{Note4}%
  \BibitemOpen
  \bibinfo {note} {Such a window is not supported by current lattice evidence
  with $2+1$ flavors of massive quarks in QCD, see Fig.~9 of Ref.~\cite
  {Bonati:2018fvg} and Fig.~1 of Ref.~\cite {Cuteri:2022vwk}.}\BibitemShut
  {Stop}%
\bibitem [{Note5()}]{Note5}%
  \BibitemOpen
  \bibinfo {note} {It is natural to expect that a conformal manifold is always
  generated by a marginal operator. See e.g.~Ref.~\cite {Komatsu:2025cai} for
  recent progress toward a proof of this relation.}\BibitemShut {Stop}%
\bibitem [{Note6()}]{Note6}%
  \BibitemOpen
  \bibinfo {note} {For $\theta _B\protect \neq 0$, we should allow
  chiral-symmetric $(\protect \mathbb {Z}_2)^2_{\protect \mathcal {C}\protect
  \mathcal {S},\protect \mathcal {R}}$-neutral $(\protect \mathbb
  {Z}_2)^3_{\protect \mathcal {C},\protect \mathcal {S},\protect \mathcal
  {R}}$-charged deformations, such as $i\lambda _4 \DOTSI \intop \ilimits@
  [\protect \mathrm {tr}(\Phi ^\dagger d \Phi )^3 + \protect \text {c.c.}]$.
  However, operators in this sector are expected to be irrelevant.}\BibitemShut
  {Stop}%
\bibitem [{Note7()}]{Note7}%
  \BibitemOpen
  \bibinfo {note} {One may speculate that the model~\protect \eqref
  {eq:Landau-Ginzburg} flows to a fixed point with emergent $U(1)_A$ symmetry,
  with the $\lambda _3$ term irrelevant~\cite
  {Pisarski:1983ms,Cohen:1996ng,Lee:1996zy,Cohen:2002st,Fukushima:2008wg,Aoki:2012yj,Pelissetto:2013hqa,Cossu:2013uua}.
  However, as discussed in the Supplemental Material~\cite [Sec.~S3]{supp}, no
  convincing evidence supports this possibility in the model~\protect \eqref
  {eq:Landau-Ginzburg}.}\BibitemShut {Stop}%
\bibitem [{Note8()}]{Note8}%
  \BibitemOpen
  \bibinfo {note} {The symmetry~\protect \eqref {eq:3D_IR_symmetry} is
  $O(4)\rtimes (\protect \mathbb {Z}_2)_{\protect \mathcal {R}}$ at $\theta
  _B=0,\pi $, and $SO(4)\rtimes (\protect \mathbb {Z}_2)_{\protect \mathcal
  {R}}$ for $\theta _B\protect \neq 0,\pi $.}\BibitemShut {Stop}%
\bibitem [{\citenamefont {Butti}\ \emph {et~al.}(2003)\citenamefont {Butti},
  \citenamefont {Pelissetto},\ and\ \citenamefont {Vicari}}]{Butti:2003nu}%
  \BibitemOpen
  \bibfield  {author} {\bibinfo {author} {\bibfnamefont {A.}~\bibnamefont
  {Butti}}, \bibinfo {author} {\bibfnamefont {A.}~\bibnamefont {Pelissetto}},\
  and\ \bibinfo {author} {\bibfnamefont {E.}~\bibnamefont {Vicari}},\
  }\bibfield  {title} {\bibinfo {title} {{On the nature of the finite
  temperature transition in QCD}},\ }\href
  {https://doi.org/10.1088/1126-6708/2003/08/029} {\bibfield  {journal}
  {\bibinfo  {journal} {JHEP}\ }\textbf {\bibinfo {volume} {08}},\ \bibinfo
  {pages} {029}},\ \Eprint {https://arxiv.org/abs/hep-ph/0307036}
  {arXiv:hep-ph/0307036} \BibitemShut {NoStop}%
\bibitem [{\citenamefont {Calabrese}\ and\ \citenamefont
  {Parruccini}(2004)}]{Calabrese:2004uk}%
  \BibitemOpen
  \bibfield  {author} {\bibinfo {author} {\bibfnamefont {P.}~\bibnamefont
  {Calabrese}}\ and\ \bibinfo {author} {\bibfnamefont {P.}~\bibnamefont
  {Parruccini}},\ }\bibfield  {title} {\bibinfo {title} {{Five loop epsilon
  expansion for U(n) x U(m) models: Finite temperature phase transition in
  light QCD}},\ }\href {https://doi.org/10.1088/1126-6708/2004/05/018}
  {\bibfield  {journal} {\bibinfo  {journal} {JHEP}\ }\textbf {\bibinfo
  {volume} {05}},\ \bibinfo {pages} {018}},\ \Eprint
  {https://arxiv.org/abs/hep-ph/0403140} {arXiv:hep-ph/0403140} \BibitemShut
  {NoStop}%
\bibitem [{\citenamefont {Adzhemyan}\ \emph {et~al.}(2022)\citenamefont
  {Adzhemyan}, \citenamefont {Ivanova}, \citenamefont {Kompaniets},
  \citenamefont {Kudlis},\ and\ \citenamefont {Sokolov}}]{Adzhemyan:2021sug}%
  \BibitemOpen
  \bibfield  {author} {\bibinfo {author} {\bibfnamefont {L.~T.}\ \bibnamefont
  {Adzhemyan}}, \bibinfo {author} {\bibfnamefont {E.~V.}\ \bibnamefont
  {Ivanova}}, \bibinfo {author} {\bibfnamefont {M.~V.}\ \bibnamefont
  {Kompaniets}}, \bibinfo {author} {\bibfnamefont {A.}~\bibnamefont {Kudlis}},\
  and\ \bibinfo {author} {\bibfnamefont {A.~I.}\ \bibnamefont {Sokolov}},\
  }\bibfield  {title} {\bibinfo {title} {{Six-loop {\ensuremath{\varepsilon}}
  expansion of three-dimensional U(n){\texttimes}U(m) models}},\ }\href
  {https://doi.org/10.1016/j.nuclphysb.2022.115680} {\bibfield  {journal}
  {\bibinfo  {journal} {Nucl. Phys. B}\ }\textbf {\bibinfo {volume} {975}},\
  \bibinfo {pages} {115680} (\bibinfo {year} {2022})},\ \Eprint
  {https://arxiv.org/abs/2104.12195} {arXiv:2104.12195 [hep-th]} \BibitemShut
  {NoStop}%
\bibitem [{\citenamefont {Fejos}(2014)}]{Fejos:2014qga}%
  \BibitemOpen
  \bibfield  {author} {\bibinfo {author} {\bibfnamefont {G.}~\bibnamefont
  {Fejos}},\ }\bibfield  {title} {\bibinfo {title} {{Fluctuation induced first
  order phase transition in $U(n)\times U(n)$ models using chiral invariant
  expansion of functional renormalization group flows}},\ }\href
  {https://doi.org/10.1103/PhysRevD.90.096011} {\bibfield  {journal} {\bibinfo
  {journal} {Phys. Rev. D}\ }\textbf {\bibinfo {volume} {90}},\ \bibinfo
  {pages} {096011} (\bibinfo {year} {2014})},\ \Eprint
  {https://arxiv.org/abs/1409.3695} {arXiv:1409.3695 [hep-ph]} \BibitemShut
  {NoStop}%
\bibitem [{\citenamefont {Resch}\ \emph {et~al.}(2019)\citenamefont {Resch},
  \citenamefont {Rennecke},\ and\ \citenamefont {Schaefer}}]{Resch:2017vjs}%
  \BibitemOpen
  \bibfield  {author} {\bibinfo {author} {\bibfnamefont {S.}~\bibnamefont
  {Resch}}, \bibinfo {author} {\bibfnamefont {F.}~\bibnamefont {Rennecke}},\
  and\ \bibinfo {author} {\bibfnamefont {B.-J.}\ \bibnamefont {Schaefer}},\
  }\bibfield  {title} {\bibinfo {title} {{Mass sensitivity of the three-flavor
  chiral phase transition}},\ }\href
  {https://doi.org/10.1103/PhysRevD.99.076005} {\bibfield  {journal} {\bibinfo
  {journal} {Phys. Rev. D}\ }\textbf {\bibinfo {volume} {99}},\ \bibinfo
  {pages} {076005} (\bibinfo {year} {2019})},\ \Eprint
  {https://arxiv.org/abs/1712.07961} {arXiv:1712.07961 [hep-ph]} \BibitemShut
  {NoStop}%
\bibitem [{\citenamefont {Gaberdiel}\ \emph {et~al.}(2009)\citenamefont
  {Gaberdiel}, \citenamefont {Konechny},\ and\ \citenamefont
  {Schmidt-Colinet}}]{Gaberdiel:2008fn}%
  \BibitemOpen
  \bibfield  {author} {\bibinfo {author} {\bibfnamefont {M.~R.}\ \bibnamefont
  {Gaberdiel}}, \bibinfo {author} {\bibfnamefont {A.}~\bibnamefont
  {Konechny}},\ and\ \bibinfo {author} {\bibfnamefont {C.}~\bibnamefont
  {Schmidt-Colinet}},\ }\bibfield  {title} {\bibinfo {title} {{Conformal
  perturbation theory beyond the leading order}},\ }\href
  {https://doi.org/10.1088/1751-8113/42/10/105402} {\bibfield  {journal}
  {\bibinfo  {journal} {J. Phys. A}\ }\textbf {\bibinfo {volume} {42}},\
  \bibinfo {pages} {105402} (\bibinfo {year} {2009})},\ \Eprint
  {https://arxiv.org/abs/0811.3149} {arXiv:0811.3149 [hep-th]} \BibitemShut
  {NoStop}%
\bibitem [{\citenamefont {Bashmakov}\ \emph {et~al.}(2017)\citenamefont
  {Bashmakov}, \citenamefont {Bertolini},\ and\ \citenamefont
  {Raj}}]{Bashmakov:2017rko}%
  \BibitemOpen
  \bibfield  {author} {\bibinfo {author} {\bibfnamefont {V.}~\bibnamefont
  {Bashmakov}}, \bibinfo {author} {\bibfnamefont {M.}~\bibnamefont
  {Bertolini}},\ and\ \bibinfo {author} {\bibfnamefont {H.}~\bibnamefont
  {Raj}},\ }\bibfield  {title} {\bibinfo {title} {{On non-supersymmetric
  conformal manifolds: field theory and holography}},\ }\href
  {https://doi.org/10.1007/JHEP11(2017)167} {\bibfield  {journal} {\bibinfo
  {journal} {JHEP}\ }\textbf {\bibinfo {volume} {11}},\ \bibinfo {pages}
  {167}},\ \Eprint {https://arxiv.org/abs/1709.01749} {arXiv:1709.01749
  [hep-th]} \BibitemShut {NoStop}%
\bibitem [{\citenamefont {Behan}(2018)}]{Behan:2017mwi}%
  \BibitemOpen
  \bibfield  {author} {\bibinfo {author} {\bibfnamefont {C.}~\bibnamefont
  {Behan}},\ }\bibfield  {title} {\bibinfo {title} {{Conformal manifolds: ODEs
  from OPEs}},\ }\href {https://doi.org/10.1007/JHEP03(2018)127} {\bibfield
  {journal} {\bibinfo  {journal} {JHEP}\ }\textbf {\bibinfo {volume} {03}},\
  \bibinfo {pages} {127}},\ \Eprint {https://arxiv.org/abs/1709.03967}
  {arXiv:1709.03967 [hep-th]} \BibitemShut {NoStop}%
\bibitem [{\citenamefont {Hollands}(2018)}]{Hollands:2017chb}%
  \BibitemOpen
  \bibfield  {author} {\bibinfo {author} {\bibfnamefont {S.}~\bibnamefont
  {Hollands}},\ }\bibfield  {title} {\bibinfo {title} {{Action principle for
  OPE}},\ }\href {https://doi.org/10.1016/j.nuclphysb.2017.11.013} {\bibfield
  {journal} {\bibinfo  {journal} {Nucl. Phys. B}\ }\textbf {\bibinfo {volume}
  {926}},\ \bibinfo {pages} {614} (\bibinfo {year} {2018})},\ \Eprint
  {https://arxiv.org/abs/1710.05601} {arXiv:1710.05601 [hep-th]} \BibitemShut
  {NoStop}%
\bibitem [{\citenamefont {Sen}\ and\ \citenamefont
  {Tachikawa}(2017)}]{Sen:2017gfr}%
  \BibitemOpen
  \bibfield  {author} {\bibinfo {author} {\bibfnamefont {K.}~\bibnamefont
  {Sen}}\ and\ \bibinfo {author} {\bibfnamefont {Y.}~\bibnamefont
  {Tachikawa}},\ }\href@noop {} {\bibinfo {title} {{First-order conformal
  perturbation theory by marginal operators}}} (\bibinfo {year} {2017}),\
  \Eprint {https://arxiv.org/abs/1711.05947} {arXiv:1711.05947 [hep-th]}
  \BibitemShut {NoStop}%
\bibitem [{\citenamefont {Giambrone}\ \emph {et~al.}(2022)\citenamefont
  {Giambrone}, \citenamefont {Guarino}, \citenamefont {Malek}, \citenamefont
  {Samtleben}, \citenamefont {Sterckx},\ and\ \citenamefont
  {Trigiante}}]{Giambrone:2021wsm}%
  \BibitemOpen
  \bibfield  {author} {\bibinfo {author} {\bibfnamefont {A.}~\bibnamefont
  {Giambrone}}, \bibinfo {author} {\bibfnamefont {A.}~\bibnamefont {Guarino}},
  \bibinfo {author} {\bibfnamefont {E.}~\bibnamefont {Malek}}, \bibinfo
  {author} {\bibfnamefont {H.}~\bibnamefont {Samtleben}}, \bibinfo {author}
  {\bibfnamefont {C.}~\bibnamefont {Sterckx}},\ and\ \bibinfo {author}
  {\bibfnamefont {M.}~\bibnamefont {Trigiante}},\ }\bibfield  {title} {\bibinfo
  {title} {{Holographic evidence for nonsupersymmetric conformal manifolds}},\
  }\href {https://doi.org/10.1103/PhysRevD.105.066018} {\bibfield  {journal}
  {\bibinfo  {journal} {Phys. Rev. D}\ }\textbf {\bibinfo {volume} {105}},\
  \bibinfo {pages} {066018} (\bibinfo {year} {2022})},\ \Eprint
  {https://arxiv.org/abs/2112.11966} {arXiv:2112.11966 [hep-th]} \BibitemShut
  {NoStop}%
\bibitem [{\citenamefont {Zinn-Justin}(2021)}]{ZinnJustin_book}%
  \BibitemOpen
  \bibfield  {author} {\bibinfo {author} {\bibfnamefont {J.}~\bibnamefont
  {Zinn-Justin}},\ }\href {https://doi.org/10.1093/oso/9780198834625.001.0001}
  {\emph {\bibinfo {title} {Quantum Field Theory and Critical Phenomena: Fifth
  Edition}}}\ (\bibinfo  {publisher} {Oxford University Press},\ \bibinfo
  {year} {2021})\BibitemShut {NoStop}%
\bibitem [{\citenamefont {Nahum}\ \emph {et~al.}(2015)\citenamefont {Nahum},
  \citenamefont {Chalker}, \citenamefont {Serna}, \citenamefont {Ortu{\~n}o},\
  and\ \citenamefont {Somoza}}]{Nahum:2015jya}%
  \BibitemOpen
  \bibfield  {author} {\bibinfo {author} {\bibfnamefont {A.}~\bibnamefont
  {Nahum}}, \bibinfo {author} {\bibfnamefont {J.~T.}\ \bibnamefont {Chalker}},
  \bibinfo {author} {\bibfnamefont {P.}~\bibnamefont {Serna}}, \bibinfo
  {author} {\bibfnamefont {M.}~\bibnamefont {Ortu{\~n}o}},\ and\ \bibinfo
  {author} {\bibfnamefont {A.~M.}\ \bibnamefont {Somoza}},\ }\bibfield  {title}
  {\bibinfo {title} {{Deconfined Quantum Criticality, Scaling Violations, and
  Classical Loop Models}},\ }\href {https://doi.org/10.1103/PhysRevX.5.041048}
  {\bibfield  {journal} {\bibinfo  {journal} {Phys. Rev. X}\ }\textbf {\bibinfo
  {volume} {5}},\ \bibinfo {pages} {041048} (\bibinfo {year} {2015})},\ \Eprint
  {https://arxiv.org/abs/1506.06798} {arXiv:1506.06798 [cond-mat.str-el]}
  \BibitemShut {NoStop}%
\bibitem [{\citenamefont {Jones}(2024)}]{Jones:2024ept}%
  \BibitemOpen
  \bibfield  {author} {\bibinfo {author} {\bibfnamefont {R.~A.}\ \bibnamefont
  {Jones}},\ }\emph {\bibinfo {title} {{Explorations in two dimensional
  strongly correlated quantum matter: from exactly solvable models to conformal
  bootstrap}}},\ \href {https://dspace.mit.edu/handle/1721.1/157573?show=full}
  {Ph.D. thesis},\ \bibinfo  {school} {MIT} (\bibinfo {year}
  {2024})\BibitemShut {NoStop}%
\bibitem [{\citenamefont {De~Cesare}\ and\ \citenamefont
  {Rychkov}(2025)}]{DeCesare:2025ukl}%
  \BibitemOpen
  \bibfield  {author} {\bibinfo {author} {\bibfnamefont {F.}~\bibnamefont
  {De~Cesare}}\ and\ \bibinfo {author} {\bibfnamefont {S.}~\bibnamefont
  {Rychkov}},\ }\bibfield  {title} {\bibinfo {title} {{Disturbing news about
  the $d=2+\epsilon$ expansion}},\ }\href
  {https://doi.org/10.1093/ptep/ptaf103} {\bibfield  {journal} {\bibinfo
  {journal} {PTEP}\ }\textbf {\bibinfo {volume} {2025}},\ \bibinfo {pages}
  {093B02} (\bibinfo {year} {2025})},\ \Eprint
  {https://arxiv.org/abs/2505.21611} {arXiv:2505.21611 [hep-th]} \BibitemShut
  {NoStop}%
\bibitem [{\citenamefont {De~Cesare}\ and\ \citenamefont
  {Rychkov}(2026)}]{DeCesare:2026dwm}%
  \BibitemOpen
  \bibfield  {author} {\bibinfo {author} {\bibfnamefont {F.}~\bibnamefont
  {De~Cesare}}\ and\ \bibinfo {author} {\bibfnamefont {S.}~\bibnamefont
  {Rychkov}},\ }\bibfield  {title} {\bibinfo {title} {{Disturbing news about
  the $d=2+\epsilon$ expansion II. Assessing the recombination scenario}},\
  }\href@noop {} {\bibfield  {journal} {\bibinfo  {journal} {arXiv}\ }
  (\bibinfo {year} {2026})},\ \Eprint {https://arxiv.org/abs/2602.10194}
  {arXiv:2602.10194 [hep-th]} \BibitemShut {NoStop}%
\bibitem [{\citenamefont {Philipsen}(2021)}]{Philipsen:2021qji}%
  \BibitemOpen
  \bibfield  {author} {\bibinfo {author} {\bibfnamefont {O.}~\bibnamefont
  {Philipsen}},\ }\bibfield  {title} {\bibinfo {title} {{Lattice Constraints on
  the QCD Chiral Phase Transition at Finite Temperature and Baryon Density}},\
  }\href {https://doi.org/10.3390/sym13112079} {\bibfield  {journal} {\bibinfo
  {journal} {Symmetry}\ }\textbf {\bibinfo {volume} {13}},\ \bibinfo {pages}
  {2079} (\bibinfo {year} {2021})},\ \Eprint {https://arxiv.org/abs/2111.03590}
  {arXiv:2111.03590 [hep-lat]} \BibitemShut {NoStop}%
\bibitem [{\citenamefont {McKane}\ and\ \citenamefont
  {Stone}(1980)}]{McKane:1979cm}%
  \BibitemOpen
  \bibfield  {author} {\bibinfo {author} {\bibfnamefont {A.}~\bibnamefont
  {McKane}}\ and\ \bibinfo {author} {\bibfnamefont {M.}~\bibnamefont {Stone}},\
  }\bibfield  {title} {\bibinfo {title} {{Nonlinear sigma models: a
  perturbative approach to symmetry restoration}},\ }\href
  {https://doi.org/10.1016/0550-3213(80)90396-X} {\bibfield  {journal}
  {\bibinfo  {journal} {Nucl. Phys. B}\ }\textbf {\bibinfo {volume} {163}},\
  \bibinfo {pages} {169} (\bibinfo {year} {1980})}\BibitemShut {NoStop}%
\bibitem [{\citenamefont {Hikami}(1981)}]{Hikami:1980hi}%
  \BibitemOpen
  \bibfield  {author} {\bibinfo {author} {\bibfnamefont {S.}~\bibnamefont
  {Hikami}},\ }\bibfield  {title} {\bibinfo {title} {{Three Loop Beta -
  Functions of Nonlinear Sigma Models on Symmetric Spaces}},\ }\href
  {https://doi.org/10.1016/0370-2693(81)90989-8} {\bibfield  {journal}
  {\bibinfo  {journal} {Phys. Lett. B}\ }\textbf {\bibinfo {volume} {98}},\
  \bibinfo {pages} {208} (\bibinfo {year} {1981})}\BibitemShut {NoStop}%
\bibitem [{\citenamefont {Hikami}(1983)}]{Hikami:1982ak}%
  \BibitemOpen
  \bibfield  {author} {\bibinfo {author} {\bibfnamefont {S.}~\bibnamefont
  {Hikami}},\ }\bibfield  {title} {\bibinfo {title} {{Isomorphism and beta
  function of the nonlinear sigma model in symmetric spaces}},\ }\href
  {https://doi.org/10.1016/0550-3213(83)90260-2} {\bibfield  {journal}
  {\bibinfo  {journal} {Nucl. Phys. B}\ }\textbf {\bibinfo {volume} {215}},\
  \bibinfo {pages} {555} (\bibinfo {year} {1983})}\BibitemShut {NoStop}%
\bibitem [{\citenamefont {Wegner}(1989)}]{Wegner:1989ss}%
  \BibitemOpen
  \bibfield  {author} {\bibinfo {author} {\bibfnamefont {F.}~\bibnamefont
  {Wegner}},\ }\bibfield  {title} {\bibinfo {title} {{Four Loop Order Beta
  Function of Nonlinear $\sigma$ Models in Symmetric Spaces}},\ }\href
  {https://doi.org/10.1016/0550-3213(89)90063-1} {\bibfield  {journal}
  {\bibinfo  {journal} {Nucl. Phys. B}\ }\textbf {\bibinfo {volume} {316}},\
  \bibinfo {pages} {663} (\bibinfo {year} {1989})}\BibitemShut {NoStop}%
\bibitem [{\citenamefont {Poland}\ \emph {et~al.}(2019)\citenamefont {Poland},
  \citenamefont {Rychkov},\ and\ \citenamefont {Vichi}}]{Poland:2018epd}%
  \BibitemOpen
  \bibfield  {author} {\bibinfo {author} {\bibfnamefont {D.}~\bibnamefont
  {Poland}}, \bibinfo {author} {\bibfnamefont {S.}~\bibnamefont {Rychkov}},\
  and\ \bibinfo {author} {\bibfnamefont {A.}~\bibnamefont {Vichi}},\ }\bibfield
   {title} {\bibinfo {title} {{The Conformal Bootstrap: Theory, Numerical
  Techniques, and Applications}},\ }\href
  {https://doi.org/10.1103/RevModPhys.91.015002} {\bibfield  {journal}
  {\bibinfo  {journal} {Rev. Mod. Phys.}\ }\textbf {\bibinfo {volume} {91}},\
  \bibinfo {pages} {015002} (\bibinfo {year} {2019})},\ \Eprint
  {https://arxiv.org/abs/1805.04405} {arXiv:1805.04405 [hep-th]} \BibitemShut
  {NoStop}%
\bibitem [{\citenamefont {Kos}\ \emph {et~al.}(2015)\citenamefont {Kos},
  \citenamefont {Poland}, \citenamefont {Simmons-Duffin},\ and\ \citenamefont
  {Vichi}}]{Kos:2015mba}%
  \BibitemOpen
  \bibfield  {author} {\bibinfo {author} {\bibfnamefont {F.}~\bibnamefont
  {Kos}}, \bibinfo {author} {\bibfnamefont {D.}~\bibnamefont {Poland}},
  \bibinfo {author} {\bibfnamefont {D.}~\bibnamefont {Simmons-Duffin}},\ and\
  \bibinfo {author} {\bibfnamefont {A.}~\bibnamefont {Vichi}},\ }\bibfield
  {title} {\bibinfo {title} {{Bootstrapping the O(N) Archipelago}},\ }\href
  {https://doi.org/10.1007/JHEP11(2015)106} {\bibfield  {journal} {\bibinfo
  {journal} {JHEP}\ }\textbf {\bibinfo {volume} {11}},\ \bibinfo {pages}
  {106}},\ \Eprint {https://arxiv.org/abs/1504.07997} {arXiv:1504.07997
  [hep-th]} \BibitemShut {NoStop}%
\bibitem [{\citenamefont {Kousvos}\ and\ \citenamefont
  {Stergiou}(2023)}]{Kousvos:2022ewl}%
  \BibitemOpen
  \bibfield  {author} {\bibinfo {author} {\bibfnamefont {S.~R.}\ \bibnamefont
  {Kousvos}}\ and\ \bibinfo {author} {\bibfnamefont {A.}~\bibnamefont
  {Stergiou}},\ }\bibfield  {title} {\bibinfo {title} {{CFTs with
  $\mathbf{U(m)\times U(n)}$ global symmetry in 3D and the chiral phase
  transition of QCD}},\ }\href {https://doi.org/10.21468/SciPostPhys.15.2.075}
  {\bibfield  {journal} {\bibinfo  {journal} {SciPost Phys.}\ }\textbf
  {\bibinfo {volume} {15}},\ \bibinfo {pages} {075} (\bibinfo {year} {2023})},\
  \Eprint {https://arxiv.org/abs/2209.02837} {arXiv:2209.02837 [hep-th]}
  \BibitemShut {NoStop}%
\bibitem [{Note9()}]{Note9}%
  \BibitemOpen
  \bibinfo {note} {The most ambitious conjecture would be that the 3D $SU(N_f)$
  PCM~\protect \eqref {eq:lowT_EFT} is asymptotically safe~\cite
  {Wilson:1973jj,Weinberg:1976xy} and gives a path-integral description of the
  3D $SU(N_f)_{\theta _B}$ field theory, such that (schematically)
  $g_c^{-2}-g^{-2}\sim \mu $ for a critical coupling $g_c$.}\BibitemShut
  {Stop}%
\bibitem [{\citenamefont {Luscher}(1982)}]{Luscher:1981zq}%
  \BibitemOpen
  \bibfield  {author} {\bibinfo {author} {\bibfnamefont {M.}~\bibnamefont
  {Luscher}},\ }\bibfield  {title} {\bibinfo {title} {{Topology of Lattice
  Gauge Fields}},\ }\href {https://doi.org/10.1007/BF02029132} {\bibfield
  {journal} {\bibinfo  {journal} {Commun. Math. Phys.}\ }\textbf {\bibinfo
  {volume} {85}},\ \bibinfo {pages} {39} (\bibinfo {year} {1982})}\BibitemShut
  {NoStop}%
\bibitem [{\citenamefont {Schramm}\ and\ \citenamefont
  {Svetitsky}(2000)}]{Schramm:2000cc}%
  \BibitemOpen
  \bibfield  {author} {\bibinfo {author} {\bibfnamefont {A.~J.}\ \bibnamefont
  {Schramm}}\ and\ \bibinfo {author} {\bibfnamefont {B.}~\bibnamefont
  {Svetitsky}},\ }\bibfield  {title} {\bibinfo {title} {{Topology and
  metastability in the lattice Skyrme model}},\ }\href
  {https://doi.org/10.1103/PhysRevD.62.114020} {\bibfield  {journal} {\bibinfo
  {journal} {Phys. Rev. D}\ }\textbf {\bibinfo {volume} {62}},\ \bibinfo
  {pages} {114020} (\bibinfo {year} {2000})},\ \Eprint
  {https://arxiv.org/abs/hep-lat/0008003} {arXiv:hep-lat/0008003} \BibitemShut
  {NoStop}%
\bibitem [{\citenamefont {Shiozaki}(2024)}]{Shiozaki:2024yrm}%
  \BibitemOpen
  \bibfield  {author} {\bibinfo {author} {\bibfnamefont {K.}~\bibnamefont
  {Shiozaki}},\ }\bibfield  {title} {\bibinfo {title} {{A discrete formulation
  for three-dimensional winding number}},\ }\Eprint
  {https://arxiv.org/abs/2403.05291} {arXiv:2403.05291 [cond-mat.mes-hall]}
  (\bibinfo {year} {2024}),\ \bibinfo {note} {preprint}\BibitemShut {NoStop}%
\bibitem [{\citenamefont {L{\'o}pez-Contreras}\ \emph
  {et~al.}(2025)\citenamefont {L{\'o}pez-Contreras}, \citenamefont
  {Garc{\'\i}a-Hern{\'a}ndez}, \citenamefont {Polanco-Eu{\'a}n},\ and\
  \citenamefont {Bietenholz}}]{Lopez-Contreras:2025iik}%
  \BibitemOpen
  \bibfield  {author} {\bibinfo {author} {\bibfnamefont {E.}~\bibnamefont
  {L{\'o}pez-Contreras}}, \bibinfo {author} {\bibfnamefont {J.~A.}\
  \bibnamefont {Garc{\'\i}a-Hern{\'a}ndez}}, \bibinfo {author} {\bibfnamefont
  {E.~N.}\ \bibnamefont {Polanco-Eu{\'a}n}},\ and\ \bibinfo {author}
  {\bibfnamefont {W.}~\bibnamefont {Bietenholz}},\ }\bibfield  {title}
  {\bibinfo {title} {{Phase diagram of chiral 2-flavor QCD based on an
  effective approach}},\ }\href@noop {} {\bibfield  {journal} {\bibinfo
  {journal} {arXiv}\ } (\bibinfo {year} {2025})},\ \Eprint
  {https://arxiv.org/abs/2511.17795} {arXiv:2511.17795 [hep-lat]} \BibitemShut
  {NoStop}%
\bibitem [{\citenamefont {Chen}(2024)}]{Chen:2024ddr}%
  \BibitemOpen
  \bibfield  {author} {\bibinfo {author} {\bibfnamefont {J.-Y.}\ \bibnamefont
  {Chen}},\ }\bibfield  {title} {\bibinfo {title} {{Instanton Density Operator
  in Lattice QCD from Higher Category Theory}},\ }\Eprint
  {https://arxiv.org/abs/2406.06673} {arXiv:2406.06673 [hep-lat]}  (\bibinfo
  {year} {2024}),\ \bibinfo {note} {preprint}\BibitemShut {NoStop}%
\bibitem [{\citenamefont {Gorbenko}\ \emph
  {et~al.}(2018{\natexlab{a}})\citenamefont {Gorbenko}, \citenamefont
  {Rychkov},\ and\ \citenamefont {Zan}}]{Gorbenko:2018ncu}%
  \BibitemOpen
  \bibfield  {author} {\bibinfo {author} {\bibfnamefont {V.}~\bibnamefont
  {Gorbenko}}, \bibinfo {author} {\bibfnamefont {S.}~\bibnamefont {Rychkov}},\
  and\ \bibinfo {author} {\bibfnamefont {B.}~\bibnamefont {Zan}},\ }\bibfield
  {title} {\bibinfo {title} {{Walking, Weak first-order transitions, and
  Complex CFTs}},\ }\href {https://doi.org/10.1007/JHEP10(2018)108} {\bibfield
  {journal} {\bibinfo  {journal} {JHEP}\ }\textbf {\bibinfo {volume} {10}},\
  \bibinfo {pages} {108}},\ \Eprint {https://arxiv.org/abs/1807.11512}
  {arXiv:1807.11512 [hep-th]} \BibitemShut {NoStop}%
\bibitem [{\citenamefont {Gorbenko}\ \emph
  {et~al.}(2018{\natexlab{b}})\citenamefont {Gorbenko}, \citenamefont
  {Rychkov},\ and\ \citenamefont {Zan}}]{Gorbenko:2018dtm}%
  \BibitemOpen
  \bibfield  {author} {\bibinfo {author} {\bibfnamefont {V.}~\bibnamefont
  {Gorbenko}}, \bibinfo {author} {\bibfnamefont {S.}~\bibnamefont {Rychkov}},\
  and\ \bibinfo {author} {\bibfnamefont {B.}~\bibnamefont {Zan}},\ }\bibfield
  {title} {\bibinfo {title} {{Walking, Weak first-order transitions, and
  Complex CFTs II. Two-dimensional Potts model at $Q>4$}},\ }\href
  {https://doi.org/10.21468/SciPostPhys.5.5.050} {\bibfield  {journal}
  {\bibinfo  {journal} {SciPost Phys.}\ }\textbf {\bibinfo {volume} {5}},\
  \bibinfo {pages} {050} (\bibinfo {year} {2018}{\natexlab{b}})},\ \Eprint
  {https://arxiv.org/abs/1808.04380} {arXiv:1808.04380 [hep-th]} \BibitemShut
  {NoStop}%
\bibitem [{\citenamefont {Bonati}\ \emph {et~al.}(2019)\citenamefont {Bonati},
  \citenamefont {Calore}, \citenamefont {D'Elia}, \citenamefont {Mesiti},
  \citenamefont {Negro}, \citenamefont {Sanfilippo}, \citenamefont {Schifano},
  \citenamefont {Silvi},\ and\ \citenamefont {Tripiccione}}]{Bonati:2018fvg}%
  \BibitemOpen
  \bibfield  {author} {\bibinfo {author} {\bibfnamefont {C.}~\bibnamefont
  {Bonati}}, \bibinfo {author} {\bibfnamefont {E.}~\bibnamefont {Calore}},
  \bibinfo {author} {\bibfnamefont {M.}~\bibnamefont {D'Elia}}, \bibinfo
  {author} {\bibfnamefont {M.}~\bibnamefont {Mesiti}}, \bibinfo {author}
  {\bibfnamefont {F.}~\bibnamefont {Negro}}, \bibinfo {author} {\bibfnamefont
  {F.}~\bibnamefont {Sanfilippo}}, \bibinfo {author} {\bibfnamefont {S.~F.}\
  \bibnamefont {Schifano}}, \bibinfo {author} {\bibfnamefont {G.}~\bibnamefont
  {Silvi}},\ and\ \bibinfo {author} {\bibfnamefont {R.}~\bibnamefont
  {Tripiccione}},\ }\bibfield  {title} {\bibinfo {title} {{Roberge-Weiss
  endpoint and chiral symmetry restoration in $N_f = 2+1$ QCD}},\ }\href
  {https://doi.org/10.1103/PhysRevD.99.014502} {\bibfield  {journal} {\bibinfo
  {journal} {Phys. Rev.}\ }\textbf {\bibinfo {volume} {D99}},\ \bibinfo {pages}
  {014502} (\bibinfo {year} {2019})},\ \Eprint
  {https://arxiv.org/abs/1807.02106} {arXiv:1807.02106 [hep-lat]} \BibitemShut
  {NoStop}%
\bibitem [{\citenamefont {Cuteri}\ \emph {et~al.}(2022)\citenamefont {Cuteri},
  \citenamefont {Goswami}, \citenamefont {Karsch}, \citenamefont {Lahiri},
  \citenamefont {Neumann}, \citenamefont {Philipsen}, \citenamefont {Schmidt},\
  and\ \citenamefont {Sciarra}}]{Cuteri:2022vwk}%
  \BibitemOpen
  \bibfield  {author} {\bibinfo {author} {\bibfnamefont {F.}~\bibnamefont
  {Cuteri}}, \bibinfo {author} {\bibfnamefont {J.}~\bibnamefont {Goswami}},
  \bibinfo {author} {\bibfnamefont {F.}~\bibnamefont {Karsch}}, \bibinfo
  {author} {\bibfnamefont {A.}~\bibnamefont {Lahiri}}, \bibinfo {author}
  {\bibfnamefont {M.}~\bibnamefont {Neumann}}, \bibinfo {author} {\bibfnamefont
  {O.}~\bibnamefont {Philipsen}}, \bibinfo {author} {\bibfnamefont
  {C.}~\bibnamefont {Schmidt}},\ and\ \bibinfo {author} {\bibfnamefont
  {A.}~\bibnamefont {Sciarra}},\ }\bibfield  {title} {\bibinfo {title} {{Toward
  the chiral phase transition in the Roberge-Weiss plane}},\ }\href
  {https://doi.org/10.1103/PhysRevD.106.014510} {\bibfield  {journal} {\bibinfo
   {journal} {Phys. Rev. D}\ }\textbf {\bibinfo {volume} {106}},\ \bibinfo
  {pages} {014510} (\bibinfo {year} {2022})},\ \Eprint
  {https://arxiv.org/abs/2205.12707} {arXiv:2205.12707 [hep-lat]} \BibitemShut
  {NoStop}%
\bibitem [{\citenamefont {Komatsu}\ \emph {et~al.}(2025)\citenamefont
  {Komatsu}, \citenamefont {Kusuki}, \citenamefont {Meineri},\ and\
  \citenamefont {Ooguri}}]{Komatsu:2025cai}%
  \BibitemOpen
  \bibfield  {author} {\bibinfo {author} {\bibfnamefont {S.}~\bibnamefont
  {Komatsu}}, \bibinfo {author} {\bibfnamefont {Y.}~\bibnamefont {Kusuki}},
  \bibinfo {author} {\bibfnamefont {M.}~\bibnamefont {Meineri}},\ and\ \bibinfo
  {author} {\bibfnamefont {H.}~\bibnamefont {Ooguri}},\ }\bibfield  {title}
  {\bibinfo {title} {{Continuous Family of Conformal Field Theories and Exactly
  Marginal Operators}},\ }\href@noop {} {\bibfield  {journal} {\bibinfo
  {journal} {arXiv}\ } (\bibinfo {year} {2025})},\ \Eprint
  {https://arxiv.org/abs/2512.11045} {arXiv:2512.11045 [hep-th]} \BibitemShut
  {NoStop}%
\bibitem [{\citenamefont {Cohen}(1996)}]{Cohen:1996ng}%
  \BibitemOpen
  \bibfield  {author} {\bibinfo {author} {\bibfnamefont {T.~D.}\ \bibnamefont
  {Cohen}},\ }\bibfield  {title} {\bibinfo {title} {{The High temperature phase
  of QCD and U(1)-A symmetry}},\ }\href
  {https://doi.org/10.1103/PhysRevD.54.R1867} {\bibfield  {journal} {\bibinfo
  {journal} {Phys. Rev. D}\ }\textbf {\bibinfo {volume} {54}},\ \bibinfo
  {pages} {R1867} (\bibinfo {year} {1996})},\ \Eprint
  {https://arxiv.org/abs/hep-ph/9601216} {arXiv:hep-ph/9601216} \BibitemShut
  {NoStop}%
\bibitem [{\citenamefont {Lee}\ and\ \citenamefont
  {Hatsuda}(1996)}]{Lee:1996zy}%
  \BibitemOpen
  \bibfield  {author} {\bibinfo {author} {\bibfnamefont {S.~H.}\ \bibnamefont
  {Lee}}\ and\ \bibinfo {author} {\bibfnamefont {T.}~\bibnamefont {Hatsuda}},\
  }\bibfield  {title} {\bibinfo {title} {{U-a(1) symmetry restoration in QCD
  with N(f) flavors}},\ }\href {https://doi.org/10.1103/PhysRevD.54.R1871}
  {\bibfield  {journal} {\bibinfo  {journal} {Phys. Rev. D}\ }\textbf {\bibinfo
  {volume} {54}},\ \bibinfo {pages} {R1871} (\bibinfo {year} {1996})},\ \Eprint
  {https://arxiv.org/abs/hep-ph/9601373} {arXiv:hep-ph/9601373} \BibitemShut
  {NoStop}%
\bibitem [{\citenamefont {Cohen}\ and\ \citenamefont
  {Glozman}(2002)}]{Cohen:2002st}%
  \BibitemOpen
  \bibfield  {author} {\bibinfo {author} {\bibfnamefont {T.~D.}\ \bibnamefont
  {Cohen}}\ and\ \bibinfo {author} {\bibfnamefont {L.~Y.}\ \bibnamefont
  {Glozman}},\ }\bibfield  {title} {\bibinfo {title} {{Does one observe chiral
  symmetry restoration in baryon spectrum?}},\ }\href
  {https://doi.org/10.1142/S0217751X02009679} {\bibfield  {journal} {\bibinfo
  {journal} {Int. J. Mod. Phys. A}\ }\textbf {\bibinfo {volume} {17}},\
  \bibinfo {pages} {1327} (\bibinfo {year} {2002})},\ \Eprint
  {https://arxiv.org/abs/hep-ph/0201242} {arXiv:hep-ph/0201242} \BibitemShut
  {NoStop}%
\bibitem [{\citenamefont {Fukushima}(2008)}]{Fukushima:2008wg}%
  \BibitemOpen
  \bibfield  {author} {\bibinfo {author} {\bibfnamefont {K.}~\bibnamefont
  {Fukushima}},\ }\bibfield  {title} {\bibinfo {title} {{Phase diagrams in the
  three-flavor Nambu-Jona-Lasinio model with the Polyakov loop}},\ }\href
  {https://doi.org/10.1103/PhysRevD.77.114028} {\bibfield  {journal} {\bibinfo
  {journal} {Phys. Rev. D}\ }\textbf {\bibinfo {volume} {77}},\ \bibinfo
  {pages} {114028} (\bibinfo {year} {2008})},\ \bibinfo {note} {[Erratum:
  Phys.Rev.D 78, 039902 (2008)]},\ \Eprint {https://arxiv.org/abs/0803.3318}
  {arXiv:0803.3318 [hep-ph]} \BibitemShut {NoStop}%
\bibitem [{\citenamefont {Aoki}\ \emph {et~al.}(2012)\citenamefont {Aoki},
  \citenamefont {Fukaya},\ and\ \citenamefont {Taniguchi}}]{Aoki:2012yj}%
  \BibitemOpen
  \bibfield  {author} {\bibinfo {author} {\bibfnamefont {S.}~\bibnamefont
  {Aoki}}, \bibinfo {author} {\bibfnamefont {H.}~\bibnamefont {Fukaya}},\ and\
  \bibinfo {author} {\bibfnamefont {Y.}~\bibnamefont {Taniguchi}},\ }\bibfield
  {title} {\bibinfo {title} {{Chiral symmetry restoration, eigenvalue density
  of Dirac operator and axial U(1) anomaly at finite temperature}},\ }\href
  {https://doi.org/10.1103/PhysRevD.86.114512} {\bibfield  {journal} {\bibinfo
  {journal} {Phys. Rev. D}\ }\textbf {\bibinfo {volume} {86}},\ \bibinfo
  {pages} {114512} (\bibinfo {year} {2012})},\ \Eprint
  {https://arxiv.org/abs/1209.2061} {arXiv:1209.2061 [hep-lat]} \BibitemShut
  {NoStop}%
\bibitem [{\citenamefont {Pelissetto}\ and\ \citenamefont
  {Vicari}(2013)}]{Pelissetto:2013hqa}%
  \BibitemOpen
  \bibfield  {author} {\bibinfo {author} {\bibfnamefont {A.}~\bibnamefont
  {Pelissetto}}\ and\ \bibinfo {author} {\bibfnamefont {E.}~\bibnamefont
  {Vicari}},\ }\bibfield  {title} {\bibinfo {title} {{Relevance of the axial
  anomaly at the finite-temperature chiral transition in QCD}},\ }\href
  {https://doi.org/10.1103/PhysRevD.88.105018} {\bibfield  {journal} {\bibinfo
  {journal} {Phys. Rev. D}\ }\textbf {\bibinfo {volume} {88}},\ \bibinfo
  {pages} {105018} (\bibinfo {year} {2013})},\ \Eprint
  {https://arxiv.org/abs/1309.5446} {arXiv:1309.5446 [hep-lat]} \BibitemShut
  {NoStop}%
\bibitem [{\citenamefont {Cossu}\ \emph {et~al.}(2013)\citenamefont {Cossu},
  \citenamefont {Aoki}, \citenamefont {Fukaya}, \citenamefont {Hashimoto},
  \citenamefont {Kaneko}, \citenamefont {Matsufuru},\ and\ \citenamefont
  {Noaki}}]{Cossu:2013uua}%
  \BibitemOpen
  \bibfield  {author} {\bibinfo {author} {\bibfnamefont {G.}~\bibnamefont
  {Cossu}}, \bibinfo {author} {\bibfnamefont {S.}~\bibnamefont {Aoki}},
  \bibinfo {author} {\bibfnamefont {H.}~\bibnamefont {Fukaya}}, \bibinfo
  {author} {\bibfnamefont {S.}~\bibnamefont {Hashimoto}}, \bibinfo {author}
  {\bibfnamefont {T.}~\bibnamefont {Kaneko}}, \bibinfo {author} {\bibfnamefont
  {H.}~\bibnamefont {Matsufuru}},\ and\ \bibinfo {author} {\bibfnamefont
  {J.-I.}\ \bibnamefont {Noaki}},\ }\bibfield  {title} {\bibinfo {title}
  {{Finite temperature study of the axial U(1) symmetry on the lattice with
  overlap fermion formulation}},\ }\href
  {https://doi.org/10.1103/PhysRevD.87.114514} {\bibfield  {journal} {\bibinfo
  {journal} {Phys. Rev. D}\ }\textbf {\bibinfo {volume} {87}},\ \bibinfo
  {pages} {114514} (\bibinfo {year} {2013})},\ \bibinfo {note} {[Erratum:
  Phys.Rev.D 88, 019901 (2013)]},\ \Eprint {https://arxiv.org/abs/1304.6145}
  {arXiv:1304.6145 [hep-lat]} \BibitemShut {NoStop}%
\bibitem [{\citenamefont {Wilson}\ and\ \citenamefont
  {Kogut}(1974)}]{Wilson:1973jj}%
  \BibitemOpen
  \bibfield  {author} {\bibinfo {author} {\bibfnamefont {K.~G.}\ \bibnamefont
  {Wilson}}\ and\ \bibinfo {author} {\bibfnamefont {J.~B.}\ \bibnamefont
  {Kogut}},\ }\bibfield  {title} {\bibinfo {title} {{The Renormalization group
  and the epsilon expansion}},\ }\href
  {https://doi.org/10.1016/0370-1573(74)90023-4} {\bibfield  {journal}
  {\bibinfo  {journal} {Phys. Rept.}\ }\textbf {\bibinfo {volume} {12}},\
  \bibinfo {pages} {75} (\bibinfo {year} {1974})}\BibitemShut {NoStop}%
\bibitem [{\citenamefont {Weinberg}(1976)}]{Weinberg:1976xy}%
  \BibitemOpen
  \bibfield  {author} {\bibinfo {author} {\bibfnamefont {S.}~\bibnamefont
  {Weinberg}},\ }\bibfield  {title} {\bibinfo {title} {{Critical Phenomena for
  Field Theorists}},\ }in\ \href {https://doi.org/10.1007/978-1-4684-0931-4_1}
  {\emph {\bibinfo {booktitle} {{14th International School of Subnuclear
  Physics: Understanding the Fundamental Constitutents of Matter}}}}\ (\bibinfo
  {year} {1976})\BibitemShut {NoStop}%
\end{thebibliography}%


\begin{thebibliography}{47}%
\makeatletter
\providecommand \@ifxundefined [1]{%
 \@ifx{#1\undefined}
}%
\providecommand \@ifnum [1]{%
 \ifnum #1\expandafter \@firstoftwo
 \else \expandafter \@secondoftwo
 \fi
}%
\providecommand \@ifx [1]{%
 \ifx #1\expandafter \@firstoftwo
 \else \expandafter \@secondoftwo
 \fi
}%
\providecommand \natexlab [1]{#1}%
\providecommand \enquote  [1]{``#1''}%
\providecommand \bibnamefont  [1]{#1}%
\providecommand \bibfnamefont [1]{#1}%
\providecommand \citenamefont [1]{#1}%
\providecommand \href@noop [0]{\@secondoftwo}%
\providecommand \href [0]{\begingroup \@sanitize@url \@href}%
\providecommand \@href[1]{\@@startlink{#1}\@@href}%
\providecommand \@@href[1]{\endgroup#1\@@endlink}%
\providecommand \@sanitize@url [0]{\catcode `\\12\catcode `\$12\catcode
  `\&12\catcode `\#12\catcode `\^12\catcode `\_12\catcode `\%12\relax}%
\providecommand \@@startlink[1]{}%
\providecommand \@@endlink[0]{}%
\providecommand \url  [0]{\begingroup\@sanitize@url \@url }%
\providecommand \@url [1]{\endgroup\@href {#1}{\urlprefix }}%
\providecommand \urlprefix  [0]{URL }%
\providecommand \Eprint [0]{\href }%
\providecommand \doibase [0]{https://doi.org/}%
\providecommand \selectlanguage [0]{\@gobble}%
\providecommand \bibinfo  [0]{\@secondoftwo}%
\providecommand \bibfield  [0]{\@secondoftwo}%
\providecommand \translation [1]{[#1]}%
\providecommand \BibitemOpen [0]{}%
\providecommand \bibitemStop [0]{}%
\providecommand \bibitemNoStop [0]{.\EOS\space}%
\providecommand \EOS [0]{\spacefactor3000\relax}%
\providecommand \BibitemShut  [1]{\csname bibitem#1\endcsname}%
\let\auto@bib@innerbib\@empty
\bibitem [{\citenamefont {Kapusta}\ and\ \citenamefont
  {Gale}(2011)}]{KapustaGale201102}%
  \BibitemOpen
  \bibfield  {author} {\bibinfo {author} {\bibfnamefont {J.~I.}\ \bibnamefont
  {Kapusta}}\ and\ \bibinfo {author} {\bibfnamefont {C.}~\bibnamefont {Gale}},\
  }\href@noop {} {\emph {\bibinfo {title} {Finite-Temperature Field Theory:
  Principles and Applications (Cambridge Monographs on Mathematical
  Physics)}}},\ \bibinfo {edition} {2nd}\ ed.\ (\bibinfo  {publisher}
  {Cambridge University Press},\ \bibinfo {year} {2011})\BibitemShut {NoStop}%
\bibitem [{\citenamefont {Roberge}\ and\ \citenamefont
  {Weiss}(1986)}]{Roberge:1986mm}%
  \BibitemOpen
  \bibfield  {author} {\bibinfo {author} {\bibfnamefont {A.}~\bibnamefont
  {Roberge}}\ and\ \bibinfo {author} {\bibfnamefont {N.}~\bibnamefont
  {Weiss}},\ }\bibfield  {title} {\bibinfo {title} {{Gauge Theories With
  Imaginary Chemical Potential and the Phases of {QCD}}},\ }\href
  {https://doi.org/10.1016/0550-3213(86)90582-1} {\bibfield  {journal}
  {\bibinfo  {journal} {Nucl. Phys.}\ }\textbf {\bibinfo {volume} {B275}},\
  \bibinfo {pages} {734} (\bibinfo {year} {1986})}\BibitemShut {NoStop}%
\bibitem [{\citenamefont {Kallosh}\ \emph {et~al.}(1995)\citenamefont
  {Kallosh}, \citenamefont {Linde}, \citenamefont {Linde},\ and\ \citenamefont
  {Susskind}}]{Kallosh:1995hi}%
  \BibitemOpen
  \bibfield  {author} {\bibinfo {author} {\bibfnamefont {R.}~\bibnamefont
  {Kallosh}}, \bibinfo {author} {\bibfnamefont {A.~D.}\ \bibnamefont {Linde}},
  \bibinfo {author} {\bibfnamefont {D.~A.}\ \bibnamefont {Linde}},\ and\
  \bibinfo {author} {\bibfnamefont {L.}~\bibnamefont {Susskind}},\ }\bibfield
  {title} {\bibinfo {title} {{Gravity and global symmetries}},\ }\href
  {https://doi.org/10.1103/PhysRevD.52.912} {\bibfield  {journal} {\bibinfo
  {journal} {Phys. Rev. D}\ }\textbf {\bibinfo {volume} {52}},\ \bibinfo
  {pages} {912} (\bibinfo {year} {1995})},\ \Eprint
  {https://arxiv.org/abs/hep-th/9502069} {arXiv:hep-th/9502069} \BibitemShut
  {NoStop}%
\bibitem [{\citenamefont {Gabadadze}(1999)}]{Gabadadze:1999na}%
  \BibitemOpen
  \bibfield  {author} {\bibinfo {author} {\bibfnamefont {G.}~\bibnamefont
  {Gabadadze}},\ }\bibfield  {title} {\bibinfo {title} {{On field / string
  theory approach to theta dependence in large n Yang-Mills theory}},\ }\href
  {https://doi.org/10.1016/S0550-3213(99)00244-8} {\bibfield  {journal}
  {\bibinfo  {journal} {Nucl. Phys. B}\ }\textbf {\bibinfo {volume} {552}},\
  \bibinfo {pages} {194} (\bibinfo {year} {1999})},\ \Eprint
  {https://arxiv.org/abs/hep-th/9902191} {arXiv:hep-th/9902191} \BibitemShut
  {NoStop}%
\bibitem [{\citenamefont {Gabadadze}\ and\ \citenamefont
  {Shifman}(2000)}]{Gabadadze:2000vw}%
  \BibitemOpen
  \bibfield  {author} {\bibinfo {author} {\bibfnamefont {G.}~\bibnamefont
  {Gabadadze}}\ and\ \bibinfo {author} {\bibfnamefont {M.~A.}\ \bibnamefont
  {Shifman}},\ }\bibfield  {title} {\bibinfo {title} {{Vacuum structure and the
  axion walls in gluodynamics and QCD with light quarks}},\ }\href
  {https://doi.org/10.1103/PhysRevD.62.114003} {\bibfield  {journal} {\bibinfo
  {journal} {Phys. Rev. D}\ }\textbf {\bibinfo {volume} {62}},\ \bibinfo
  {pages} {114003} (\bibinfo {year} {2000})},\ \Eprint
  {https://arxiv.org/abs/hep-ph/0007345} {arXiv:hep-ph/0007345} \BibitemShut
  {NoStop}%
\bibitem [{\citenamefont {Gabadadze}\ and\ \citenamefont
  {Shifman}(2002)}]{Gabadadze:2002ff}%
  \BibitemOpen
  \bibfield  {author} {\bibinfo {author} {\bibfnamefont {G.}~\bibnamefont
  {Gabadadze}}\ and\ \bibinfo {author} {\bibfnamefont {M.}~\bibnamefont
  {Shifman}},\ }\bibfield  {title} {\bibinfo {title} {{QCD vacuum and axions:
  What's happening?}},\ }\href {https://doi.org/10.1142/S0217751X02011357}
  {\bibfield  {journal} {\bibinfo  {journal} {Int. J. Mod. Phys. A}\ }\textbf
  {\bibinfo {volume} {17}},\ \bibinfo {pages} {3689} (\bibinfo {year}
  {2002})},\ \Eprint {https://arxiv.org/abs/hep-ph/0206123}
  {arXiv:hep-ph/0206123} \BibitemShut {NoStop}%
\bibitem [{\citenamefont {Dvali}(2005)}]{Dvali:2005an}%
  \BibitemOpen
  \bibfield  {author} {\bibinfo {author} {\bibfnamefont {G.}~\bibnamefont
  {Dvali}},\ }\bibfield  {title} {\bibinfo {title} {{Three-form gauging of
  axion symmetries and gravity}},\ }\href@noop {} {\bibfield  {journal}
  {\bibinfo  {journal} {{}}\ } (\bibinfo {year} {2005})},\ \Eprint
  {https://arxiv.org/abs/hep-th/0507215} {arXiv:hep-th/0507215} \BibitemShut
  {NoStop}%
\bibitem [{\citenamefont {Burgess}\ \emph {et~al.}(2024)\citenamefont
  {Burgess}, \citenamefont {Choi},\ and\ \citenamefont
  {Quevedo}}]{Burgess:2023ifd}%
  \BibitemOpen
  \bibfield  {author} {\bibinfo {author} {\bibfnamefont {C.~P.}\ \bibnamefont
  {Burgess}}, \bibinfo {author} {\bibfnamefont {G.}~\bibnamefont {Choi}},\ and\
  \bibinfo {author} {\bibfnamefont {F.}~\bibnamefont {Quevedo}},\ }\bibfield
  {title} {\bibinfo {title} {{UV and IR effects in axion quality control}},\
  }\href {https://doi.org/10.1007/JHEP03(2024)051} {\bibfield  {journal}
  {\bibinfo  {journal} {JHEP}\ }\textbf {\bibinfo {volume} {03}},\ \bibinfo
  {pages} {051}},\ \Eprint {https://arxiv.org/abs/2301.00549} {arXiv:2301.00549
  [hep-th]} \BibitemShut {NoStop}%
\bibitem [{\citenamefont {Hellerman}\ \emph {et~al.}(2007)\citenamefont
  {Hellerman}, \citenamefont {Henriques}, \citenamefont {Pantev}, \citenamefont
  {Sharpe},\ and\ \citenamefont {Ando}}]{Hellerman:2006zs}%
  \BibitemOpen
  \bibfield  {author} {\bibinfo {author} {\bibfnamefont {S.}~\bibnamefont
  {Hellerman}}, \bibinfo {author} {\bibfnamefont {A.}~\bibnamefont
  {Henriques}}, \bibinfo {author} {\bibfnamefont {T.}~\bibnamefont {Pantev}},
  \bibinfo {author} {\bibfnamefont {E.}~\bibnamefont {Sharpe}},\ and\ \bibinfo
  {author} {\bibfnamefont {M.}~\bibnamefont {Ando}},\ }\bibfield  {title}
  {\bibinfo {title} {{Cluster decomposition, T-duality, and gerby CFT's}},\
  }\href {https://doi.org/10.4310/ATMP.2007.v11.n5.a2} {\bibfield  {journal}
  {\bibinfo  {journal} {Adv. Theor. Math. Phys.}\ }\textbf {\bibinfo {volume}
  {11}},\ \bibinfo {pages} {751} (\bibinfo {year} {2007})},\ \Eprint
  {https://arxiv.org/abs/hep-th/0606034} {arXiv:hep-th/0606034} \BibitemShut
  {NoStop}%
\bibitem [{\citenamefont {Sharpe}(2014)}]{Sharpe:2014tca}%
  \BibitemOpen
  \bibfield  {author} {\bibinfo {author} {\bibfnamefont {E.}~\bibnamefont
  {Sharpe}},\ }\bibfield  {title} {\bibinfo {title} {{Decomposition in diverse
  dimensions}},\ }\href {https://doi.org/10.1103/PhysRevD.90.025030} {\bibfield
   {journal} {\bibinfo  {journal} {Phys. Rev.}\ }\textbf {\bibinfo {volume}
  {D90}},\ \bibinfo {pages} {025030} (\bibinfo {year} {2014})},\ \Eprint
  {https://arxiv.org/abs/1404.3986} {arXiv:1404.3986 [hep-th]} \BibitemShut
  {NoStop}%
\bibitem [{\citenamefont {Tanizaki}\ and\ \citenamefont
  {Unsal}(2020)}]{Tanizaki:2019rbk}%
  \BibitemOpen
  \bibfield  {author} {\bibinfo {author} {\bibfnamefont {Y.}~\bibnamefont
  {Tanizaki}}\ and\ \bibinfo {author} {\bibfnamefont {M.}~\bibnamefont
  {Unsal}},\ }\bibfield  {title} {\bibinfo {title} {{Modified instanton sum in
  QCD and higher-groups}},\ }\href {https://doi.org/10.1007/JHEP03(2020)123}
  {\bibfield  {journal} {\bibinfo  {journal} {JHEP}\ }\textbf {\bibinfo
  {volume} {03}},\ \bibinfo {pages} {123}},\ \Eprint
  {https://arxiv.org/abs/1912.01033} {arXiv:1912.01033 [hep-th]} \BibitemShut
  {NoStop}%
\bibitem [{\citenamefont {Komargodski}\ \emph {et~al.}(2021)\citenamefont
  {Komargodski}, \citenamefont {Ohmori}, \citenamefont {Roumpedakis},\ and\
  \citenamefont {Seifnashri}}]{Komargodski:2020mxz}%
  \BibitemOpen
  \bibfield  {author} {\bibinfo {author} {\bibfnamefont {Z.}~\bibnamefont
  {Komargodski}}, \bibinfo {author} {\bibfnamefont {K.}~\bibnamefont {Ohmori}},
  \bibinfo {author} {\bibfnamefont {K.}~\bibnamefont {Roumpedakis}},\ and\
  \bibinfo {author} {\bibfnamefont {S.}~\bibnamefont {Seifnashri}},\ }\bibfield
   {title} {\bibinfo {title} {{Symmetries and strings of adjoint QCD$_{2}$}},\
  }\href {https://doi.org/10.1007/JHEP03(2021)103} {\bibfield  {journal}
  {\bibinfo  {journal} {JHEP}\ }\textbf {\bibinfo {volume} {03}},\ \bibinfo
  {pages} {103}},\ \Eprint {https://arxiv.org/abs/2008.07567} {arXiv:2008.07567
  [hep-th]} \BibitemShut {NoStop}%
\bibitem [{\citenamefont {Cherman}\ and\ \citenamefont
  {Jacobson}(2021)}]{Cherman:2020cvw}%
  \BibitemOpen
  \bibfield  {author} {\bibinfo {author} {\bibfnamefont {A.}~\bibnamefont
  {Cherman}}\ and\ \bibinfo {author} {\bibfnamefont {T.}~\bibnamefont
  {Jacobson}},\ }\bibfield  {title} {\bibinfo {title} {{Lifetimes of near
  eternal false vacua}},\ }\href {https://doi.org/10.1103/PhysRevD.103.105012}
  {\bibfield  {journal} {\bibinfo  {journal} {Phys. Rev. D}\ }\textbf {\bibinfo
  {volume} {103}},\ \bibinfo {pages} {105012} (\bibinfo {year} {2021})},\
  \Eprint {https://arxiv.org/abs/2012.10555} {arXiv:2012.10555 [hep-th]}
  \BibitemShut {NoStop}%
\bibitem [{\citenamefont {Lin}\ \emph {et~al.}(2025)\citenamefont {Lin},
  \citenamefont {Robbins},\ and\ \citenamefont {Roy}}]{Lin:2025oml}%
  \BibitemOpen
  \bibfield  {author} {\bibinfo {author} {\bibfnamefont {L.}~\bibnamefont
  {Lin}}, \bibinfo {author} {\bibfnamefont {D.}~\bibnamefont {Robbins}},\ and\
  \bibinfo {author} {\bibfnamefont {S.}~\bibnamefont {Roy}},\ }\bibfield
  {title} {\bibinfo {title} {{Decomposition and (non-invertible)
  ({\ensuremath{-}}1)-form symmetries from the symmetry topological field
  theory}},\ }\href {https://doi.org/10.1007/JHEP09(2025)131} {\bibfield
  {journal} {\bibinfo  {journal} {JHEP}\ }\textbf {\bibinfo {volume} {09}},\
  \bibinfo {pages} {131}},\ \Eprint {https://arxiv.org/abs/2503.21862}
  {arXiv:2503.21862 [hep-th]} \BibitemShut {NoStop}%
\bibitem [{\citenamefont {Sharpe}(2024)}]{Sharpe:2022ene}%
  \BibitemOpen
  \bibfield  {author} {\bibinfo {author} {\bibfnamefont {E.}~\bibnamefont
  {Sharpe}},\ }\bibinfo {title} {{An introduction to decomposition}}\ (\bibinfo
  {year} {2024})\ \Eprint {https://arxiv.org/abs/2204.09117} {arXiv:2204.09117
  [hep-th]} \BibitemShut {NoStop}%
\bibitem [{\citenamefont {Seiberg}\ and\ \citenamefont
  {Seifnashri}(2025)}]{Seiberg:2025bqy}%
  \BibitemOpen
  \bibfield  {author} {\bibinfo {author} {\bibfnamefont {N.}~\bibnamefont
  {Seiberg}}\ and\ \bibinfo {author} {\bibfnamefont {S.}~\bibnamefont
  {Seifnashri}},\ }\bibfield  {title} {\bibinfo {title} {{Symmetry
  transmutation and anomaly matching}},\ }\href
  {https://doi.org/10.1007/JHEP09(2025)014} {\bibfield  {journal} {\bibinfo
  {journal} {JHEP}\ }\textbf {\bibinfo {volume} {09}},\ \bibinfo {pages}
  {014}},\ \Eprint {https://arxiv.org/abs/2505.08618} {arXiv:2505.08618
  [hep-th]} \BibitemShut {NoStop}%
\bibitem [{\citenamefont {Kobayashi}\ \emph {et~al.}(2023)\citenamefont
  {Kobayashi}, \citenamefont {Yokokura},\ and\ \citenamefont
  {Yonekura}}]{Kobayashi:2023ajk}%
  \BibitemOpen
  \bibfield  {author} {\bibinfo {author} {\bibfnamefont {S.~K.}\ \bibnamefont
  {Kobayashi}}, \bibinfo {author} {\bibfnamefont {T.}~\bibnamefont
  {Yokokura}},\ and\ \bibinfo {author} {\bibfnamefont {K.}~\bibnamefont
  {Yonekura}},\ }\bibfield  {title} {\bibinfo {title} {{The QCD phase diagram
  in the space of imaginary chemical potential via {\textquoteright}t Hooft
  anomalies}},\ }\href {https://doi.org/10.1007/JHEP08(2023)132} {\bibfield
  {journal} {\bibinfo  {journal} {JHEP}\ }\textbf {\bibinfo {volume} {08}},\
  \bibinfo {pages} {132}},\ \Eprint {https://arxiv.org/abs/2305.01217}
  {arXiv:2305.01217 [hep-th]} \BibitemShut {NoStop}%
\bibitem [{\citenamefont {Cohen}(1996)}]{Cohen:1996ng}%
  \BibitemOpen
  \bibfield  {author} {\bibinfo {author} {\bibfnamefont {T.~D.}\ \bibnamefont
  {Cohen}},\ }\bibfield  {title} {\bibinfo {title} {{The High temperature phase
  of QCD and U(1)-A symmetry}},\ }\href
  {https://doi.org/10.1103/PhysRevD.54.R1867} {\bibfield  {journal} {\bibinfo
  {journal} {Phys. Rev. D}\ }\textbf {\bibinfo {volume} {54}},\ \bibinfo
  {pages} {R1867} (\bibinfo {year} {1996})},\ \Eprint
  {https://arxiv.org/abs/hep-ph/9601216} {arXiv:hep-ph/9601216} \BibitemShut
  {NoStop}%
\bibitem [{\citenamefont {Lee}\ and\ \citenamefont
  {Hatsuda}(1996)}]{Lee:1996zy}%
  \BibitemOpen
  \bibfield  {author} {\bibinfo {author} {\bibfnamefont {S.~H.}\ \bibnamefont
  {Lee}}\ and\ \bibinfo {author} {\bibfnamefont {T.}~\bibnamefont {Hatsuda}},\
  }\bibfield  {title} {\bibinfo {title} {{U-a(1) symmetry restoration in QCD
  with N(f) flavors}},\ }\href {https://doi.org/10.1103/PhysRevD.54.R1871}
  {\bibfield  {journal} {\bibinfo  {journal} {Phys. Rev. D}\ }\textbf {\bibinfo
  {volume} {54}},\ \bibinfo {pages} {R1871} (\bibinfo {year} {1996})},\ \Eprint
  {https://arxiv.org/abs/hep-ph/9601373} {arXiv:hep-ph/9601373} \BibitemShut
  {NoStop}%
\bibitem [{\citenamefont {Cohen}\ and\ \citenamefont
  {Glozman}(2002)}]{Cohen:2002st}%
  \BibitemOpen
  \bibfield  {author} {\bibinfo {author} {\bibfnamefont {T.~D.}\ \bibnamefont
  {Cohen}}\ and\ \bibinfo {author} {\bibfnamefont {L.~Y.}\ \bibnamefont
  {Glozman}},\ }\bibfield  {title} {\bibinfo {title} {{Does one observe chiral
  symmetry restoration in baryon spectrum?}},\ }\href
  {https://doi.org/10.1142/S0217751X02009679} {\bibfield  {journal} {\bibinfo
  {journal} {Int. J. Mod. Phys. A}\ }\textbf {\bibinfo {volume} {17}},\
  \bibinfo {pages} {1327} (\bibinfo {year} {2002})},\ \Eprint
  {https://arxiv.org/abs/hep-ph/0201242} {arXiv:hep-ph/0201242} \BibitemShut
  {NoStop}%
\bibitem [{\citenamefont {Fukushima}(2008)}]{Fukushima:2008wg}%
  \BibitemOpen
  \bibfield  {author} {\bibinfo {author} {\bibfnamefont {K.}~\bibnamefont
  {Fukushima}},\ }\bibfield  {title} {\bibinfo {title} {{Phase diagrams in the
  three-flavor Nambu-Jona-Lasinio model with the Polyakov loop}},\ }\href
  {https://doi.org/10.1103/PhysRevD.77.114028} {\bibfield  {journal} {\bibinfo
  {journal} {Phys. Rev. D}\ }\textbf {\bibinfo {volume} {77}},\ \bibinfo
  {pages} {114028} (\bibinfo {year} {2008})},\ \bibinfo {note} {[Erratum:
  Phys.Rev.D 78, 039902 (2008)]},\ \Eprint {https://arxiv.org/abs/0803.3318}
  {arXiv:0803.3318 [hep-ph]} \BibitemShut {NoStop}%
\bibitem [{\citenamefont {Aoki}\ \emph {et~al.}(2012)\citenamefont {Aoki},
  \citenamefont {Fukaya},\ and\ \citenamefont {Taniguchi}}]{Aoki:2012yj}%
  \BibitemOpen
  \bibfield  {author} {\bibinfo {author} {\bibfnamefont {S.}~\bibnamefont
  {Aoki}}, \bibinfo {author} {\bibfnamefont {H.}~\bibnamefont {Fukaya}},\ and\
  \bibinfo {author} {\bibfnamefont {Y.}~\bibnamefont {Taniguchi}},\ }\bibfield
  {title} {\bibinfo {title} {{Chiral symmetry restoration, eigenvalue density
  of Dirac operator and axial U(1) anomaly at finite temperature}},\ }\href
  {https://doi.org/10.1103/PhysRevD.86.114512} {\bibfield  {journal} {\bibinfo
  {journal} {Phys. Rev. D}\ }\textbf {\bibinfo {volume} {86}},\ \bibinfo
  {pages} {114512} (\bibinfo {year} {2012})},\ \Eprint
  {https://arxiv.org/abs/1209.2061} {arXiv:1209.2061 [hep-lat]} \BibitemShut
  {NoStop}%
\bibitem [{\citenamefont {Pelissetto}\ and\ \citenamefont
  {Vicari}(2013)}]{Pelissetto:2013hqa}%
  \BibitemOpen
  \bibfield  {author} {\bibinfo {author} {\bibfnamefont {A.}~\bibnamefont
  {Pelissetto}}\ and\ \bibinfo {author} {\bibfnamefont {E.}~\bibnamefont
  {Vicari}},\ }\bibfield  {title} {\bibinfo {title} {{Relevance of the axial
  anomaly at the finite-temperature chiral transition in QCD}},\ }\href
  {https://doi.org/10.1103/PhysRevD.88.105018} {\bibfield  {journal} {\bibinfo
  {journal} {Phys. Rev. D}\ }\textbf {\bibinfo {volume} {88}},\ \bibinfo
  {pages} {105018} (\bibinfo {year} {2013})},\ \Eprint
  {https://arxiv.org/abs/1309.5446} {arXiv:1309.5446 [hep-lat]} \BibitemShut
  {NoStop}%
\bibitem [{\citenamefont {Cossu}\ \emph {et~al.}(2013)\citenamefont {Cossu},
  \citenamefont {Aoki}, \citenamefont {Fukaya}, \citenamefont {Hashimoto},
  \citenamefont {Kaneko}, \citenamefont {Matsufuru},\ and\ \citenamefont
  {Noaki}}]{Cossu:2013uua}%
  \BibitemOpen
  \bibfield  {author} {\bibinfo {author} {\bibfnamefont {G.}~\bibnamefont
  {Cossu}}, \bibinfo {author} {\bibfnamefont {S.}~\bibnamefont {Aoki}},
  \bibinfo {author} {\bibfnamefont {H.}~\bibnamefont {Fukaya}}, \bibinfo
  {author} {\bibfnamefont {S.}~\bibnamefont {Hashimoto}}, \bibinfo {author}
  {\bibfnamefont {T.}~\bibnamefont {Kaneko}}, \bibinfo {author} {\bibfnamefont
  {H.}~\bibnamefont {Matsufuru}},\ and\ \bibinfo {author} {\bibfnamefont
  {J.-I.}\ \bibnamefont {Noaki}},\ }\bibfield  {title} {\bibinfo {title}
  {{Finite temperature study of the axial U(1) symmetry on the lattice with
  overlap fermion formulation}},\ }\href
  {https://doi.org/10.1103/PhysRevD.87.114514} {\bibfield  {journal} {\bibinfo
  {journal} {Phys. Rev. D}\ }\textbf {\bibinfo {volume} {87}},\ \bibinfo
  {pages} {114514} (\bibinfo {year} {2013})},\ \bibinfo {note} {[Erratum:
  Phys.Rev.D 88, 019901 (2013)]},\ \Eprint {https://arxiv.org/abs/1304.6145}
  {arXiv:1304.6145 [hep-lat]} \BibitemShut {NoStop}%
\bibitem [{\citenamefont {Pisarski}\ and\ \citenamefont
  {Rennecke}(2024)}]{Pisarski:2024esv}%
  \BibitemOpen
  \bibfield  {author} {\bibinfo {author} {\bibfnamefont {R.~D.}\ \bibnamefont
  {Pisarski}}\ and\ \bibinfo {author} {\bibfnamefont {F.}~\bibnamefont
  {Rennecke}},\ }\bibfield  {title} {\bibinfo {title} {{Conjectures about the
  Chiral Phase Transition in QCD from Anomalous Multi-Instanton
  Interactions}},\ }\href {https://doi.org/10.1103/PhysRevLett.132.251903}
  {\bibfield  {journal} {\bibinfo  {journal} {Phys. Rev. Lett.}\ }\textbf
  {\bibinfo {volume} {132}},\ \bibinfo {pages} {251903} (\bibinfo {year}
  {2024})},\ \Eprint {https://arxiv.org/abs/2401.06130} {arXiv:2401.06130
  [hep-ph]} \BibitemShut {NoStop}%
\bibitem [{\citenamefont {Nakayama}\ and\ \citenamefont
  {Ohtsuki}(2015)}]{Nakayama:2014sba}%
  \BibitemOpen
  \bibfield  {author} {\bibinfo {author} {\bibfnamefont {Y.}~\bibnamefont
  {Nakayama}}\ and\ \bibinfo {author} {\bibfnamefont {T.}~\bibnamefont
  {Ohtsuki}},\ }\bibfield  {title} {\bibinfo {title} {{Bootstrapping phase
  transitions in QCD and frustrated spin systems}},\ }\href
  {https://doi.org/10.1103/PhysRevD.91.021901} {\bibfield  {journal} {\bibinfo
  {journal} {Phys. Rev. D}\ }\textbf {\bibinfo {volume} {91}},\ \bibinfo
  {pages} {021901} (\bibinfo {year} {2015})},\ \Eprint
  {https://arxiv.org/abs/1407.6195} {arXiv:1407.6195 [hep-th]} \BibitemShut
  {NoStop}%
\bibitem [{\citenamefont {Henriksson}\ \emph {et~al.}(2020)\citenamefont
  {Henriksson}, \citenamefont {Kousvos},\ and\ \citenamefont
  {Stergiou}}]{Henriksson:2020fqi}%
  \BibitemOpen
  \bibfield  {author} {\bibinfo {author} {\bibfnamefont {J.}~\bibnamefont
  {Henriksson}}, \bibinfo {author} {\bibfnamefont {S.~R.}\ \bibnamefont
  {Kousvos}},\ and\ \bibinfo {author} {\bibfnamefont {A.}~\bibnamefont
  {Stergiou}},\ }\bibfield  {title} {\bibinfo {title} {{Analytic and Numerical
  Bootstrap of CFTs with $O(m)\times O(n)$ Global Symmetry in 3D}},\ }\href
  {https://doi.org/10.21468/SciPostPhys.9.3.035} {\bibfield  {journal}
  {\bibinfo  {journal} {SciPost Phys.}\ }\textbf {\bibinfo {volume} {9}},\
  \bibinfo {pages} {035} (\bibinfo {year} {2020})},\ \Eprint
  {https://arxiv.org/abs/2004.14388} {arXiv:2004.14388 [hep-th]} \BibitemShut
  {NoStop}%
\bibitem [{\citenamefont {Yabunaka}\ and\ \citenamefont
  {Delamotte}(2017)}]{Yabunaka:2017uox}%
  \BibitemOpen
  \bibfield  {author} {\bibinfo {author} {\bibfnamefont {S.}~\bibnamefont
  {Yabunaka}}\ and\ \bibinfo {author} {\bibfnamefont {B.}~\bibnamefont
  {Delamotte}},\ }\bibfield  {title} {\bibinfo {title} {{Surprises in $O(N)$
  Models: Nonperturbative Fixed Points, Large $N$ Limits, and
  Multicriticality}},\ }\href {https://doi.org/10.1103/PhysRevLett.119.191602}
  {\bibfield  {journal} {\bibinfo  {journal} {Phys. Rev. Lett.}\ }\textbf
  {\bibinfo {volume} {119}},\ \bibinfo {pages} {191602} (\bibinfo {year}
  {2017})},\ \Eprint {https://arxiv.org/abs/1707.04383} {arXiv:1707.04383
  [cond-mat.stat-mech]} \BibitemShut {NoStop}%
\bibitem [{\citenamefont {Yabunaka}\ and\ \citenamefont
  {Delamotte}(2018)}]{Yabunaka:2018mju}%
  \BibitemOpen
  \bibfield  {author} {\bibinfo {author} {\bibfnamefont {S.}~\bibnamefont
  {Yabunaka}}\ and\ \bibinfo {author} {\bibfnamefont {B.}~\bibnamefont
  {Delamotte}},\ }\bibfield  {title} {\bibinfo {title} {{Why Might the Standard
  Large $N$ Analysis Fail in the O($N$) Model: The Role of Cusps in the Fixed
  Point Potentials}},\ }\href {https://doi.org/10.1103/PhysRevLett.121.231601}
  {\bibfield  {journal} {\bibinfo  {journal} {Phys. Rev. Lett.}\ }\textbf
  {\bibinfo {volume} {121}},\ \bibinfo {pages} {231601} (\bibinfo {year}
  {2018})},\ \Eprint {https://arxiv.org/abs/1807.04681} {arXiv:1807.04681
  [cond-mat.stat-mech]} \BibitemShut {NoStop}%
\bibitem [{\citenamefont {Sorokin}(2021)}]{Sorokin:2021jwf}%
  \BibitemOpen
  \bibfield  {author} {\bibinfo {author} {\bibfnamefont {A.~O.}\ \bibnamefont
  {Sorokin}},\ }\bibfield  {title} {\bibinfo {title} {{First-order and
  pseudo-first-order transition in the high dimensional $O(N)\otimes O(M)$
  model}},\ }\href@noop {} {\bibfield  {journal} {\bibinfo  {journal} {arXiv}\
  } (\bibinfo {year} {2021})},\ \Eprint {https://arxiv.org/abs/2105.00072}
  {arXiv:2105.00072 [cond-mat.str-el]} \BibitemShut {NoStop}%
\bibitem [{\citenamefont {Sorokin}(2022)}]{Sorokin:2022zwh}%
  \BibitemOpen
  \bibfield  {author} {\bibinfo {author} {\bibfnamefont {A.~O.}\ \bibnamefont
  {Sorokin}},\ }\bibfield  {title} {\bibinfo {title} {{Phase transition in the
  three-dimensional $O(N)\otimes O(M)$ model: a Monte Carlo study}},\
  }\href@noop {} {\bibfield  {journal} {\bibinfo  {journal} {arXiv}\ }
  (\bibinfo {year} {2022})},\ \Eprint {https://arxiv.org/abs/2205.07199}
  {arXiv:2205.07199 [hep-lat]} \BibitemShut {NoStop}%
\bibitem [{\citenamefont {S{\'a}nchez-Villalobos}\ \emph
  {et~al.}(2025)\citenamefont {S{\'a}nchez-Villalobos}, \citenamefont
  {Delamotte},\ and\ \citenamefont {Wschebor}}]{Sanchez-Villalobos:2024vmd}%
  \BibitemOpen
  \bibfield  {author} {\bibinfo {author} {\bibfnamefont {C.~A.}\ \bibnamefont
  {S{\'a}nchez-Villalobos}}, \bibinfo {author} {\bibfnamefont {B.}~\bibnamefont
  {Delamotte}},\ and\ \bibinfo {author} {\bibfnamefont {N.}~\bibnamefont
  {Wschebor}},\ }\bibfield  {title} {\bibinfo {title} {{O(N){\texttimes}O(2)
  scalar models: Including O({\ensuremath{\partial}}2) corrections in the
  functional renormalization group analysis}},\ }\href
  {https://doi.org/10.1103/PhysRevE.111.034104} {\bibfield  {journal} {\bibinfo
   {journal} {Phys. Rev. E}\ }\textbf {\bibinfo {volume} {111}},\ \bibinfo
  {pages} {034104} (\bibinfo {year} {2025})},\ \Eprint
  {https://arxiv.org/abs/2411.02616} {arXiv:2411.02616 [cond-mat.stat-mech]}
  \BibitemShut {NoStop}%
\bibitem [{\citenamefont {Pisarski}\ and\ \citenamefont
  {Wilczek}(1984)}]{Pisarski:1983ms}%
  \BibitemOpen
  \bibfield  {author} {\bibinfo {author} {\bibfnamefont {R.~D.}\ \bibnamefont
  {Pisarski}}\ and\ \bibinfo {author} {\bibfnamefont {F.}~\bibnamefont
  {Wilczek}},\ }\bibfield  {title} {\bibinfo {title} {{Remarks on the Chiral
  Phase Transition in Chromodynamics}},\ }\href
  {https://doi.org/10.1103/PhysRevD.29.338} {\bibfield  {journal} {\bibinfo
  {journal} {Phys. Rev. D}\ }\textbf {\bibinfo {volume} {29}},\ \bibinfo
  {pages} {338} (\bibinfo {year} {1984})}\BibitemShut {NoStop}%
\bibitem [{\citenamefont {Butti}\ \emph {et~al.}(2003)\citenamefont {Butti},
  \citenamefont {Pelissetto},\ and\ \citenamefont {Vicari}}]{Butti:2003nu}%
  \BibitemOpen
  \bibfield  {author} {\bibinfo {author} {\bibfnamefont {A.}~\bibnamefont
  {Butti}}, \bibinfo {author} {\bibfnamefont {A.}~\bibnamefont {Pelissetto}},\
  and\ \bibinfo {author} {\bibfnamefont {E.}~\bibnamefont {Vicari}},\
  }\bibfield  {title} {\bibinfo {title} {{On the nature of the finite
  temperature transition in QCD}},\ }\href
  {https://doi.org/10.1088/1126-6708/2003/08/029} {\bibfield  {journal}
  {\bibinfo  {journal} {JHEP}\ }\textbf {\bibinfo {volume} {08}},\ \bibinfo
  {pages} {029}},\ \Eprint {https://arxiv.org/abs/hep-ph/0307036}
  {arXiv:hep-ph/0307036} \BibitemShut {NoStop}%
\bibitem [{\citenamefont {Calabrese}\ and\ \citenamefont
  {Parruccini}(2004)}]{Calabrese:2004uk}%
  \BibitemOpen
  \bibfield  {author} {\bibinfo {author} {\bibfnamefont {P.}~\bibnamefont
  {Calabrese}}\ and\ \bibinfo {author} {\bibfnamefont {P.}~\bibnamefont
  {Parruccini}},\ }\bibfield  {title} {\bibinfo {title} {{Five loop epsilon
  expansion for U(n) x U(m) models: Finite temperature phase transition in
  light QCD}},\ }\href {https://doi.org/10.1088/1126-6708/2004/05/018}
  {\bibfield  {journal} {\bibinfo  {journal} {JHEP}\ }\textbf {\bibinfo
  {volume} {05}},\ \bibinfo {pages} {018}},\ \Eprint
  {https://arxiv.org/abs/hep-ph/0403140} {arXiv:hep-ph/0403140} \BibitemShut
  {NoStop}%
\bibitem [{\citenamefont {Adzhemyan}\ \emph {et~al.}(2022)\citenamefont
  {Adzhemyan}, \citenamefont {Ivanova}, \citenamefont {Kompaniets},
  \citenamefont {Kudlis},\ and\ \citenamefont {Sokolov}}]{Adzhemyan:2021sug}%
  \BibitemOpen
  \bibfield  {author} {\bibinfo {author} {\bibfnamefont {L.~T.}\ \bibnamefont
  {Adzhemyan}}, \bibinfo {author} {\bibfnamefont {E.~V.}\ \bibnamefont
  {Ivanova}}, \bibinfo {author} {\bibfnamefont {M.~V.}\ \bibnamefont
  {Kompaniets}}, \bibinfo {author} {\bibfnamefont {A.}~\bibnamefont {Kudlis}},\
  and\ \bibinfo {author} {\bibfnamefont {A.~I.}\ \bibnamefont {Sokolov}},\
  }\bibfield  {title} {\bibinfo {title} {{Six-loop {\ensuremath{\varepsilon}}
  expansion of three-dimensional U(n){\texttimes}U(m) models}},\ }\href
  {https://doi.org/10.1016/j.nuclphysb.2022.115680} {\bibfield  {journal}
  {\bibinfo  {journal} {Nucl. Phys. B}\ }\textbf {\bibinfo {volume} {975}},\
  \bibinfo {pages} {115680} (\bibinfo {year} {2022})},\ \Eprint
  {https://arxiv.org/abs/2104.12195} {arXiv:2104.12195 [hep-th]} \BibitemShut
  {NoStop}%
\bibitem [{\citenamefont {Fejos}(2014)}]{Fejos:2014qga}%
  \BibitemOpen
  \bibfield  {author} {\bibinfo {author} {\bibfnamefont {G.}~\bibnamefont
  {Fejos}},\ }\bibfield  {title} {\bibinfo {title} {{Fluctuation induced first
  order phase transition in $U(n)\times U(n)$ models using chiral invariant
  expansion of functional renormalization group flows}},\ }\href
  {https://doi.org/10.1103/PhysRevD.90.096011} {\bibfield  {journal} {\bibinfo
  {journal} {Phys. Rev. D}\ }\textbf {\bibinfo {volume} {90}},\ \bibinfo
  {pages} {096011} (\bibinfo {year} {2014})},\ \Eprint
  {https://arxiv.org/abs/1409.3695} {arXiv:1409.3695 [hep-ph]} \BibitemShut
  {NoStop}%
\bibitem [{\citenamefont {Resch}\ \emph {et~al.}(2019)\citenamefont {Resch},
  \citenamefont {Rennecke},\ and\ \citenamefont {Schaefer}}]{Resch:2017vjs}%
  \BibitemOpen
  \bibfield  {author} {\bibinfo {author} {\bibfnamefont {S.}~\bibnamefont
  {Resch}}, \bibinfo {author} {\bibfnamefont {F.}~\bibnamefont {Rennecke}},\
  and\ \bibinfo {author} {\bibfnamefont {B.-J.}\ \bibnamefont {Schaefer}},\
  }\bibfield  {title} {\bibinfo {title} {{Mass sensitivity of the three-flavor
  chiral phase transition}},\ }\href
  {https://doi.org/10.1103/PhysRevD.99.076005} {\bibfield  {journal} {\bibinfo
  {journal} {Phys. Rev. D}\ }\textbf {\bibinfo {volume} {99}},\ \bibinfo
  {pages} {076005} (\bibinfo {year} {2019})},\ \Eprint
  {https://arxiv.org/abs/1712.07961} {arXiv:1712.07961 [hep-ph]} \BibitemShut
  {NoStop}%
\bibitem [{\citenamefont {Fejos}(2022)}]{Fejos:2022mso}%
  \BibitemOpen
  \bibfield  {author} {\bibinfo {author} {\bibfnamefont {G.}~\bibnamefont
  {Fejos}},\ }\bibfield  {title} {\bibinfo {title} {{Second-order chiral phase
  transition in three-flavor quantum chromodynamics?}},\ }\href
  {https://doi.org/10.1103/PhysRevD.105.L071506} {\bibfield  {journal}
  {\bibinfo  {journal} {Phys. Rev. D}\ }\textbf {\bibinfo {volume} {105}},\
  \bibinfo {pages} {L071506} (\bibinfo {year} {2022})},\ \Eprint
  {https://arxiv.org/abs/2201.07909} {arXiv:2201.07909 [hep-ph]} \BibitemShut
  {NoStop}%
\bibitem [{\citenamefont {Fejos}\ and\ \citenamefont
  {Hatsuda}(2024)}]{Fejos:2024bgl}%
  \BibitemOpen
  \bibfield  {author} {\bibinfo {author} {\bibfnamefont {G.}~\bibnamefont
  {Fejos}}\ and\ \bibinfo {author} {\bibfnamefont {T.}~\bibnamefont
  {Hatsuda}},\ }\bibfield  {title} {\bibinfo {title} {{Order of the
  SU(Nf){\texttimes}SU(Nf) chiral transition via the functional renormalization
  group}},\ }\href {https://doi.org/10.1103/PhysRevD.110.016021} {\bibfield
  {journal} {\bibinfo  {journal} {Phys. Rev. D}\ }\textbf {\bibinfo {volume}
  {110}},\ \bibinfo {pages} {016021} (\bibinfo {year} {2024})},\ \Eprint
  {https://arxiv.org/abs/2404.00554} {arXiv:2404.00554 [hep-ph]} \BibitemShut
  {NoStop}%
\bibitem [{\citenamefont {Hikami}(1981)}]{Hikami:1980hi}%
  \BibitemOpen
  \bibfield  {author} {\bibinfo {author} {\bibfnamefont {S.}~\bibnamefont
  {Hikami}},\ }\bibfield  {title} {\bibinfo {title} {{Three Loop Beta -
  Functions of Nonlinear Sigma Models on Symmetric Spaces}},\ }\href
  {https://doi.org/10.1016/0370-2693(81)90989-8} {\bibfield  {journal}
  {\bibinfo  {journal} {Phys. Lett. B}\ }\textbf {\bibinfo {volume} {98}},\
  \bibinfo {pages} {208} (\bibinfo {year} {1981})}\BibitemShut {NoStop}%
\bibitem [{\citenamefont {Hikami}(1983)}]{Hikami:1982ak}%
  \BibitemOpen
  \bibfield  {author} {\bibinfo {author} {\bibfnamefont {S.}~\bibnamefont
  {Hikami}},\ }\bibfield  {title} {\bibinfo {title} {{Isomorphism and beta
  function of the nonlinear sigma model in symmetric spaces}},\ }\href
  {https://doi.org/10.1016/0550-3213(83)90260-2} {\bibfield  {journal}
  {\bibinfo  {journal} {Nucl. Phys. B}\ }\textbf {\bibinfo {volume} {215}},\
  \bibinfo {pages} {555} (\bibinfo {year} {1983})}\BibitemShut {NoStop}%
\bibitem [{\citenamefont {Wegner}(1989)}]{Wegner:1989ss}%
  \BibitemOpen
  \bibfield  {author} {\bibinfo {author} {\bibfnamefont {F.}~\bibnamefont
  {Wegner}},\ }\bibfield  {title} {\bibinfo {title} {{Four Loop Order Beta
  Function of Nonlinear $\sigma$ Models in Symmetric Spaces}},\ }\href
  {https://doi.org/10.1016/0550-3213(89)90063-1} {\bibfield  {journal}
  {\bibinfo  {journal} {Nucl. Phys. B}\ }\textbf {\bibinfo {volume} {316}},\
  \bibinfo {pages} {663} (\bibinfo {year} {1989})}\BibitemShut {NoStop}%
\bibitem [{\citenamefont {McKane}\ and\ \citenamefont
  {Stone}(1980)}]{McKane:1979cm}%
  \BibitemOpen
  \bibfield  {author} {\bibinfo {author} {\bibfnamefont {A.}~\bibnamefont
  {McKane}}\ and\ \bibinfo {author} {\bibfnamefont {M.}~\bibnamefont {Stone}},\
  }\bibfield  {title} {\bibinfo {title} {{Nonlinear sigma models: a
  perturbative approach to symmetry restoration}},\ }\href
  {https://doi.org/10.1016/0550-3213(80)90396-X} {\bibfield  {journal}
  {\bibinfo  {journal} {Nucl. Phys. B}\ }\textbf {\bibinfo {volume} {163}},\
  \bibinfo {pages} {169} (\bibinfo {year} {1980})}\BibitemShut {NoStop}%
\bibitem [{\citenamefont {Henriksson}(2023)}]{Henriksson:2022rnm}%
  \BibitemOpen
  \bibfield  {author} {\bibinfo {author} {\bibfnamefont {J.}~\bibnamefont
  {Henriksson}},\ }\bibfield  {title} {\bibinfo {title} {{The critical O(N)
  CFT: Methods and conformal data}},\ }\href
  {https://doi.org/10.1016/j.physrep.2022.12.002} {\bibfield  {journal}
  {\bibinfo  {journal} {Phys. Rept.}\ }\textbf {\bibinfo {volume} {1002}},\
  \bibinfo {pages} {1} (\bibinfo {year} {2023})},\ \Eprint
  {https://arxiv.org/abs/2201.09520} {arXiv:2201.09520 [hep-th]} \BibitemShut
  {NoStop}%
\bibitem [{\citenamefont {Dyson}(1952)}]{Dyson:1952tj}%
  \BibitemOpen
  \bibfield  {author} {\bibinfo {author} {\bibfnamefont {F.~J.}\ \bibnamefont
  {Dyson}},\ }\bibfield  {title} {\bibinfo {title} {{Divergence of perturbation
  theory in quantum electrodynamics}},\ }\href
  {https://doi.org/10.1103/PhysRev.85.631} {\bibfield  {journal} {\bibinfo
  {journal} {Phys. Rev.}\ }\textbf {\bibinfo {volume} {85}},\ \bibinfo {pages}
  {631} (\bibinfo {year} {1952})}\BibitemShut {NoStop}%
\bibitem [{\citenamefont {Aniceto}\ \emph {et~al.}(2019)\citenamefont
  {Aniceto}, \citenamefont {Basar},\ and\ \citenamefont
  {Schiappa}}]{Aniceto:2018bis}%
  \BibitemOpen
  \bibfield  {author} {\bibinfo {author} {\bibfnamefont {I.}~\bibnamefont
  {Aniceto}}, \bibinfo {author} {\bibfnamefont {G.}~\bibnamefont {Basar}},\
  and\ \bibinfo {author} {\bibfnamefont {R.}~\bibnamefont {Schiappa}},\
  }\bibfield  {title} {\bibinfo {title} {{A Primer on Resurgent Transseries and
  Their Asymptotics}},\ }\href {https://doi.org/10.1016/j.physrep.2019.02.003}
  {\bibfield  {journal} {\bibinfo  {journal} {Phys. Rept.}\ }\textbf {\bibinfo
  {volume} {809}},\ \bibinfo {pages} {1} (\bibinfo {year} {2019})},\ \Eprint
  {https://arxiv.org/abs/1802.10441} {arXiv:1802.10441 [hep-th]} \BibitemShut
  {NoStop}%
\end{thebibliography}%

\end{document}


\title{Supplemental Material for ``Does hot QCD have a conformal manifold in the chiral limit?''}

\author{Shi Chen \orcidlink{0000-0002-8554-5098}}
\affiliation{School of Physics and Astronomy, University of Minnesota, Minneapolis, MN 55455, USA}
\author{Aleksey Cherman \orcidlink{0000-0002-1039-8476}}
\affiliation{School of Physics and Astronomy, University of Minnesota, Minneapolis, MN 55455, USA}
\author{Robert D. Pisarski \orcidlink{0000-0002-7862-4759}}
\affiliation{Physics Department, Brookhaven National Laboratory,
Upton, NY 11973-5000}

\date{\today}

\maketitle

\onecolumngrid
\begin{center}
\vspace{-2\baselineskip}
\href{mailto:s.chern.phys@gmail.com}{s.chern.phys@gmail.com}, \href{mailto:acherman@umn.edu}{acherman@umn.edu}, \href{mailto:pisarski@bnl.gov}{pisarski@bnl.gov}
\end{center}
\renewcommand{\thesection}{S\arabic{section}}
\numberwithin{equation}{section}

\section[Effective theory in the high-T gapped phase]{Effective theory in the high-$T$ gapped phase}
\label{sec:high_T_EFT}

According to perturbative calculations~\cite{KapustaGale201102, Roberge:1986mm}, the high-$T$ (i.e.~small-$\beta$) phase of massless QCD is gapped for all $\theta_B$. 
The free energy density $f_{\beta}(\theta_B)$ of Eq.~\eqref{main-eq:free_energy} at small $\beta$ is an even $2\pi$-periodic analytic function for every $\theta_B$ except at $\theta_B=\pi$ (mod $2\pi$), where it has a cusp-like maximum. 
All the propagating degrees of freedom are gapped out at distance scales $\ell \gg \beta$, so a 3D effective field theory at $\ell \gg \beta$ describes the colorless vacua only.
Such a 3D QFT is not completely trivial because the theory is not uniformly trivially gapped for all $\theta_B$.
It has a path-integral description in terms of two colorless fields, a real scalar field $\varphi$ and a 2-form $U(1)$ gauge field $b$ \cite{Kallosh:1995hi,Gabadadze:1999na,Gabadadze:2000vw,Gabadadze:2002ff,Dvali:2005an,Burgess:2023ifd}:
\begin{equation}\label{eq:high_T_EFT}
    \int\calD b\calD\varphi \exp\left\{ -\int\d^3x\,
    \bar{f}_{\beta}(\varphi) + \frac{i}{2\pi} \int\varphi\d b\right\}\,.
\end{equation}
The above $\bar{f}_{\beta}(\varphi)$ is an even analytic function of $\varphi$ such that
\begin{equation}
    \bar{f}_{\beta}(\varphi) 
    \begin{cases}
        = f_{\beta}(\varphi), & -\pi\le\varphi\le\pi \\
        > f_{\beta}(\varphi), & \text{for other }\varphi
    \end{cases}.
\end{equation}
This 3D model has a $2$-form $U(1)$ symmetry on $b$'s electric side and a $(-1)$-form $U(1)$ symmetry on $b$'s magnetic side.
With a background $3$-form $U(1)$ gauge field $C$ and a background $0$-form $U(1)$ gauge field (i.e.~a background $2\pi$-periodic scalar) $\theta_B$, the path integral reads
\begin{equation}\label{eq:high_T_EFT_background}
    \int\calD b\calD\varphi \exp\left\{ -\int\d^3x\,
    \bar{f}_{\beta}(\varphi) + \frac{i}{2\pi} \int (\varphi - \theta_B)(\d b-C)\right\}\,.
\end{equation}
This 3D model has no propagating degrees of freedom and describes infinitely many vacua (sometimes called ``universes''~\cite{Hellerman:2006zs,Sharpe:2014tca,Tanizaki:2019rbk,Komargodski:2020mxz,Cherman:2020cvw,Lin:2025oml,Sharpe:2022ene}) labeled by the $2$-form $U(1)$ symmetry charge $n\in\Z$.
For constant $\theta_B$, each vacuum has the energy density given by $\bar{f}_{\beta}(2\pi n + \theta_B)$, and two vacua cross at $\theta_B=\pi$.
The $2$-form $U(1)$ symmetry has to be emergent and only the lowest vacua are relevant for high-$T$ massless QCD.

To match the 3D 't Hooft anomaly~\eqref{main-eq:SPT_3D}, the high-$T$ effective theory~\eqref{eq:high_T_EFT} is supposed to see the full 3D symmetry~\eqref{main-eq:3D_IR_symmetry}.
The actions of discrete symmetries on the high-$T$ effective theory~\eqref{eq:high_T_EFT} are evident:
\begin{equation}
    \varphi(x)\xrightarrow{\calC}-\varphi(x)\,,\ 
    b(x)\xrightarrow{\calC}-b(x)\,,\qquad
    \varphi(x)\xrightarrow{\calS}-\varphi(x)\,,\ 
    b(x)\xrightarrow{\calS}-b(x)\,,\qquad
    \varphi(x)\xrightarrow{\calR}\varphi(\overline{x})\,,\ 
    b(x)\xrightarrow{\calR}-\overline{b}(\overline{x})\,.
\end{equation}
At first glance, it seems that there is no room for the $SU(N_f)_{L,R}$ symmetry to act, since Eq.~\eqref{eq:high_T_EFT_background} has no continuous 0-form symmetry of any sort.  
Nevertheless, the emergent $2$-form $U(1)$ symmetry can be a homomorphism target of the $0$-form $SU(N_f)_{L,R}$ symmetry.
Namely, the $A_{L,R}$ gauge fields can be coupled via the substitution 
\begin{equation}\label{eq:C=A^2}
    C = \frac{1}{4\pi}\tr\!\left( A_L \d A_L - A_R \d A_R 
    - \i\frac{2}{3} A_L^3 + \i\frac{2}{3} A_R^3 \right).
\end{equation}
The emergent $2$-form $U(1)$ symmetry has a mixed 't Hooft anomaly with the $(-1)$-form $U(1)$ symmetry, described by the 4D invertible theory
\begin{equation}\label{eq:high_T_anomaly}
    \exp\left\{\frac{i}{2\pi}\int\theta_B\d C\right\}.
\end{equation}
This emergent anomaly precisely reproduces the 3D anomaly~\eqref{main-eq:SPT_3D} via the substitution~\eqref{eq:C=A^2}.
Mathematically, such a homomorphism from $0$-form $SU(N_f)_{L,R}$ to $2$-form $U(1)$ is encoded in the classifying-space fibration $BSU(N_f)\to B^3U(1)$ that induces an isomorphism from $\pi_4(BSU(N_f))=\Z$ to $\pi_4(B^3U(1))=\Z$.

The above way of matching anomalies via flowing to an emergent higher-form symmetry was recently packaged into the notion of symmetry transmutation~\cite{Seiberg:2025bqy}.
So we can say that the $0$-form $SU(N_f)_{L,R}$ chiral symmetry of QCD is transmuted into a $2$-form $U(1)$ symmetry in its high-$T$ effective theory.

\section[Second-order phase transition at theta=pi in the high-T phase]{Second-order phase transition at $\theta_B=\pi$ in the high-$T$ phase}
\label{sec:anomalons}

In the high-$T$ phase, a window of second-order phase transitions at $\theta_B=\pi$ would have a chance to be realized at relatively lower $T$, if a 3D CFT consistent with symmetry and anomaly existed.
The upper edge of this gapless window would be a tricritical point, while the lower edge could be connected to the DQCP or the $SU(N_f)_{\pi}$ CFT on the critical line of chiral transitions.

One may speculate that such an appropriate 3D CFT can simply be provided by colorless free fermions, because the 3D anomaly~\eqref{main-eq:SPT_3D} can be matched by an $N_f$-flavor free massless 4D Dirac fermion at finite temperatures.
At $\theta_B=\pi$, one Matsubara mode becomes massless, leading to the 3D effective theory at $\ell\gg\beta$:
\begin{align}\label{eq:anomalon}
    \int\calD\psi\calD\bar{\psi_{1}}\calD\psi_{2}\calD\bar{\psi_{2}} 
    \ \exp\left\{-\int d^{3}x\, \Big(\bar{\psi}_1\slashed{\partial}\psi_1 +\bar{\psi_{2}}\slashed{\partial}\psi_{2}\Big)\right\}\,,
    \qquad(\theta_B=\pi)
\end{align}
with two $N_f$-flavor free massless 3D Dirac fermions $\psi_1$ and $\psi_2$.
At $\theta_B\neq\pi$, all the Matsubara modes are massive and these 3D fermions are gapped out at $\ell\gg\beta$.
This speculation has an interplay with the Landau-DQCP scenario.
Around the chiral transition, these 3D fermions have to talk to the chiral condensate $\mathcal{X}$ through Yukawa couplings,
\begin{equation}
    y_1\int d^3x \left(\bar{\psi}_{1} \mathcal{X} \mathcal{X}^{\dag} \psi_{1} - \bar{\psi_{2}} \mathcal{X}^{\dag} \mathcal{X} \psi_{2}\right)\,,
    \qquad
    y_2\int d^3x \left(\bar{\psi}_{1}\mathcal{X}\psi_{2} + \bar{\psi_{2}}\mathcal{X}^{\dag}\psi_{1}\right)\,,
    \qquad\text{etc.}
\end{equation}
Therefore, the DQCP at $\theta_B=\pi$ might be described by a certain 3D Gross-Neveu-Yukawa universality class, as long as the gapless window exists and has a free-fermion description.

Nevertheless, as already pointed out by Ref.~\cite{Kobayashi:2023ajk}, although it matches the anomaly~\eqref{main-eq:SPT_3D} of QCD, the free-fermion CFT~\eqref{eq:anomalon} does not match the symmetry~\eqref{main-eq:QCD_symmetry} of QCD.
A colorless fermionic object in the fundamental representation of $SU(N_f)_L$ or in the anti-fundamental representation of $SU(N_f)_R$ can exist in QCD only when 
\begin{equation}
    N_c = 1 \mod N_f
    \qquad\text{and}\qquad
    N_c = 1 \mod 2
\end{equation}
When this happens, this colorless fermionic object necessarily carries a baryon number $q_B$ satisfying $q_B = 1 \mod N_f$.
For example, with physical $N_c=3$, the above conditions are satisfied by $N_f=2$, but are violated by $N_f=3$.
Therefore, a gapless window at $\theta_B=\pi$ described by the free-fermion CFT~\eqref{eq:anomalon} is at least conceivable for $N_f=2$, but is immediately excluded for $N_f=3$.
Of course, a more complicated free-fermion CFT might simultaneously match the symmetry and the anomaly, and we leave a detailed exploration of such possibilities to the future.

\section[The fate of emergent U(1)A symmetry at the chiral transition]{The fate of emergent $U(1)_A$ symmetry at the chiral transition}
\label{sec:U1A}

It is conceivable that a $U(1)_A$ symmetry emerges on the critical line in Landau scenarios~\cite{Cohen:1996ng,Lee:1996zy,Cohen:2002st,Fukushima:2008wg,Aoki:2012yj,Aoki:2012yj,Pelissetto:2013hqa,Cossu:2013uua,Pisarski:2024esv}.
For this to happen, two requirements are needed: (I) the Ginzburg-Landau model~\eqref{main-eq:Landau-Ginzburg} with $\lambda_3=0$ must have an IR fixed point and (II) $\det\mathcal{X}$ must be an irrelevant operator at that fixed point.
Establishing (I) is already highly nontrivial since the $\lambda_2$ fluctuations destabilize the mean-field $O(2N_f^2)$ critical point.

When $N_f=2$, the model~\eqref{main-eq:Landau-Ginzburg} with $\lambda_3=0$ may flow to some Wilson-Fisher-like $O(4)\times O(2)$ fixed point~\cite{Pelissetto:2013hqa,Nakayama:2014sba,Henriksson:2020fqi} (see also~Refs.~\cite{Yabunaka:2017uox,Yabunaka:2018mju,Sorokin:2021jwf,Sorokin:2022zwh,Sanchez-Villalobos:2024vmd}).
But $\det\mathcal{X}\sim\mathcal{X}^2$ is relevant at that fixed point, since it
is literally a mass term.
When $N_f\ge3$, there is no convincing evidence for the existence of fixed points in the model~\eqref{main-eq:Landau-Ginzburg} with $\lambda_3=0$~\cite{Pisarski:1983ms,Butti:2003nu,Calabrese:2004uk,Adzhemyan:2021sug,Fejos:2014qga,Resch:2017vjs,Fejos:2022mso,Fejos:2024bgl}. 
We thus regard a second-order chiral transition with emergent $U(1)_A$ as disfavored in massless QCD for all $N_f\ge 2$ given current evidence.

Further searches for fixed points in the model~\eqref{main-eq:Landau-Ginzburg} with $\lambda_3=0$ using lattice Monte Carlo and other methods would still be valuable.
This is because if such a fixed point existed, $\det\mathcal{X}\sim\mathcal{X}^{N_f}$ should be irrelevant for sufficiently large $N_f$, and thus $U(1)_A$ would have a chance to emerge at QCD chiral transition with sufficiently large $N_f$.
Nevertheless, it is hard to imagine that anything like this can happen for $N_f=3$, since it seems very difficult to get $\det\mathcal{X}\sim\mathcal{X}^3$ to be irrelevant.

\section[Extrapolating from (2+epsilon)D to 3D]{Extrapolating from $(2+\epsilon)$D to $3$D}
\label{sec:epsilon_extrapolation}
The beta function of the $(2+\epsilon)$D $SU(N_f)$ PCM is known up to the four-loop order~\cite{Hikami:1980hi,Hikami:1982ak,Wegner:1989ss} (their coupling $t$ in the Lagrangian is related to our $g^2$ through $g^2=4t$):
\begin{align}
\begin{split}
    \beta(g^2) \ =&\: \ \epsilon g^2 -\frac{1}{4} N_f g^4
    -\frac{1}{32} N_f^2 g^6-\frac{3}{512} N_f^3 g^8 - \frac{1}{1024}\left[\frac{19}{12}N_f^4 + 3\zeta_3N_f^2 \right]g^{10}
    + \calO\left(g^{12}\right)\,,
\end{split}
\end{align} 
where $\zeta_3\approx 1.202$ is the Riemann zeta function evaluated at $3$.
The UV fixed point is thus located at 
\begin{equation}
    g_*^2\ =\ \frac{4}{N_f}\epsilon - \frac{2}{N_f}\epsilon^2 + \frac{1}{2N_f}\epsilon^3 
    - \left(\frac{1}{3N_f}
    +\frac{3\zeta_3}{N_f^3}\right)\epsilon^4 
    + \calO(\epsilon^5)
\end{equation}
The scaling dimension of the energy operator $\calO_E$ is related to the critical exponent $\nu$ that governs the divergence of the correlation length as the UV fixed point:
\begin{equation}\label{eq:Delta_energy}
\begin{split}
    \Delta_E(\epsilon)
    \ =\ 
    2+\epsilon - \frac{1}{\nu}
    \ =\ 
    2 + \epsilon +\beta'(g^2_{*}) \ =\ 
    2 - \frac{1}{2}\epsilon^2 - \frac{1}{4}\epsilon^3 
    - \left( \frac{5}{16} + \frac{9\zeta_3}{4N_f^2} \right)\epsilon^4
    + \calO(\epsilon^5)
    \,.
\end{split}
\end{equation}
This series appears to be non-alternating. Curiously, $N_f$-dependence only
shows up from the fourth order.

The anomalous dimension of the vertex operator $\calO_{U}$ in the $\boldsymbol{N_f}\otimes\overline{\boldsymbol{N_f}}$ representation of $SU(N_f)_L\times SU(N_f)_R$ is known up to the three-loop order~\cite{McKane:1979cm} (their coupling $g^2$ in the Lagrangian is the same as ours):
\begin{align}
    \gamma_{U}(g^2) \  = \ \frac{\left(N_f^2-1\right)}{4 N_f} \left(g^2 
    + \frac{3 N_f^2}{128}  g^6\right) +\calO\left(g^8\right)\,.
\end{align}
This $\gamma_{U}$ is the anomalous dimension with respect to the engineering dimension $0$.
Therefore, this $\gamma_{U}$ is equivalent to $\frac{1}{2}\zeta$ used in Ref.~\cite[Section~4.1.3]{Henriksson:2022rnm} (see especially their Footnote~35) for the $O(N)$ nonlinear sigma model.
Hence the scaling dimension of $\calO_{U}$ at the UV fixed point is given by
\begin{align}
    \label{eq:Delta_U}
    \Delta_{U}(\epsilon)
    \ =\ \gamma_{U}(g_*^2)
    \ = \ \left(1-\frac{1}{N_f^2}\right)\left[\epsilon 
    - \frac{1}{2}  \epsilon^2 +  \frac{1}{2} \epsilon^3\right] 
    +\calO\left(\epsilon ^4\right)  \,.
\end{align}
This series appears to be alternating.
The factorization of the $N_f$-dependence is unlikely to hold from the fourth order.

To extrapolate $\Delta_E(\epsilon)$ and $\Delta_{U}(\epsilon)$ to $\epsilon=1$, we need to deal with a standard challenge.
Since there are no non-trivial fixed points for $\epsilon<0$, Dyson's
argument~\cite{Dyson:1952tj} suggests that the perturbative expansion in $\epsilon$
should be asymptotic.  Simply plugging $\epsilon = 1$ into a truncated series representation of $\Delta_E(\epsilon)$ and $\Delta_{U}(\epsilon)$ is thus likely to be a very poor approximation.  However, it is natural to expect observables like
$\Delta_{E}(\epsilon), \Delta_{U}(\epsilon)$ to have resurgent transseries
representations around $\epsilon = 0$, see e.g.~Ref.~\cite{Aniceto:2018bis} for
a review.  The standard way to extrapolate asymptotic series is via Borel
summation of an analytic continuation of the Borel transform of the series. Here
we will adopt the
simplest and most widely used analytic continuation, which is via Pad\'e
approximants, which are rational functions that match the series expansion to a
given order.  We will use this Pad\'e-Borel (PB) resummation approach to estimate the values of $\Delta_E(\epsilon)$ and $\Delta_{U}(\epsilon)$ at $\epsilon=1$.  To estimate the errors on our estimates, we will use combinations of the variation of our results as we vary the choice of Pad\'e approximant together  with the non-perturbative ambiguities in the resummations (when such ambiguities appear).

\subsection[Extrapolation of Delta-energy]{Extrapolation of $\Delta_E$}
\label{sec:delta_energy}
Replacing $\epsilon^n$ with $\delta^n/n!$, we obtain the Borel transform of the truncated $\Delta_E(\epsilon)$ series~\eqref{eq:Delta_energy}:
\begin{align}\label{eq:Borel_energy}
    \calB[\Delta_E](\delta)\ = 2-\frac{\delta^2}{4}
    - \frac{\delta^3}{24}
    - \left( \frac{5}{16} + \frac{9\zeta_3}{4N_f^2} \right)\frac{\delta^4}{24}
    +\calO\left(\delta^5\right)\,.
\end{align}
If $\Delta_E(\epsilon)$ is indeed non-alternating, we would expect $\calB[\Delta_E](\delta)$ to have a leading singularity on the positive real axis.
Such a singularity leads to imaginary ambiguities in the Borel integral that must be canceled by appropriate nonperturbative contributions in a resurgent transseries representation of $\Delta_E(\epsilon)$.

An $[m/n]$ Pad\'e approximant to $\calB[\Delta_E](\delta)$ is a degree-$(m,n)$ rational function that matches the series expansion of $\calB[\Delta_E](\delta)$ to order $\delta^{m+n}$. 
The near-diagonal Pad\'e approximants at order $\delta^4$ are
\begin{subequations}
\begin{align}
    \calB[\Delta_E](\delta)\big|_{[3/1]} &= 
    \frac{2 - \Bigl(\frac{5}{8}\!+\!\frac{9\zeta_3}{2N_f^2}\Bigr)\delta - \frac{1}{4}\delta^2 + \Bigl(\frac{7}{192}\!+\!\frac{9\zeta_3}{16N_f^2}\Bigr)\delta^3 }
    {1 - \Bigl(\frac{5}{16}\!+\!\frac{9\zeta_3}{4N_f^2}\Bigr)\delta}\,,\\
    \calB[\Delta_E](\delta)\big|_{[2/2]} &= 
    \frac{2 - \frac{1}{3}\delta - \Bigl(\frac{43}{144}\!+\!\frac{3\zeta_3}{4N_f^2}\Bigr)\delta^2}
    {1 - \frac{1}{6}\delta - \Bigl(\frac{7}{288}\!+\!\frac{3\zeta_3}{8N_f^2}\Bigr)\delta^2}\,,\\
    \calB[\Delta_E](\delta)\big|_{[1/3]} &= 
    \frac{2 - \Bigl(\frac{17}{8}\!+\!\frac{9\zeta_3}{2N_f^2}\Bigr)\delta }
    {1 - \Bigl(\frac{17}{16}\!+\!\frac{9\zeta_3}{4N_f^2}\Bigr)\delta + \frac{1}{8}\delta^2 - \Bigl(\frac{43}{384}\!+\!\frac{9\zeta_3}{32N_f^2}\Bigr)\delta^3}\,.
\end{align}
\end{subequations}
All of them have poles on the positive real axis.
We then obtain the $[m/n]$ PB-resummed function $\Delta_E(\epsilon)$ by performing the Borel integral,
\begin{align}
    \Delta_E(\epsilon)\big|_{[m/n]} = \frac{1}{\epsilon}\int_0^{\infty} d \delta\, \exp\left(-\delta e^{i \theta}/\epsilon \right)\, \calB[\Delta_E]( \delta e^{i \theta} )\big|_{[m/n]}\,.
\end{align}
where $\theta \ll 1$.  In a resurgent transseries representation of $\Delta_E(\epsilon)$, $\im \Delta_E(\epsilon)\big|_{[m/n]}$ would be canceled by a corresponding imaginary part of a non-perturbative term in the transseries, which would also give a real contribution of approximately the same magnitude.  We thus use the magnitude of the imaginary ambiguities in the Pad\'e-Borel resummations as an estimate of the error on our extrapolations.

\begin{figure*}[ht]
  \centering
  \includegraphics[width=0.5\textwidth]{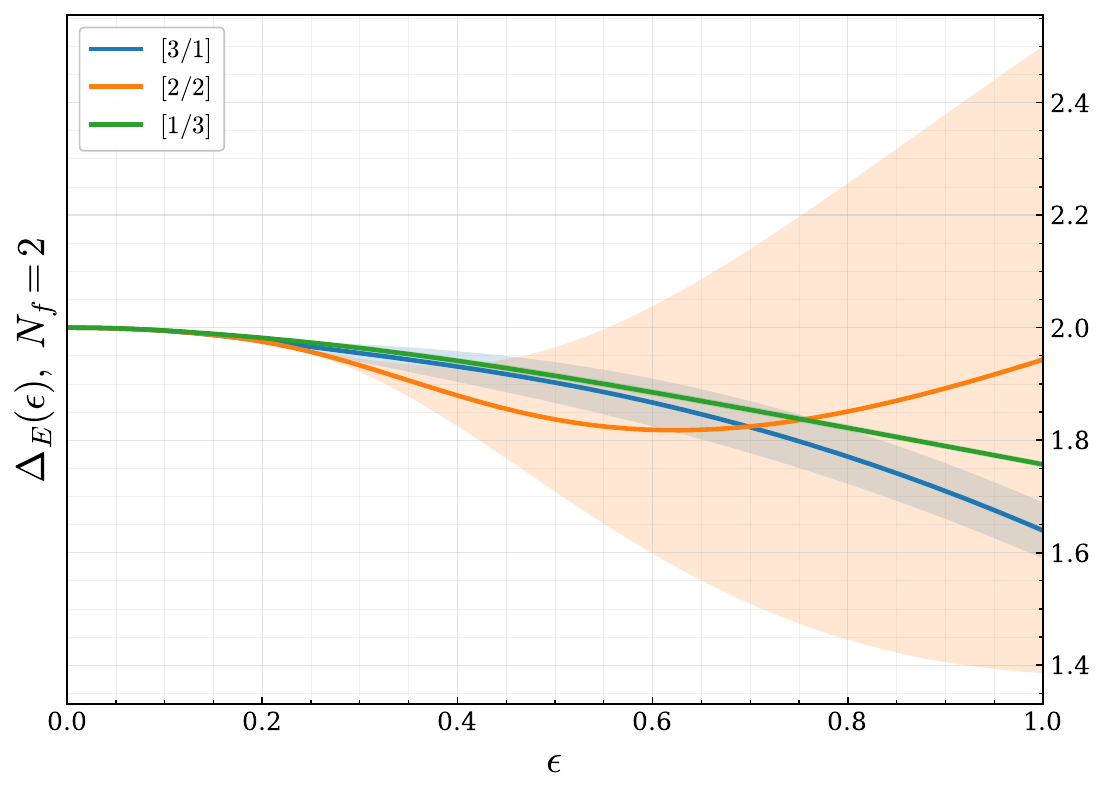}%
  \hfill
  \includegraphics[width=0.5\textwidth]{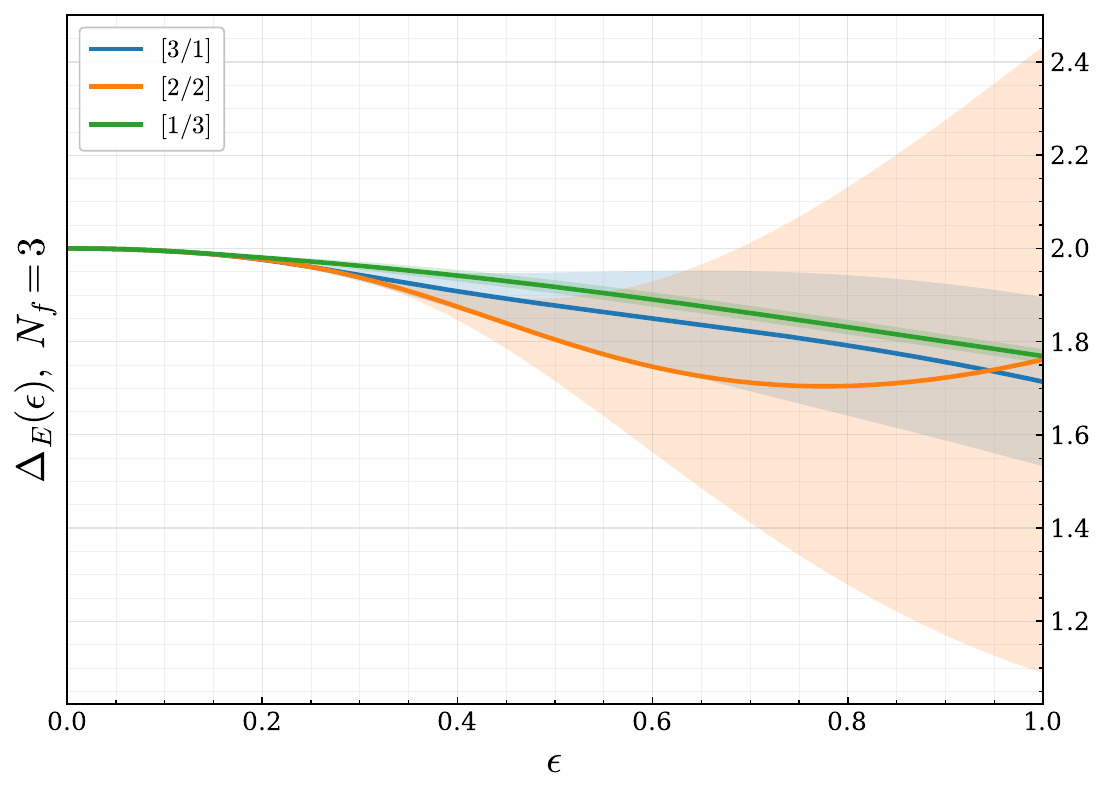}%
  \hfill
  \\
  \includegraphics[width=0.5\textwidth]{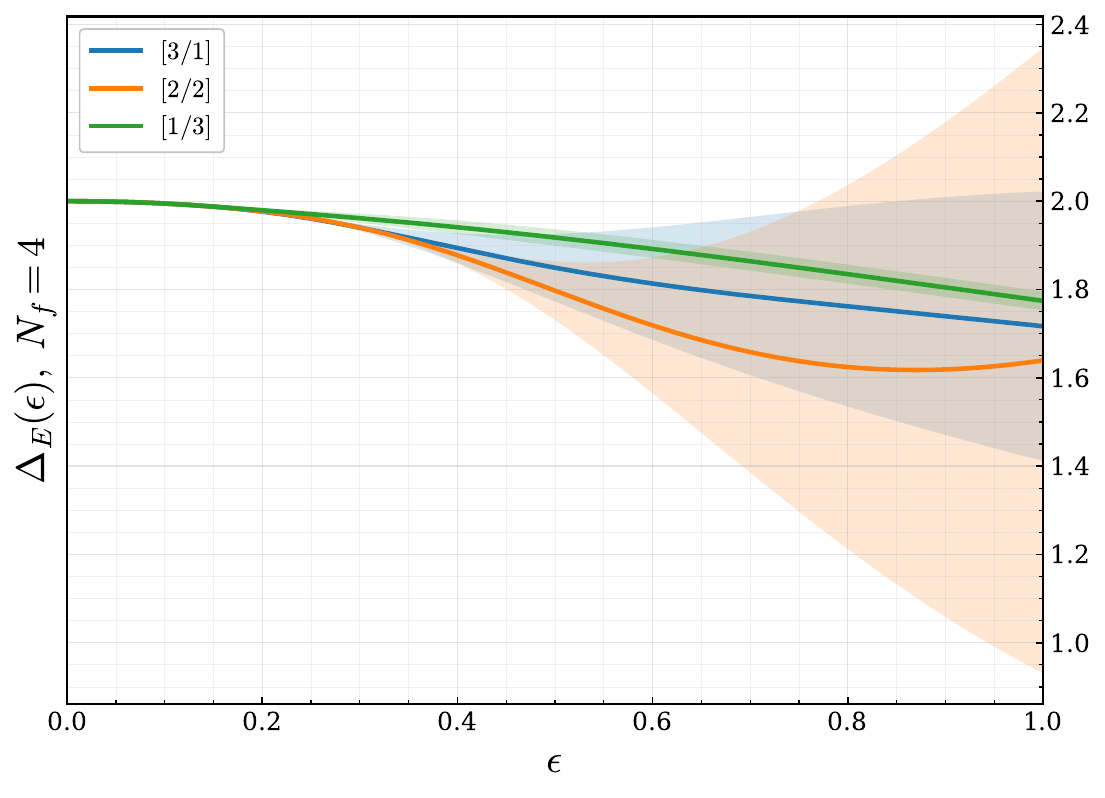}%
  \hfill
  \includegraphics[width=0.5\textwidth]{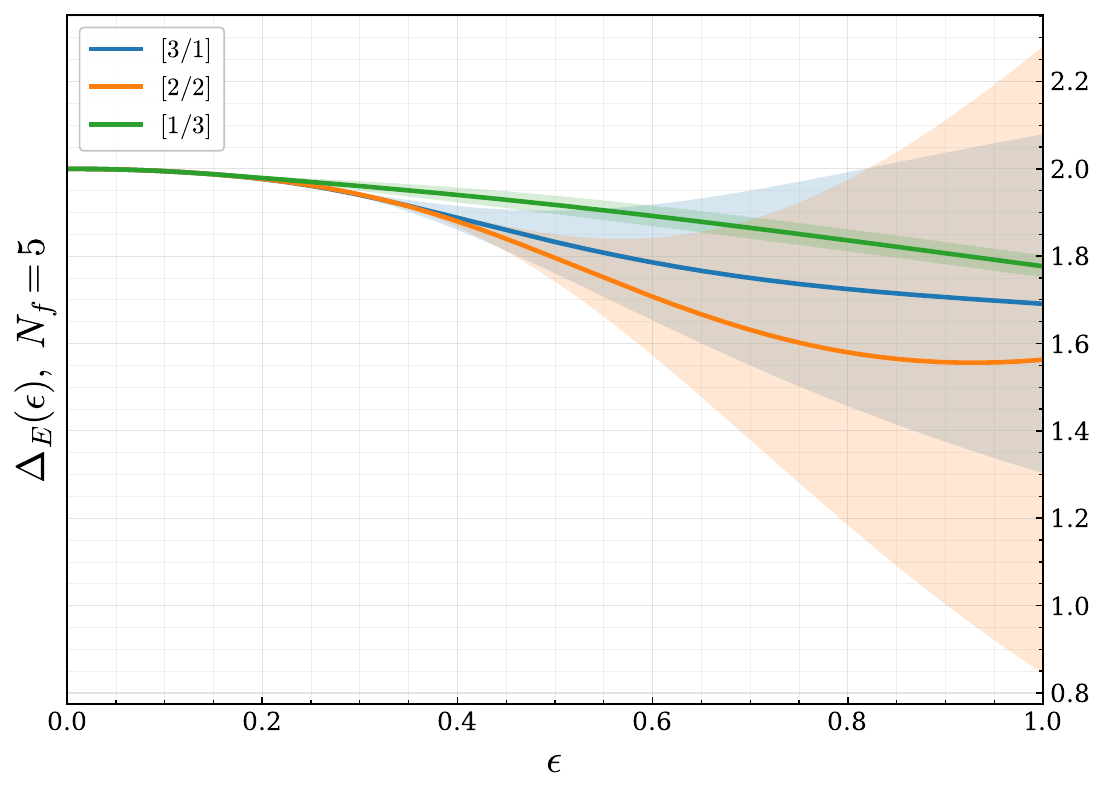}%
  \hfill
  \\
  \includegraphics[width=0.5\textwidth]{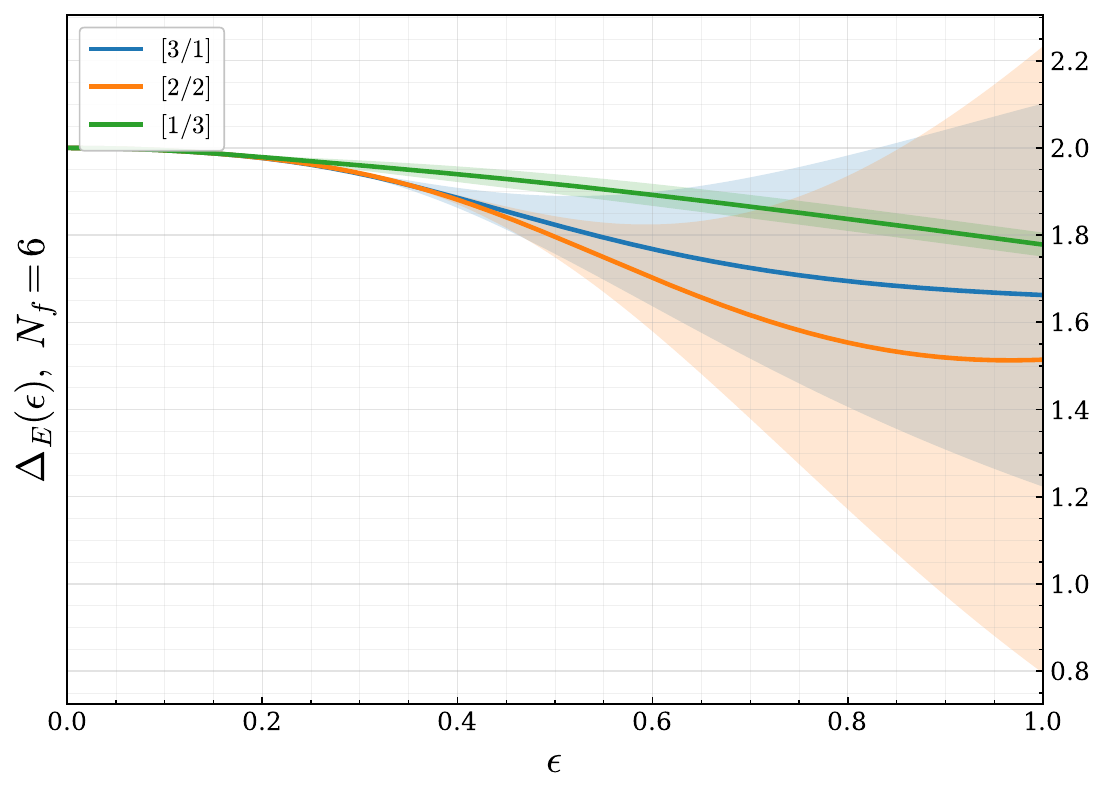}%
  \hfill
  \includegraphics[width=0.5\textwidth]{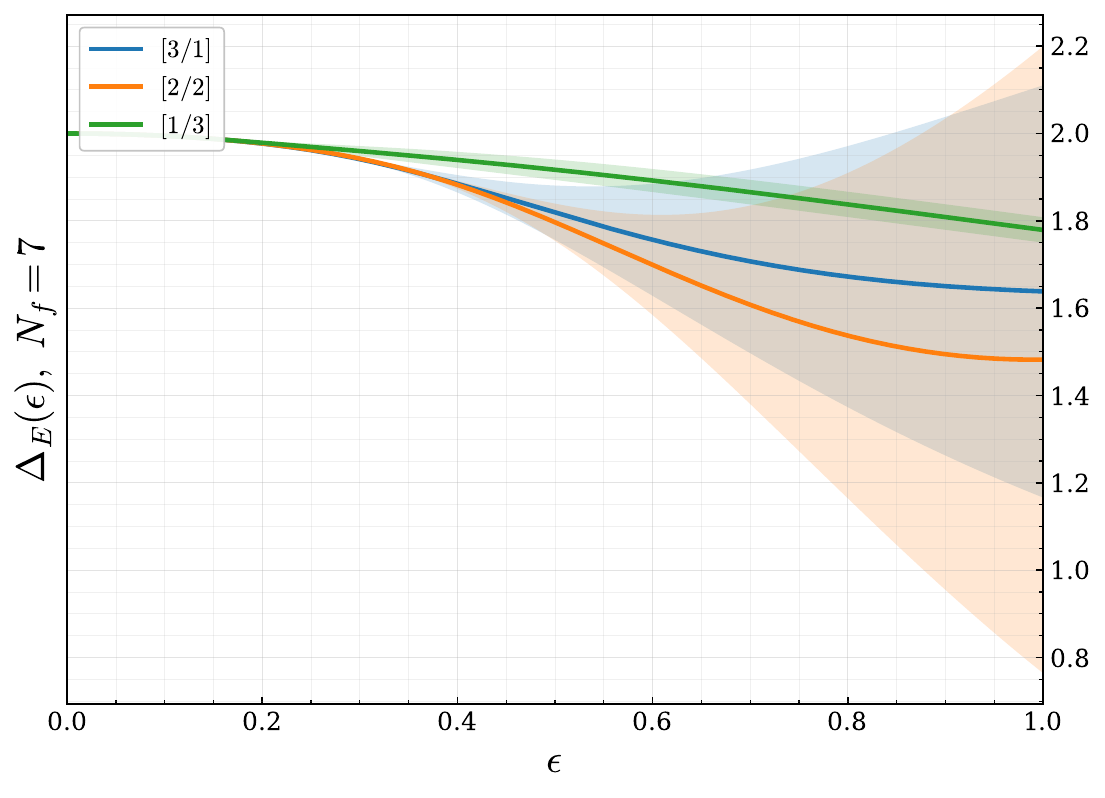}%
  
  \caption{Evaluations of the PB-resummed function $\Delta_E(\epsilon)$ with $N_f=2\sim7$. The solid lines show $\mathrm{Re}\,\Delta_E(\epsilon)$ while the shadow bands indicate error bands determined from the imaginary ambiguities in PB resummation.}
  \label{fig:Pade_Borel_E}  
\end{figure*}

We show the functions $\Delta_E(\epsilon)|_{[m/n]}$ with $N_f=2\sim7$ in Fig.~\ref{fig:Pade_Borel_E}.
The solid lines show the real parts while the extrapolation errors estimated using the procedure described above. We summarize our $\epsilon=1$ results as a function of $[m/n]$ and $N_f$ in Table~\ref{tab:Pade_sums_E}.

\begin{table}[ht]
    \centering
    \renewcommand{\arraystretch}{1.5}
    \begin{tabular}{|c|c|c|c|c|}
        \hline
        $[m/n]$ & $[3/1]$ & $[2/2]$ & $[1/3]$ \\
        \hline
        $\Delta_E(1),\ {N_f=2}$ & $1.64\pm \,0.05$ & $1.94\pm \,0.56$ & $1.76\pm \,0.01$ \\
        \hline
        $\Delta_E(1),\  {N_f=3}$ & $1.71\pm \,0.18$ & $1.76\pm \,0.67$ & $1.77\pm \,0.01$ \\
        \hline
        $\Delta_E(1),\  {N_f=4}$ & $1.72\pm \,0.31$ & $1.64\pm \,0.71$ & $1.77\pm \,0.02$ \\
        \hline
        $\Delta_E(1),\  {N_f=5}$ & $1.69\pm \,0.39$ & $1.56\pm \,0.72$ & $1.78\pm \,0.03$ \\
        \hline
        $\Delta_E(1),\  {N_f=6}$ & $1.66\pm \,0.44$ & $1.51\pm \,0.72$ & $1.78\pm \,0.03$ \\
        \hline
        $\Delta_E(1),\  {N_f=7}$ & $1.64\pm \,0.47$ & $1.48\pm \,0.72$ & $1.78\pm \,0.03$ \\
        \hline
    \end{tabular}
    \caption{Evaluations of PB-resummed $\Delta_E(1)$ with $N_f=2\sim7$. } 
    \label{tab:Pade_sums_E}
\end{table}

We use the union of the envelopes of $\Delta_E(1)|_{[3/1]}$, $\Delta_E(1)|_{[2/2]}$ to determine the error on our extrapolation of $\Delta_E$ in $D=3$, as we summarize in Table~\ref{tab:Pade_sums_E'}.
The extrapolation errors are large with the currently available four-loop data, and thus higher-loop data would be very valuable.

\begin{table}[ht]
    \centering
    \renewcommand{\arraystretch}{1.5}
    \begin{tabular}{|c|c|c|c|c|c|c|}
        \hline
        $N_f$ & $2$ & $3$ & $4$ & $5$ & $6$ & $7$\\
        \hline
        $\Delta_E$ & $1.94\pm 0.56$ & $1.76\pm 0.67$ & $1.64\pm 0.71$ & $1.56\pm 0.72$ & $1.51\pm 0.72$ & $1.48\pm 0.72$ \\
        \hline
    \end{tabular}
    \caption{Estimates of 3D $\Delta_E$ obtained from the PB resummation of the four-loop truncated $(2+\epsilon)$D perturbative expansion, with $N_f=2\sim7$.} 
    \label{tab:Pade_sums_E'}
\end{table}

\subsection[Extrapolation of Delta-phi]{Extrapolation of $\Delta_{U}$}
\label{sec:delta_U}
\begin{figure*}[ht]
  \centering
  \includegraphics[width=0.5\textwidth]{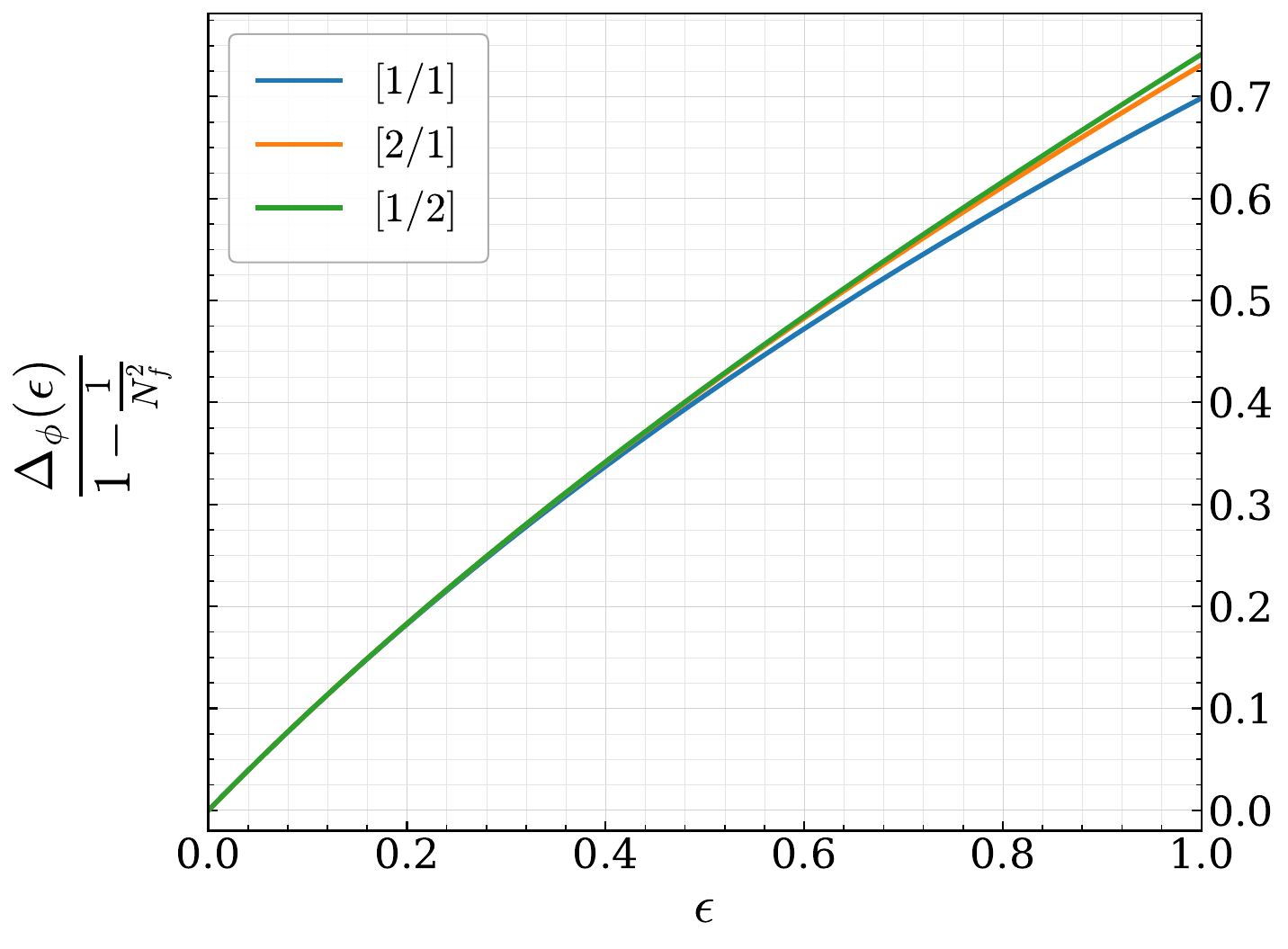}%
  
  \caption{Evaluations of PB-resummed $\Delta_{U}(\epsilon)\big/\bigl(1-\frac{1}{N_f^2}\bigr)$. The solid lines show $\mathrm{Re}\,\Delta_{U}(\epsilon)\big/\bigl(1-\frac{1}{N_f^2}\bigr)$ and the shadow bands indicate $\pm\mathrm{Im}\,\Delta_{U}(\epsilon)\big/\bigl(1-\frac{1}{N_f^2}\bigr)$.  The imaginary ambiguities only appear for $[1/2]$ and are too small to be visible.}
  \label{fig:Pade_Borel_U}  
\end{figure*}

Replacing $\epsilon^n$ with $\delta^n/n!$, we obtain the Borel transform of the truncated $\Delta_{U}(\epsilon)$ series~\eqref{eq:Delta_U}:
\begin{align}
    \calB[\Delta_{U}](\delta)\ =\ \left(1-\frac{1}{N_f^2}\right)\left[\delta-\frac{\delta^2}{4}
    +\frac{\delta^3}{12}\right]+\calO\left(\delta^4\right) \,.
\end{align}
If $\Delta_{U}(\epsilon)$ is indeed alternating, we would expect $\calB[\Delta_{U}](\delta)$ to have a leading singularity on the negative real axis, and the Borel integral might be free of imaginary ambiguities. 
The near-diagonal Pad\'e approximants up to order $\delta^3$ are
\begin{subequations}
\begin{align}
    \calB[\Delta_{U}](\delta)\big|_{[1/1]} &= 
    \left(1-\frac{1}{N_f^2}\right)
    \frac{\delta}{1+\frac{1}{4}\delta}\,,\\
    \calB[\Delta_{U}](\delta)\big|_{[2/1]} &= 
    \left(1-\frac{1}{N_f^2}\right)
    \frac{\delta + \frac{1}{12}\delta^2}{1+\frac{1}{3}\delta}\,,\\
    \calB[\Delta_{U}](\delta)\big|_{[1/2]} &= 
    \left(1-\frac{1}{N_f^2}\right)
    \frac{\delta}{1+\frac{1}{4}\delta-\frac{1}{48}\delta^2}\,.
\end{align}
\end{subequations}
The $[1/1]$ and $[2/1]$ Pad\'e approximants have no singularities on the positive real axis, while the $[1/2]$ Pad\'e approximant has a pole at $\delta\approx 15.2 \gg 0$.
We then obtain the $[m/n]$ PB-resummed function $\Delta_{U}(\epsilon)$ by performing the Borel integral,
\begin{align}
    \Delta_{U}(\epsilon)\big|_{[m/n]} =  \frac{1}{\epsilon}  \int_0^{\infty} d \delta\, e^{-\delta/\epsilon}\, \calB[\Delta_{U}](\delta)\big|_{[m/n]}\,.
\end{align}
We show the functions $\Delta_{U}(\epsilon)|_{[m/n]}\big/\bigl(1-\frac{1}{N_f^2}\bigr)$ in Fig.~\ref{fig:Pade_Borel_U}.
The solid lines show the real parts while the shadow bands indicate the imaginary ambiguities, which only appear for $[1/2]$ and are essentially invisible.  (The imaginary ambiguities only become appreciable for $\epsilon \gtrsim 2$.) 
We summarize the results at $\epsilon=1$ in Table~\ref{tab:Pade_sums_U}.

\begin{table}[ht]
    \centering
    \renewcommand{\arraystretch}{1.5}
    \begin{tabular}{|c|c|c|c|c|}
        \hline
        $[m/n]$ & $[1/1]$ & $[2/1]$ & $[1/2]$ \\
        \hline
        $\Delta_{U}(1)\big/\bigl(1-\frac{1}{N_f^2}\bigr)$ & $0.54\pm 0.02$ & $0.64\pm 0.02$ & $0.67\pm 0.02$\\
        \hline
    \end{tabular}
    \caption{Evaluations of PB-resummed $\Delta_{U}(1)\big/\bigl(1-\frac{1}{N_f^2}\bigr)$. } 
    \label{tab:Pade_sums_U}
\end{table}

We take the union of the envelopes of $\Delta_{U}(1)|_{[1/1]}$,
$\Delta_{U}(1)|_{[2/1]}$, and $\Delta_{U}(1)|_{[2/1]}$ as our final
estimates of 3D $\Delta_{U}$, as we summarize in
Table~\ref{tab:Pade_sums_U}.
We would need more terms beyond the three-loop order in the $(2+\epsilon)$D expansion to get more accurate results.

\begin{table}[ht]
    \centering
    \renewcommand{\arraystretch}{1.5}
    \begin{tabular}{|c|c|c|c|c|c|c|}
        \hline
        $N_f$ & $2$ & $3$ & $4$ & $5$ & $6$ & $7$\\
        \hline
        $\Delta_{U}$ & $0.54\pm 0.02$ & $0.64\pm 0.02$ & $0.67\pm 0.02$ & $0.69\pm 0.02$ & $0.70\pm 0.02$ & $0.71\pm 0.02$ \\
        \hline
    \end{tabular}
    \caption{Estimates of 3D $\Delta_{U}$ obtained from the PB resummation of the three-loop truncated $(2+\epsilon)$D perturbative expansion, with $N_f=2\sim7$.} 
    \label{tab:Pade_sums_phi'}
\end{table}

\bibliography{non_inv}